\documentclass[10pt,aps,prc,twocolumn,floatfix,showpacs]{revtex4-1} 
\usepackage{graphicx}
\usepackage{sidecap}
\usepackage{array}
\usepackage{subfigure}
\usepackage{color}
\usepackage{lineno}
\usepackage{placeins}

\newcommand{\GeVc}{\mbox{$\mathrm{GeV} / c$}}

\begin{document}%

\title{Elliptic flow of identified hadrons in Au+Au collisions at $\sqrt{s_{NN}}=$ 7.7--62.4 GeV}

\author{
L.~Adamczyk$^{1}$,
J.~K.~Adkins$^{23}$,
G.~Agakishiev$^{21}$,
M.~M.~Aggarwal$^{34}$,
Z.~Ahammed$^{53}$,
I.~Alekseev$^{19}$,
J.~Alford$^{22}$,
C.~D.~Anson$^{31}$,
A.~Aparin$^{21}$,
D.~Arkhipkin$^{4}$
% ...
(STAR Collaboration)
}

\address{$^{1}$AGH University of Science and Technology, Cracow, Poland}
% ...
\address{$^{4}$Brookhaven National Laboratory, Upton, New York 11973, USA}
% ...
\address{$^{19}$Alikhanov Institute for Theoretical and Experimental Physics, Moscow, Russia}
% ...
\address{$^{21}$Joint Institute for Nuclear Research, Dubna, 141 980, Russia}
\address{$^{22}$Kent State University, Kent, Ohio 44242, USA}
\address{$^{23}$University of Kentucky, Lexington, Kentucky, 40506-0055, USA}
% ...
\address{$^{31}$Ohio State University, Columbus, Ohio 43210, USA}
% ...
\address{$^{34}$Panjab University, Chandigarh 160014, India}
% ...
\address{$^{53}$Variable Energy Cyclotron Centre, Kolkata 700064, India}
% ...



\begin{abstract}
Measurements of the elliptic flow, $v_{2}$, of identified hadrons ($\pi^{\pm}$, $K^{\pm}$, $K_{s}^{0}$, $p$, $\bar{p}$,
$\phi$, $\Lambda$, $\overline{\Lambda}$, $\Xi^{-}$, $\overline{\Xi}^{+}$, $\Omega^{-}$, $\overline{\Omega}^{+}$) in Au+Au collisions 
at $\sqrt{s_{NN}}=$ 7.7, 11.5, 19.6, 27, 39 and 62.4 GeV are presented.  
The measurements were done at mid-rapidity using the Time Projection Chamber
and the Time-of-Flight detectors of the STAR experiment during the Beam Energy Scan program at RHIC.
A significant difference in the $v_{2}$ values for particles and the corresponding
anti-particles was observed at all transverse momenta for the first time. The difference increases with decreasing
center-of-mass energy, $\sqrt{s_{NN}}$ (or increasing baryon chemical potential, $\mu_{B}$) and is
larger for the baryons as compared to the mesons. This implies that particles and
anti-particles are no longer consistent with the universal number-of-constituent quark (NCQ) scaling of
$v_{2}$ that was observed at $\sqrt{s_{NN}}=$ 200 GeV. However, for the group of particles NCQ scaling at $(m_{T}-m_{0})/n_{q}>$ 0.4 GeV/$c^{2}$ is not violated within $\pm$10\%. The $v_{2}$ values for $\phi$
mesons at 7.7 and 11.5 GeV are approximately two standard deviations from the trend defined by the other hadrons at the
highest measured $p_{T}$ values.
\end{abstract}

\maketitle
%

\section{Introduction} 
\label{sec_intro}

One of the main goals of the heavy ion collision program at the
Relativistic Heavy Ion Collider (RHIC) facility is to produce a state of
deconfined quarks and gluons, called the Quark-Gluon Plasma (QGP), and to
study its properties. An experimental way to understand the formation of the
QGP is by varying collision energies and studying
observables as a function of collision centrality, transverse
momentum, $p_{T}$, and rapidity, $y$.  This also allows one to study 
the structure of the QCD phase diagram. With these goals, the Beam Energy Scan
(BES) program was started in the years 2010 and 2011 at RHIC~\cite{Aggarwal:2010cw} where
Au+Au collisions were recorded at $\sqrt{s_{NN}}$ = 7.7, 11.5, 19.6, 27, 39 and 62.4 GeV. 
This paper reports the azimuthal anisotropy of identified
particles produced in collisions at BES energies, measured using the STAR detector at RHIC.

The azimuthal anisotropy of produced particles is one of the most
widely studied observables. In non-central heavy ion collisions, the
overlap region of the colliding nuclei is almond-shaped
and perpendicular to the plane defined by the impact
parameter vector and the beam axis. This plane is called the reaction plane. Due to finite number fluctuations of participating nucleons in reactions with the same centrality, 
the geometric symmetry plane in each event is not necessarily the same as the reaction plane, and is often called the participant plane. 
This plane is defined by the 
nucleons which participated in the reaction~\cite{Kharzeev:2000ph}. In a hydrodynamic
approach with local thermalization, the initial spatial anisotropy and subsequent interactions
among the constituents result in pressure gradients that are larger in
the direction of the participant plane compared to out of this plane. This results in an azimuthal
anisotropy of the momenta of the produced particles~\cite{Snellings:2011sz}. The second
harmonic parameter, $v_{2}$, of the Fourier decomposition of the 
azimuthal particle distribution relative to the event plane is called the
elliptic flow~\cite{Voloshin:1994mz,Poskanzer:1998yz}. It is experimentally measured using final-state particle momenta. 
The event plane is an approximation to the participant plane. The elliptical anisotropy with respect to the event plane is not necessarily equal to the elliptic flow with respect to the participant plane. In the literature, the magnitude of the second flow harmonic is called $v_2$ whether this quantity is calculated from the participant (or reaction) plane or the event plane. 
The expansion of the system and subsequent decrease of the spatial anisotropy leads to a self-quenching process for
$v_{2}$, thereby making it a sensitive probe of the early stage of
heavy ion collisions~\cite{Ollitrault:1992bk,Sorge:1998mk}.

Using the data from the top RHIC energy of 200 GeV several
interesting observations related to $v_{2}$ have been reported in the past decade~\cite{Voloshin:2008dg,Borghini:2004ra,Sorensen:2009cz,Adams:2005dq,Adcox:2004mh}. Large values of the elliptic flow were found to be
compatible with ideal hydrodynamic calculations~\cite{Huovinen:2001cy,Nonaka:2003hx,Hirano:2003pw} or viscous
hydrodynamic calculations~\cite{Shen:2012vn,Schenke} with a small shear viscosity to entropy density
ratio.  At low transverse momentum ($p_{T} < 2$ \GeVc), a mass ordering
of the $v_{2}$ values was observed~\cite{Adler:2001nb,Adams:2003am,Adams:2005zg}, which could readily be understood within a
hydrodynamic framework. At intermediate $p_{T}$ values ($2 < p_{T} < 6$ \GeVc),  a Number-of-Constituent Quark (NCQ) scaling of $v_{2}$ for the identified
hadrons was observed. This observation, coupled with the comparable values of the 
elliptic flow measured for multi-strange hadrons ($\phi$ and $\Xi$) and light quark 
hadrons, was used to conclude that the relevant degrees of freedom in the systems formed at the top
RHIC energy are quarks and gluons~\cite{Abelev:2007rw,Abelev:2010tr,Adams:2005zg,Abelev:2007qg,Voloshin:2008dg}. 
It was also concluded that a substantial amount of $v_{2}$ was generated during the partonic stage of these collisions. 
This was further corroborated by comparing the measurements to model
calculations with and without partonic interactions.

It is generally expected that the system will spend less time in the partonic phase
as the beam energy is lowered, and that at the lowest BES energies the system might not reach the QGP regime.
In such a scenario, it is expected that NCQ scaling of $v_{2}$
of produced particles would be broken~\cite{BodNu}. Furthermore, with decreasing
beam energy, the baryon chemical potential of the system at chemical
freeze-out increases.
These aspects could lead to new trends in the identified hadron $v_{2}$ in the BES program at RHIC which was
performed at the BES energies with unmatched statistics and particle identification capabilities.
In this paper, the STAR measurements of the beam
energy and $p_{T}$ dependence of the elliptic flow, $v_{2}$, at mid-rapidity for $\pi^{\pm}$, 
$K^{\pm}$ , $K_{s}^{0}$, $p$, $\bar{p}$, $\phi$, $\Lambda$, $\overline{\Lambda}$, $\Xi^{-}$, $\overline{\Xi}^{+}$, $\Omega^{-}$ and
$\overline{\Omega}^{+}$ in minimum bias (0--80\%) Au+Au collisions are presented. The corresponding
results for the inclusive charged particles were reported in Ref.~\cite{Adamczyk:2012ku}.

This paper is organized as follows. Section~\ref{sec_experiment} gives a brief description
of the experimental setup and the event and centrality selection. In 
Sections~\ref{sec_PID},~\ref{sec_EP_rec}, and~\ref{sec_v2sig}, the various particle identification
methods, the event plane reconstruction, $v_{2}$ signal extraction, and
systematic uncertainty estimation are discussed. In Section~\ref{sec_results}, the energy-and-momentum-dependent $v_{2}$ results are presented. In Section~\ref{sec_discuss}, comparisons to models are discussed. Finally, the summary is presented in Section~\ref{sec_summary}.

\section{Experimental setup} 
\label{sec_experiment}

The Solenoidal Tracker At RHIC (STAR) is a multi-purpose experiment at the RHIC facility at Brookhaven National Laboratory. It consists of a solenoidal magnet and an array of detectors for triggering, particle identification, and event categorization. A detailed description can be found in Ref.~\cite{STAR_TPC}. The primary detectors used for the present results are summarized in the following subsections. 

\subsection{Time-projection chamber (TPC)}
The TPC has a full azimuthal, $\phi$, coverage and a pseudorapidity, $\eta$, acceptance of $-1.8<\eta<1.8$~\cite{Anderson:2003ur}. The TPC is split into two halves along the beam direction by a central membrane. A maximum of 45 hit points per track can be reconstructed within the TPC radius limits of $0.5 < r < 2$ m. The primary collision vertex of an event is fitted using the reconstructed particle tracks. For $\sim$1000 such tracks, a primary vertex resolution of 350 $\mu m$ can be achieved. The primary vertex position is used in a subsequent track refitting for particles like $\pi$, $K$ and $p$ to improve the momentum resolution. The relative momentum resolution for pions is $\sim$3\% at $p_{T} = 1$ \GeVc. The specific energy loss ($dE/dx$) information, also provided by the TPC, can be used for particle identification ({\it cf.} Section~\ref{sec_PID}). 

\subsection{Time-of-Flight (TOF)}
The time-of-flight system is based on Multi-gap Resistive Plate Chambers (MRPCs) and was fully installed in STAR in the year 2010~\cite{TOF}. The system has an intrinsic timing resolution of $\sim$85 ps. It covers the full azimuth and a pseudorapidity range of $-0.9 < \eta < 0.9$. The matching efficiency of a TPC-reconstructed track to an MRPC cell is $\sim$90\%, which results in a total efficiency (acceptance$\times$efficiency) of $\sim$65\%. The particle mass-squared, ${\it m}^{2}$,  can be calculated using the measured time-of-flight and the reconstructed momentum from the TPC. Examples of the ${\it m}^{2}$ distributions are shown in Section~\ref{sec_PID}.

\subsection{Trigger and event selection}
In the years 2010 and 2011, Au+Au collisions at the six energies, $\sqrt{s_{NN}}$, of 7.7, 11.5, 19.5, 27, 39 and 62.4 GeV were measured. The minimum bias trigger condition for all six energies was based on a coincidence of the signals from the Zero Degree Calorimeters (ZDC), Vertex Position Detectors (VPD), and/or Beam-Beam Counters (BBC). Most of the triggered events at the lowest beam energies did not originate from Au+Au collisions, but were rather Au plus beam pipe (or other material) collisions. This was the result of the large beam emittance at the lowest beam energies. The radius of the beam pipe is 3.95 cm. The background due to these ``fixed target" events was efficiently removed in the present analysis by requiring that the primary vertex position was within a radius $r$ of less than 2 cm. The $z$-position of the primary vertices was limited to the values listed in Table.~\ref{table_1}. These values depend on the offline $z$-vertex trigger conditions which were different for the different energies. These vertex cuts were studied and optimized during the data-taking using the online vertex reconstruction performed by the High Level Trigger (HLT).

To remove pile-up events, it was required that at least two tracks from the primary vertex were matched to the cells of the time-of-flight detector. Furthermore, an extensive quality assurance of the events was performed based on the mean transverse momenta, the mean vertex position, the mean interaction rate, and the mean multiplicity in the detector. Run periods were removed if one of those quantities was several $\sigma$ away from the global mean value. The accepted number of minimum bias events for each of the six energies are also listed in Table.~\ref{table_1}.

\subsection{Centrality definition}  

\begin{figure*}[]
\centering
\resizebox{17cm}{!}{%
\includegraphics{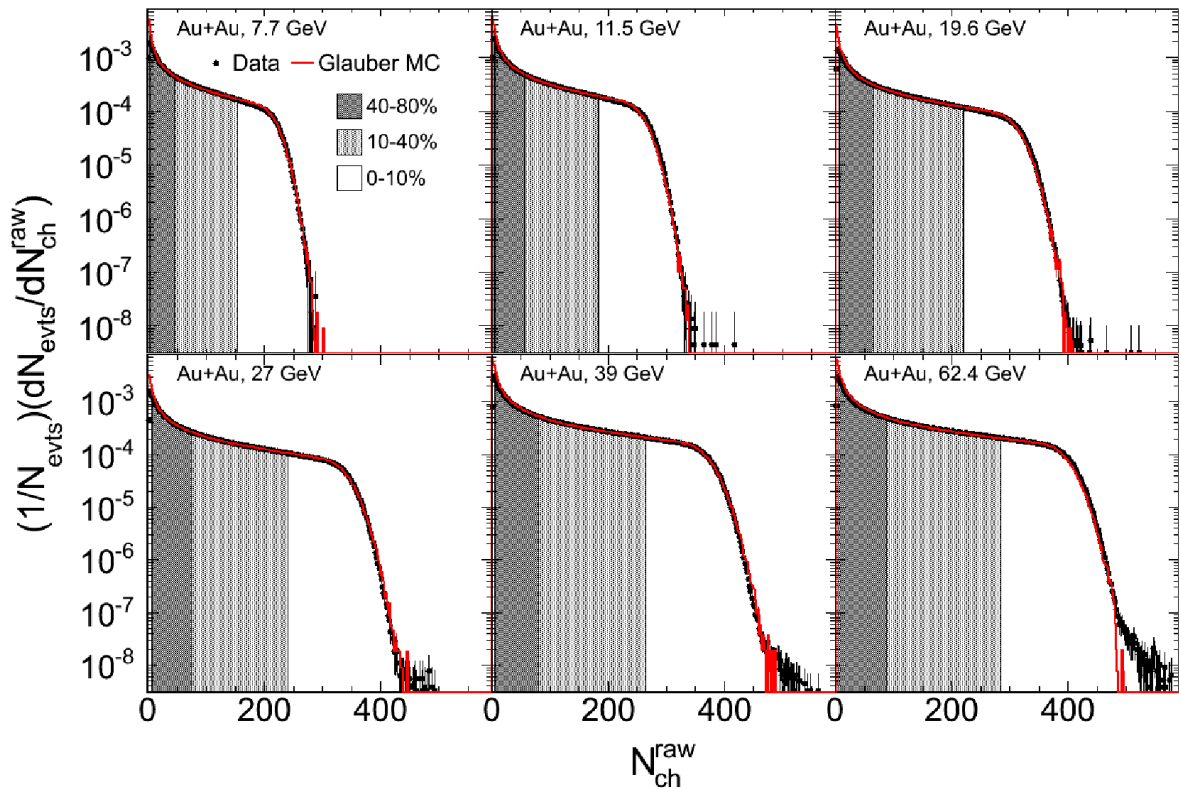}}
\caption{(Color online) The uncorrected multiplicity, $N^{\rm raw}_{\rm ch}$, distribution of reconstructed charged particles per unit pseudorapidity interval at mid-rapidity for the six different beam energies. The filled black points depict the measured data and a Glauber Monte Carlo simulation is overlayed as the solid curve. Three different centrality classes are indicated by the different shaded regions.}
\label{fRefMult}       
\end{figure*}

The centrality selection of the events was chosen to be 0--80\% of the total reaction cross section. The centrality definition was based on the uncorrected multiplicity distribution, $dN_{\rm evts}/dN^{\rm raw}_{\rm ch}$, of reconstructed charged particle tracks within a pseudorapidity range of $|\eta|$ $<$0.5. The distributions for all energies can be accurately described by a 2 component model calculation~\cite{Kharzeev:2000ph} as shown in Fig.~\ref{fRefMult}. Some of the most peripheral events were not recorded due to trigger inefficiencies. This results in a significant difference between the measured $dN_{\rm evts}/dN^{\rm raw}_{\rm ch}$ and the Glauber Monte Carlo (MC) simulation. To correct for this effect, the ratio of the simulation to the data was used as a weighting factor for the particle yields. The correction drops from a maximum of 30\% to 5\%
by the 70\% most central bin, and is negligible for the more central data. In addition to the trigger inefficiency corrections, two additional corrections were also applied to account for the $z$-vertex dependent inefficiencies. These corrections treated the acceptance and detector inefficiencies and the time dependent changes in $dN_{\rm evts}/dN^{\rm raw}_{\rm ch}$ resulting, {\it e.g.}, from minor changes in the trigger configuration.

\begin{table*}[htc]
\caption{\label{table_1} The total number of minimum-bias (MB) events used, and the $z$-vertex acceptance, for the different energies.}
{
\footnotesize\rm
\begin{tabular}{ >{\centering\arraybackslash}m{2in}  >{\centering\arraybackslash}m{2in} >{\centering\arraybackslash}m{2in} } 
\hline
$\sqrt{s_{NN}}$ (GeV) & MB events ($10^{6}$   )&$z$-vertex range (cm)\\
\hline \hline
7.7&4.3&[-70,70]\\
11.5&11.7&[-50,50]\\
19.6&35.8&[-70,70]\\
27&70.4&[-70,70]\\
39&130.4&[-40,40]\\
62.4&67.3&[-40,40]\\
\hline
\end{tabular}
}
\end{table*}

\section{Particle identification and signal extraction}  
\label{sec_PID}

\begin{figure}[]
\resizebox{8cm}{!}{%
  \includegraphics[bb = 0 0 558 780,clip]{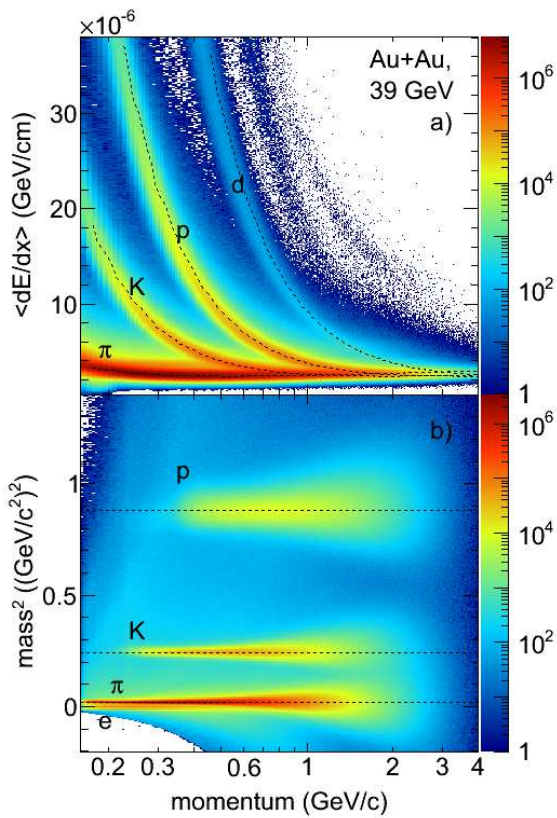}}
\caption{(Color online) The mean specific energy loss, $\langle dE/dx \rangle$, of reconstructed tracks within a pseudorapidity range of $|\eta| < 1$ in the TPC a), and the mass-squared, ${\it m}^{2}$, as a function of momentum b). The Bichsel functions ~\cite{bichsel} used to determine the $n\sigma_{\rm particle}$ values ({\it cf.} Eq.~(\ref{form_dEdx})) are shown
in a) as the dashed curves. The horizontal dashed lines in b) correspond to the nominal particle masses of $\pi$, $K$ and $p$.}
\label{fTPC_tof_pid}       
\end{figure}

\begin{figure}[]
\resizebox{6cm}{!}{%
  \includegraphics[bb = 18 44 283 675,clip]{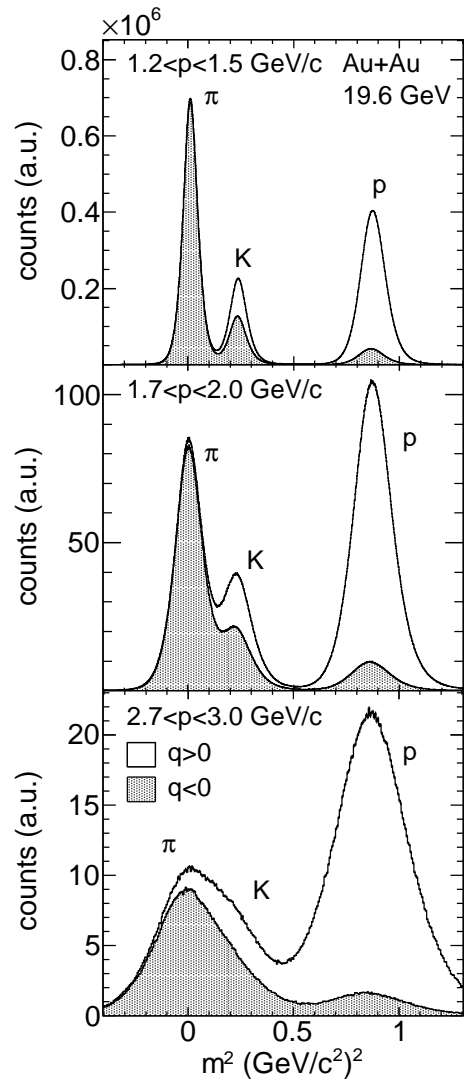}}
\caption{The mass-squared, ${\it m}^{2}$, distributions for reconstructed positive (q$>$0) and negative (q$<$0) charged particles from 0--80\% central Au+Au collisions at the beam energy of 19.6 GeV. Three different momentum ranges are shown.}
\label{fm2_19GeV}       
\end{figure}

Particle identification in the STAR experiment can be done in multiple ways. Long-lived charged particles, {\it e.g.} $\pi$, $K$ and $p$, were directly identified and reconstructed, within a pseudorapidity range of $|\eta|<1$, using the time-of-flight information and/or the specific energy loss in the TPC depending on the reconstructed track momentum. For weakly-decaying particles, {\it e.g.} $\Lambda$ and $\Xi$, the invariant mass technique and topological reconstruction methods were used. They are reconstructed within a rapidity range of $|y|<1$. The cleanest event-by-event particle identification is obtained at the lowest momenta and/or when using tight topology cuts. Statistical signal extraction methods were used to obtain the yields of the particles at higher momenta.
Up to momenta of $\sim$1.5 \GeVc\, a clean separation of $\pi$, $K$ and $p$ was obtained when combining the information from the TPC and TOF detectors. At higher momenta, the $\pi$ and $K$ signals begin to overlap. The protons still can be separated event-by-event up to $\sim$3.0 \GeVc\ by using the time-of-flight information alone. Figure~\ref{fTPC_tof_pid} shows the mean specific energy loss, $\langle dE/dx \rangle$, in the TPC and the mass-squared from the TOF as a function of the momentum. The proton, pion, and kaon $dE/dx$ bands merge for momenta above $\sim$1 \GeVc. The separation in $m^{2}$ of $\pi$, $K$ and $p$ at a beam energy of $\sqrt{s_{NN}}=$ 19.6 GeV is shown for three different momentum ranges in Fig.~\ref{fm2_19GeV}. 

In order to avoid fake tracks in the TPC and to improve the average momentum and energy loss resolution, the following track quality cuts were applied: the number of total hit points was larger than 15, and the ratio of the number of reconstructed hits to the maximum possible number of hits for each track was larger than 0.52. The momentum of each particle was limited to $0.15<p<10$ \GeVc. 
The deviation in units of $\sigma_{\rm particle}$ of $\langle dE/dx \rangle$ of a particle species from its theoretical energy loss, calculated with a Bichsel function~\cite{bichsel}, can be expressed as,
\begin{equation}
n\sigma_{\rm particle} \propto ln \left[\left< \frac{dE}{dx}\right>_{\rm particle} / \left< \frac{dE}{dx}\right>_{\rm Bichsel} \right].
\label{form_dEdx} 
\end{equation}
The distribution of $\sigma_{\rm particle}$ is nearly Gaussian for a given momentum and is properly calibrated to be centered at zero for each particle species with a width of unity. 

\subsection{Signal extraction for $\pi^{\pm}$, $K^{\pm}$, $p$, and $\bar{p}$}
\label{sec_pid_pikp}
\begin{figure*}[]
\centering
\resizebox{15cm}{!}{%
  \includegraphics{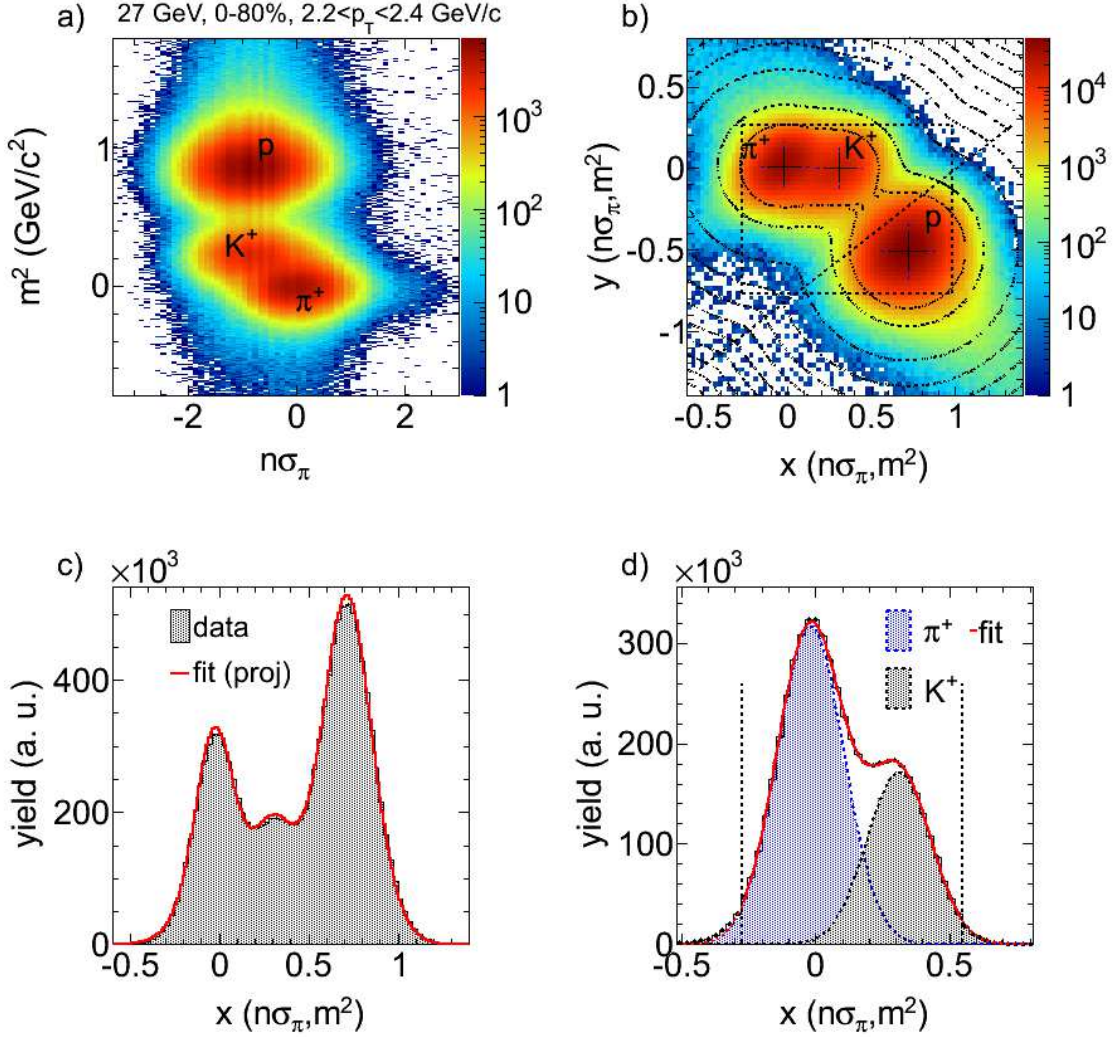}}
\caption{a) The mass-squared, $m^{2}$, versus $n\sigma_{\pi}$ and b) $x,y(n\sigma_{\pi},{\it m^{2}})$ (see Eqs.~(\ref{form_transA}) to (\ref{form_transB})) distributions for $2.2 < p_{T} < 2.4$ \GeVc\ from 0--80\% central Au+Au collisions at 27 GeV. The black dashed contour lines in b) depict the result of a simultaneous fit with three 2$\times$2D Gaussians. The diagonal dashed line depicts a cut to remove the remaining proton contamination (see text). c) The projected distribution to the $x(n\sigma_{\pi},{\it m^{2}})$ axis. The red solid curve shows the projection of the 2$\times$2D Gaussian fits. d) The same as c), but after the 2$\times$2D Gaussian of the protons was removed. The red solid line shows the sum of the two 1D Gaussian fits. The fit range is indicated by the two vertical dashed lines.}
\label{f2D_fit}       
\end{figure*}

Protons and anti-protons are identified primarily using the TOF mass-squared information. To suppress remnant contributions from pions and kaons, an additional $dE/dx$ cut of $| n\sigma_{p}|<3$ was applied. At low transverse momenta ($p_{T} <$2 \GeVc), the separation of protons relative to pions and kaons was sufficient such that all protons in a  range of $\sim 3\sigma$ around the center of the $n\sigma_{p}$ distribution are counted. At high $p_{T}$, the tails on the low mass range of the proton distributions were excluded to avoid contamination from pions and kaons. Thus, the ${\it m}^{2}$ cuts increased with the transverse momentum, $p_{T}$.

For the analysis of $\pi^{\pm}$ and $K^{\pm}$, a new technique was employed to extract the yields for each $p_{T}$ bin. This was based on a transformation of the combined TOF ${\it m}^{2}$ and TPC $dE/dx$ $n\sigma_{\pi}$ information. The goal of this transformation was to have a maximal separation between kaons and pions by transforming to a new set of variables $x,y(n\sigma_{\pi},{\it m^{2}})$ such that the widths of the particle peaks in $x$ and $y$ were identical and for which the pion and kaon peaks were aligned with the horizontal axis.
Each particle was described by two two-dimensional (2$\times$2D) Gaussians ($x,y(n\sigma_{\pi},{\it m^{2}})$), where the first Gaussian fits the peak and the second Gaussian shares the same position as the first, but the width was larger to account for the broad tail. The $\pi,K$ and $p$ peaks of the ${\it m}^{2}$ {\it vs.} $n\sigma_{\pi}$ distributions are fit simultaneously, individually for each $p_{T}$ bin with three 2$\times$2D Gaussians. The non-Gaussian tails of the $\pi,K$ and $p$ peaks along both axis were excluded from the fits. The resulting fit parameters, widths $\sigma({\it m^{2}})(\pi)$ and $\sigma(n\sigma_{\pi})(\pi)$, and peak positions, $\mu({\it m^{2}})(\pi,K)$ and $\mu(n\sigma_{\pi})(\pi,K)$, were used to first normalize the $m^{2}$ axis to the $n\sigma$ axis and then to perform a transformation which consists of a shift and a rotation. The transformations are listed in Eqs.~(\ref{form_transA} to \ref{form_transB}),

\begin{eqnarray}
f_{\rm scale} & = & \sigma_{\pi}(n\sigma_{\pi})/\sigma_{\pi}({\it m^{2}}), \label{form_transA} \\
\alpha & = & -\tanh\left( \frac{\mu_{K}({\it m^{2}})-\mu_{\pi}({\it m^{2}})}{(\mu_{K}(n\sigma_{\pi})-\mu_{\pi}(n\sigma_{\pi}))/f_{\rm scale}} \right), \\
x\prime & = & (n\sigma_{\pi}-\mu_{\pi}(n\sigma_{\pi}))/f_{\rm scale},  \\
y\prime & = & {\it m^{2}}-\mu_{\pi}({\it m^{2}}),
\end{eqnarray}

\begin{equation}
 \left( \begin{array}{c}
x(n\sigma_{\pi},{\it m^{2}}) \\
y(n\sigma_{\pi},{\it m^{2}}) 
\end{array} \right) =
\left( \begin{array}{cc}
\cos(\alpha) & -\sin(\alpha) \\
\sin(\alpha) & \cos(\alpha)
\end{array} \right)
\left( \begin{array}{c}
x\prime \\
y\prime
\end{array} \right).
\label{form_transB}
\end{equation}

Figure~\ref{f2D_fit} a) shows an example of the $m^{2}$ versus $n\sigma_{\pi}$ distributions and frame b) shows the new $x,y(n\sigma_{\pi},{\it m^{2}})$ distribution after the transformation for an intermediate transverse momentum range of $2.2<p_{T}<2.4$ \GeVc. The protons were treated as background in the $\pi^{\pm}$, $K^{\pm}$ analysis and were removed first. For this, the distributions in the new $x,y(n\sigma_{\pi},{\it m^{2}})$ frame were fitted with three 2$\times$2D Gaussians in a way analogous to that described above. To stabilize the procedure, several iterations were performed. In the final fit, only the area 2.5 $\sigma(x,y)$ away from the pion and proton peak positions was considered. The fit range and the resulting fit are shown as a dashed box and dash-dotted contour lines, respectively, in Fig.~\ref{f2D_fit} b). The projection to the $x(n\sigma_{\pi},{\it m^{2}})$ axis of the data and the fit are shown in Fig.~\ref{f2D_fit} c). The data can be well described for all $p_{T}$ bins with the fit function, which allows one to subtract the 2D proton fit function from the distribution. In addition to the fit subtraction, a mass-squared cut of ${\it m}^{2}<0.65$ (\GeVc$^{2}$)$^{2}$ was applied, as shown in Fig.~\ref{f2D_fit} b) as a diagonal line. The latter cut removes the remnant non-Gaussian tails from the protons. The corresponding result after the proton subtraction is shown in Fig.~\ref{f2D_fit} d). This distribution was fitted with two  Gaussians (2$\times$1D) to extract the pion and kaon yields. The goal of this representation, the increased separation power between the pions and kaons along the transformed horizontal axis, was reached. 

\subsection{The signal extraction for $\phi$, $\Lambda$, $\overline{\Lambda}$, $K_{s}^{0}$, $\Xi^{-}$, $\overline{\Xi}^{+}$, $\Omega^{-}$, and $\overline{\Omega}^{+}$}
\label{sec_sig_V0}
\begin{figure*}[]
\resizebox{18cm}{!}{%
  \includegraphics[angle=-90]{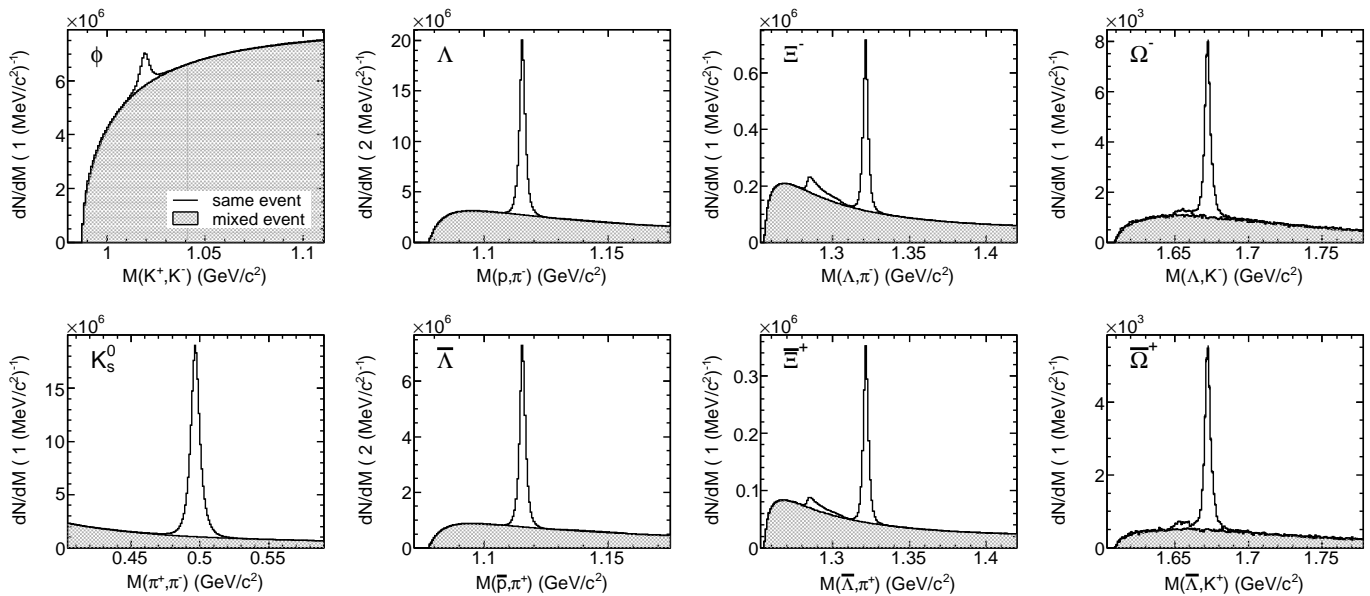}}
\caption{Examples of the invariant mass distributions at $\sqrt{s_{NN}} = $ 62.4 GeV for $\phi$, $K_{s}^{0}$, $\Lambda$, $\overline{\Lambda}$, $\Xi^{-}$, $\overline{\Xi}^{+}$, $\Omega^{-}$, and $\overline{\Omega}^{+}$. 
The combinatorial background is described by the mixed event technique which is shown as a grey shaded histogram. }
\label{fInvMass_62GeV}       
\end{figure*}

Short-lived weakly-decaying particles, generically called ${\it V^{0}}$ particles, such as $\Lambda$, $\phi$ and $\Xi$, decay into a pair of oppositely charged particles and were reconstructed using the invariant mass technique. The combinatorial background from uncorrelated particles was reduced by a direct identification of the daughter particles using the specific energy loss and/or mass-squared (${\it m}^{2}$) information and selection criteria based on the topology of the specific decay. Depending on the particle species and the magnitude of the background, $n\sigma$ cuts of $\pm 2$ or $\pm 3$ were applied to the normalized $dE/dx$ of the daughter particle tracks. Since the time-of-flight information is only available for about 65\% of the tracks within the accepted pseudorapidity range of $-1 < \eta < 1$, a general cut on the mass-squared ${\it m}^{2}$, as for $dE/dx$, was not applied. Instead, a cut on ${\it m}^{2}$ was only applied if the time-of-flight information for the track was available or the misidentification rate at a certain momentum range, when using only the $dE/dx$ information, was large. For most of the ${\it V^{0}}$ particles, the combinatorial background can be efficiently reduced with topology cuts as will be described below. In these cases, a lack of TOF information was compensated for by using tighter topology cuts. For the $\phi$ meson, the time-of-flight information was always required for daughter tracks at higher momenta, typically at $p>0.65$ \GeVc\, where the $dE/dx$ information alone was insufficient to remove the bulk of the misidentified tracks. In general, a $3\sigma$ cut on the ${\it m}^{2}$ distributions of the particles was applied. At higher momenta, the $\pi$, $K$ and $p$ distributions begin to overlap. Here, tighter and/or asymmetric cuts were used.

For the topological reconstruction of ${\it V^{0}}$ particles, geometrical information on the decays was also used, {\it e.g.} the primary and secondary/tertiary decay vertex positions, the distance of closest approach (dca) of the daughter particles to the primary vertex, the dca of the mother particle(s) to the primary vertex, and the dca between the daughter tracks. This information was determined from the helix parameterizations of the TPC reconstructed tracks. The following topology cuts were applied:
\begin{itemize}
\item dca between daughter tracks (primary and secondary daughters in case of $\Xi$ and $\Omega$),
\item dca between daughter tracks and primary vertex,
\item dca between mother particle and primary vertex,
\item dca between $\Lambda$ candidate and primary vertex (for $\Xi$ and $\Omega$), and
\item distance between primary and secondary (tertiary in case of $\Xi$ and $\Omega$) vertex.
\end{itemize}
A cut on the invariant mass of $1.108 < M(p,\pi)<1.122$ \GeVc$^{2}$ was applied to enhance the $\Lambda$ and $\overline{\Lambda}$ candidates for the $\Xi^{-}$, $\overline{\Xi}^{+}$, and $\Omega^{-}$, $\overline{\Omega}^{+}$ analyses. The particle identification and topology cuts were systematically optimized for the best significance by varying several tens of thousands of cut combinations for each particle species.

The misidentification of the daughter particles, which is more probable at the higher momenta, can result in an additional correlated background. Such a correlated background, for example from the $\Lambda$, can appear in the $\pi^{+} \pi^{-}$ ($K_{s}^{0}$) invariant mass distribution if the proton was misidentified as a $\pi^{+}$. Such a correlated background does not create a peak in the invariant mass distribution of the particles of interest since the daughter particle masses are chosen to be the nominal ones ({\it e.g.} $\pi$ mass instead of proton mass), but it appears as a broad distribution which can significantly affect the signal extraction. To remove this correlated background, additional invariant mass spectra with identical track combinations, but different daughter mass values, {\it e.g.} $(p,\pi^{-})$ and $(\pi^{+},\pi^{-})$ were investigated. The background was removed by applying invariant mass cuts on the corresponding unwanted peaks in the misidentified invariant mass distributions. Usually, the correlated background from particle misidentification increases with the $p_{T}$ values of the mother particle.
  
The remaining uncorrelated combinatorial background was described and later subtracted with the mixed event technique. Event classes were defined to mix only events with similar global properties; the classes consisted of 9 centrality ranges, 14 $z$-vertex ranges, and 10 event plane angle ranges. The event buffer depth varied between 3 and 15. The mixed event distributions were normalized at least $3\sigma$ away from the mass peak on both sides. The mixed event distributions so obtained were in excellent agreement for all particle species and energies with the combinatorial background shown in Fig.~\ref{fInvMass_62GeV} for 0--80\% centrality Au+Au collisions at $\sqrt{s_{NN}}$=62.4 GeV for $\phi$, $\Lambda$, $\overline{\Lambda}$, $K_{s}^{0}$, $\Xi^{-}$, $\overline{\Xi}^{+}$, $\Omega^{-}$, and $\overline{\Omega}^{+}$.
The correlated background to the left of the $\Xi^{-}$, $\overline{\Xi}^{+}$, $\Omega^{-}$ and $\overline{\Omega}^{+}$ peaks in Fig.~\ref{fInvMass_62GeV} is a result of a self-correlation between the three daughter particles. In the case of the $\Xi$, two $\pi$ mesons with the same charge are in the final state and both combinations ($\Lambda$($p$,$\pi_{1}$)+$\pi_{2}$ and $\Lambda$($p$,$\pi_{2}$)+$\pi_{1}$) result in similar invariant mass values. These wrong combinations appear as a bump structure to the left of the true peak. The two structures were separated by an invariant mass cut. In the case of the $\Omega$, a double misidentification of the $\pi$ and $K$ resulted in a similar effect.

\section{Event plane reconstruction}  
\label{sec_EP_rec}

\begin{figure*}[]
\centering
\resizebox{16cm}{!}{%
\includegraphics{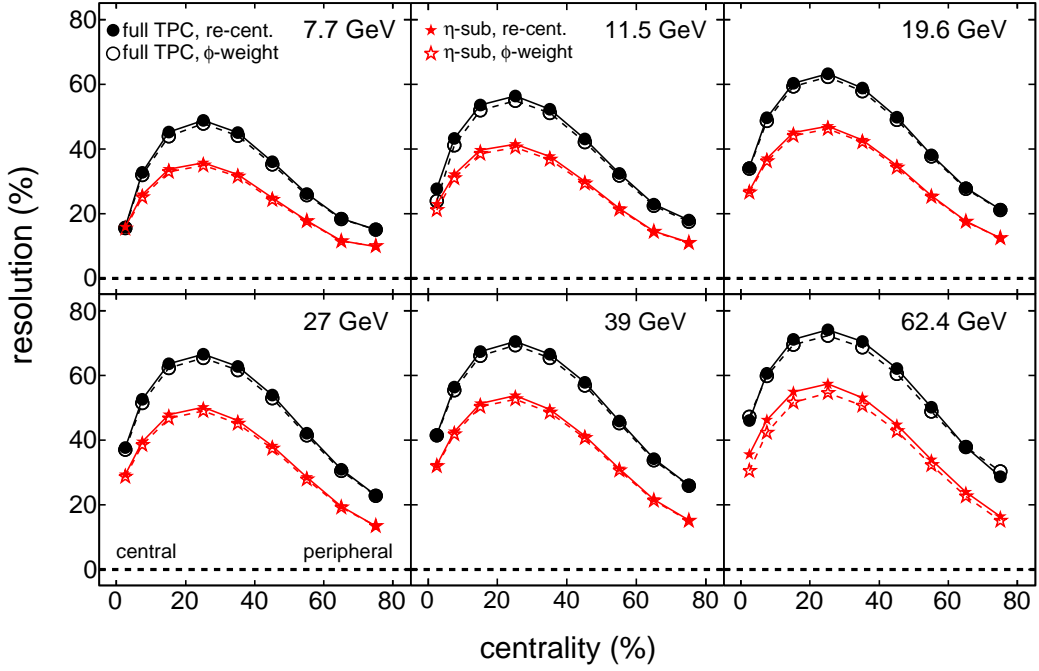}}
\caption{The event plane resolution for the full TPC event plane (circles), and the $\eta$-sub event plane (stars), as a function of the centrality for two different and independent flattening methods.}
\label{fEP_res}       
\end{figure*}

The event plane (EP) is obtained from the angles of the reconstructed particles and the beam line. It is an estimate of the participant plane which is defined by the participating nucleons in the collision. The event plane was reconstructed using the flow of the measured particles, as discussed in Ref.~\cite{Poskanzer:1998yz}. To achieve the best resolution for $v_{2}$, the second harmonic event plane angle $\Psi_{2}$ was calculated as:
\begin{equation}
\Psi_{2} = \tan^{-1} \left( \frac{\sum_{i}w_{i}\sin(2\phi_{i})}{\sum_{i}w_{i}\cos(2\phi_{i})} \right)/2,
\label{form_PsiAngle}
\end{equation}
where $\phi_{i}$ is the azimuthal angle of particle $i$ and $w_{i}$ is its weight. The weight in units of \GeVc\ was chosen to be linear with $p_{T}$ up to 2 \GeVc\ and then constant at a value of 2 for higher momenta. Only those particles with a momentum between 0.15 and 5 \GeVc, $|\eta| < 1$, dca $< 1$ cm and having more than 15 hits in the TPC were used for this calculation. Two different event planes were reconstructed: one using all of the reconstructed tracks in the TPC (``full TPC" method) and one using only those tracks in the opposite pseudorapidity hemisphere to the particle track of interest (``$\eta$-sub" method). In the full TPC case, self-correlations were avoided by removing the particle of interest from the tracks used for the event plane reconstruction. In the $\eta$-sub method, an additional pseudorapidity gap of $\pm 0.05$ was applied to reject some tracks for the event plane reconstruction. In general, the $\eta$-sub method reduces the effect of ``non-flow," which includes the decay of resonances to several charged daughter particles, Hanbury-Brown Twiss correlations, and jets~\cite{Voloshin:2008dg}. However, the resolution is lower and therefore the correction to obtain $v_{2}$ is larger.

An azimuthally non-homogeneous acceptance or efficiency of the detectors can introduce a bias in the event plane reconstruction which would yield a non-uniform $\Psi_{2}$ angle distribution in the laboratory coordinate system. To flatten the $\Psi_{2}$ distribution, the recentering or $\phi$-weight methods, in combination with the shift method, were used~\cite{Voloshin:2008dg}. In the $\phi$-weight method, a track-by-track correction is applied. It is based on the $\phi$ angle distributions, $dN/d\phi(z,p_{T},\eta,t,q)$ which were determined for five $z$-vertex ranges, four $p_{T}$ ranges, six $\eta$ ranges, and for both charge signs, $q$. Furthermore, the distributions were determined for different real time, $t$, periods during the data collection, each of which spanned approximately one day. Each track used for the event plane reconstruction was weighted in the $\phi$-weight method with the inverse value of the corresponding value of the $dN/d\phi(z,p_{T},\eta,t,q)$ distribution. Large gaps in the $\phi$ angle distribution cannot be corrected with this method. The $\phi$-weight corrected event plane angles will be denoted as $\Psi_{2,\phi}$.

The recentering method applies a correction on an event-by-event basis and is therefore more robust in the case of acceptance holes. The numerator and denominator of Eq.~(\ref{form_PsiAngle}) can be used to define the vector
\begin{equation}
\vec{Q}_{\rm raw} = \frac{1}{N} \sum_{i}^{N}
\left( \begin{array}{c}
 w_{i}\cos(2\phi_{i}) \\
 w_{i}\sin(2\phi_{i})
\end{array} \right),
\label{form_defQvec}
\end{equation}
where {\it N} is the number of tracks used for the event plane reconstruction in each event. In order to get a uniform $\Psi_{2}$ angle distribution, this $Q$-vector must be centered at (0,0). To achieve this objective, the average of the $Q$-vector over many events was subtracted event-by-event:
\begin{equation}
\vec{Q}_{\rm rc} =  \vec{Q}_{\rm raw} - \left\langle \vec{Q}_{\rm raw} \right\rangle. \nonumber 
\label{form_Qvec_rc}
\end{equation}
These averaged $Q$-vectors were determined for ten $z$-vertex ranges and for each pseudorapidity hemisphere in a real-time dependent manner, and were then parametrized as a function of the event multiplicity. The new $\Psi_{2,\rm{rc}}$ angles were then calculated from the corrected $Q$-vectors.

If the $\Psi_{2}$ angle distribution was not flat after the $\phi$-weight or the recentering corrections, an additional correction with the shift method was used to force the $\Psi_{2}$ angle distribution to be flat~\cite{Voloshin:2008dg}.  A shift angle $\Psi_{2,\rm{shift}}$ was calculated event-by-event for each event plane method in the following way:
\begin{eqnarray}
\Psi_{2,\rm{shift}} & = & -c_{2}\cos(2\Psi_{2}) + s_{2} \sin(2\Psi_{2})  \nonumber \\
& + & 0.5(-c_{4}\cos(4\Psi_{2}) + s_{4}\sin(4\Psi_{2})).
\label{form_Shift_corr}
\end{eqnarray}
The $c_{2,4}, s_{2,4}$ parameters were obtained from fits to the averaged $\phi$-weight or recentering corrected $\Psi_{2}$ angle distributions. The shift-corrected event plane angle $\Psi_{2,\rm{corr}}$ was obtained as:
\begin{equation}
\Psi_{2,\rm{corr}} = \Psi_{2,\rm{rc},\phi} + \Psi_{2,\rm{shift}},
\end{equation}
where $\Psi_{2,\rm{rc},\phi}$ is the recentering corrected event plane angle. After the shift correction a flat $\Psi_{2,\rm{corr}}$ distribution for all energies and event plane methods was thus achieved. 

To calculate the event plane (EP) resolution, independent sub-samples of randomly selected tracks (full TPC) or tracks in independent pseudorapidity hemispheres ($\eta$-sub) were used \cite{Voloshin:2008dg}. Figure~\ref{fEP_res} shows the event plane resolution for the four different reconstructed event plane types and the six beam energies. The event plane resolution is used below ({\it cf.} Section~\ref{EP_res_corr}) to correct the observed $v_{2}^{\rm obs}$ signals. The event plane resolution is approximately proportional to the flow coefficient times the square-root of the multiplicity~\cite{Poskanzer:1998yz}. It decreases with decreasing beam energy due to the lower particle multiplicities. It has a maximum for each beam energy at about 30\% centrality. For more peripheral events, the relatively low multiplicity is responsible for the decreasing resolution whereas for more central events the small flow signal is responsible. The $\phi$-weight corrected EP has a slightly smaller resolution compared to the recentering method which could be connected to the smaller number of centrality bins used for the $\phi$-weight correction. At 62.4 GeV, a significantly larger difference between the two correction methods is observed compared to all of the other beam energies. This is due to a missing TPC sector during the collection of the 62.4 GeV data. The resulting gap in the $\Psi_{2}$ angle distribution cannot be fully corrected with the $\phi$-weight method as described above. 

In general, the $\eta$-sub method has a smaller EP resolution compared to the full TPC method. This is mainly due to the factor of $\sim$2 fewer tracks used for the EP reconstruction in the former. For the most central collisions and the lowest energies, 7.7 and 11.5 GeV, a similar EP resolution for the two methods is observed. This might be an indication of a strong negative non-flow signal at the lower energies when the full TPC method is used. The negative non-flow, which originates primarily from resonance decays, results in an anticorrelation between the random sub-events used for the EP resolution calculation. The $\eta$-sub EP method reduces the non-flow by using spatially independent regions in the TPC. Therefore, in the following only the results based on the $\eta$-sub EP method will be presented. 

\section{$v_{2}$ signal extraction and systematic uncertainties} 
\label{sec_v2sig}
The azimuthal emission pattern of the particles relative to the event plane can be decomposed into a Fourier sum of cosine terms:
\begin{equation}
\frac{dN}{d(\phi-\Psi_{m})} \propto 1 + 2 \sum_{n\ge1} v_{n} \cos\left[ n(\phi-\Psi_m) \right],
\label{fPsi_phi}
\end{equation}
where $\phi$ is the azimuthal angle of the particle, $\Psi_{m}$ is the event plane angle, $v_{n}$ is the Fourier coefficient of harmonic $n$, and $m$ is the harmonic of the event plane~\cite{Poskanzer:1998yz}. In the following, only the elliptic flow coefficient $v_{2}$, will be considered,
\begin{equation}
\frac{dN}{d(\phi-\Psi_2)} \propto 1 + 2 v_{2} \cos\left[ 2(\phi-\Psi_2) \right].
\label{fPsi2_phi}
\end{equation}
\subsection{Event plane and invariant mass methods}
\label{sec_v2_extract}

\begin{figure}[]
\includegraphics[width=0.4\textwidth]{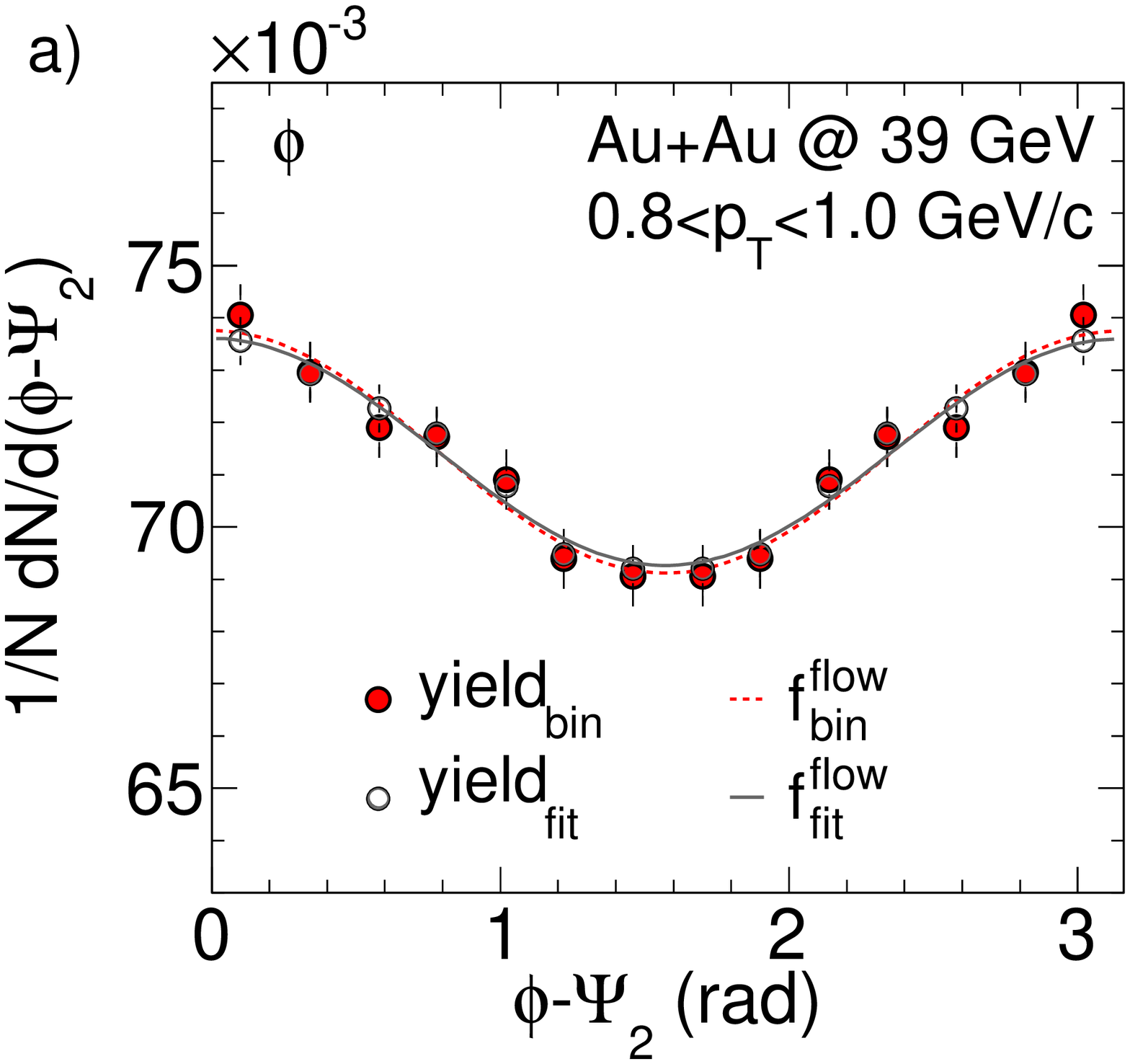} %
\includegraphics[width=0.4\textwidth]{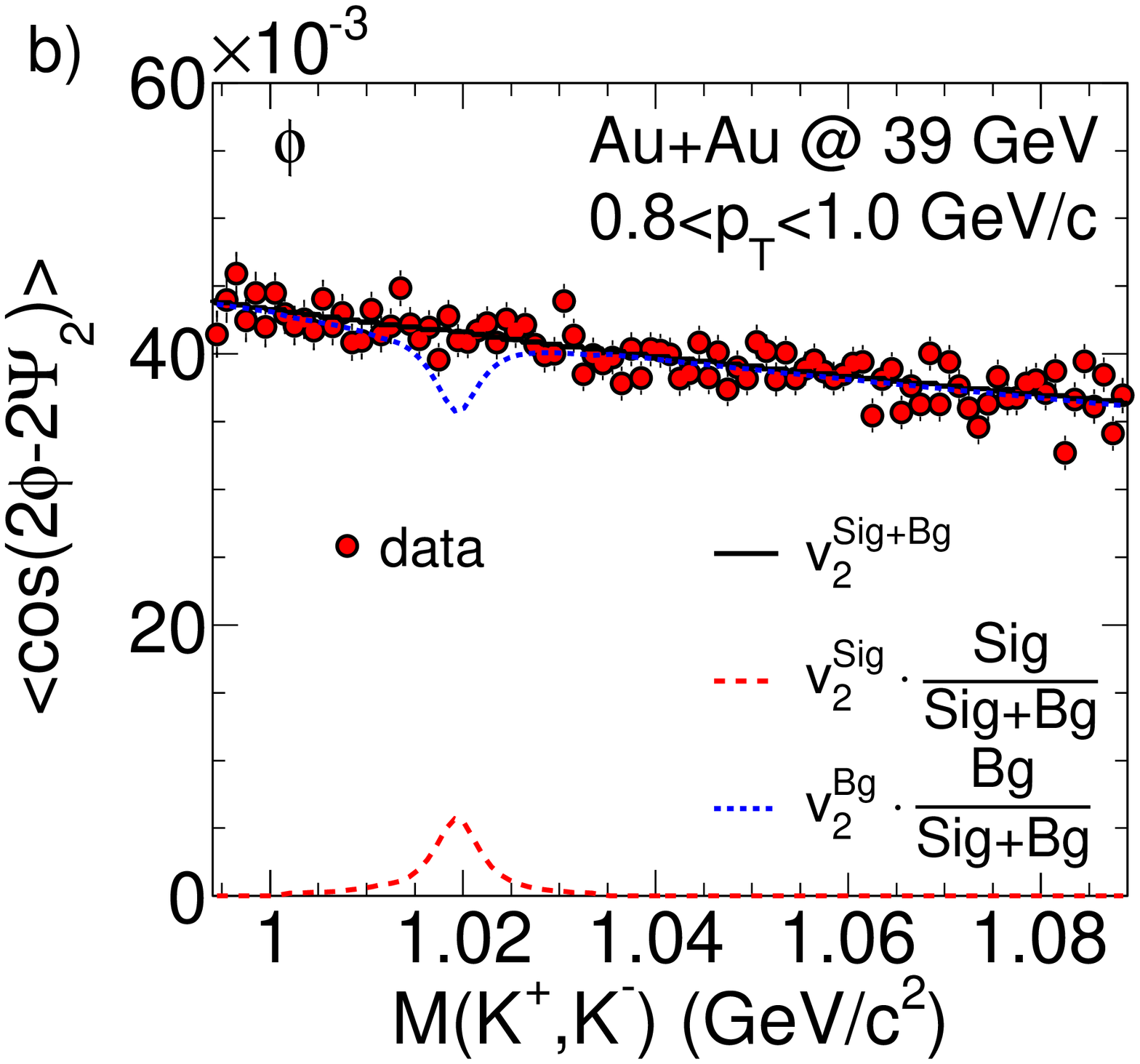} %
 \caption{(Color online) Two examples of the $v_{2}$ signal extraction for $\phi$ mesons at 39 GeV in the transverse momentum range of $0.8<p_{T}<1.0$ \GeVc. The event plane method a) and the invariant mass method b) give almost identical results. a) The $\phi-\Psi_{2}$ data points are reflected at $\pi/2$. A fit with Eq.~(\ref{fPsi2_phi}) to the data obtained by integrating the fit is shown as a solid black line. The red dashed line shows the fit result to the data obtained by counting the particles in each bin. b) The solid black curve is the fit from Eq.~(\ref{finvmass_v2_fit_B}). The dashed red curve is the signal part of that equation and the dashed blue curve is the background part.}
 \label{fv2_extract}
\end{figure}

Two techniques were used to calculate $v_{2}$: the event plane method and the invariant mass method~\cite{Borghini:2004ra}. The latter method was used in addition only for the $\phi$ mesons. Both methods give identical results, but are technically different, especially when the signal-to-background ratio is small. In the standard event plane method, the particles were first identified, then their yields were determined as a function of the relative angle $\phi-\Psi_2$. In the invariant mass method, the mean values $\left\langle \cos\left[2(\phi-\Psi_2) \right] \right\rangle$ were calculated as a function of invariant mass and then the correlation at the invariant mass peak of interest was isolated after the background subtraction.

For the event plane method, the $v_{2}$ coefficients were obtained by fits to the yield distributions with Eq.~(\ref{fPsi2_phi}). An example of such a fit is shown in Fig.~\ref{fv2_extract} a). For most of the particles, the yields were determined in two different ways: by counting the particles in bins within reasonable mass ranges and/or by integrating a fit to the corresponding mass distribution. The fits can have the form of a Gaussian or a Breit-Wigner distribution. A Breit-Wigner distribution was used for the $\phi$ mesons, and a Gaussian was used for the $\Lambda$ and $K_{s}^{0}$ particles. The yields were only determined in the $\phi-\Psi_{2}$ range of 0 to $\pi/2$. In Fig.~\ref{fv2_extract} a), the reflected data points are also shown for reference.

For the invariant mass method, the mean values $\left\langle ... \right\rangle$ (see also Ref.~\cite{Masui:2012zh}) were by definition the $v_{2}$ values of the analyzed particles ({\it cf.} Eq.~(\ref{fPsi2_phi})). Since the background cannot be distinguished from the signal on an event-by-event basis, the resulting $v_{2}$ value was the sum of signal and background as formulated in Eq.~(\ref{finvmass_v2}). The $v_{2}^{\rm Sig+Bg}(M_{\rm inv})$ can be decomposed into a signal and a background term as shown in Eq.~(\ref{finvmass_v2_fit_B}). Each term is multiplied by a statistical weight which was extracted from the same event invariant mass and the combinatorial background distributions. The background elliptic flow $v_{2}^{\rm Bg}(M_{\rm inv})$ was parameterized with the polynomial defined in Eq.~(\ref{finvmass_v2_fit_C}). Figure~\ref{fv2_extract} b) shows an example of an invariant mass fit with the total fit (black solid line), the signal term (red dashed line), and the background term (blue dashed line). In this particular case, the $v_{2}^{\rm Bg}$ is nearly identical to $v_{2}^{\rm Sig}$ which results in a monotonic distribution around the signal region.

\begin{eqnarray}
v_{2}^{\rm Sig+Bg}(M_{\rm inv}) & = & \left\langle \cos\left[2(\phi-\Psi_2) \right]_{M_{inv}} \right\rangle \label{finvmass_v2_fit_A} \label{finvmass_v2} \\
v_{2}^{\rm Sig+Bg}(M_{\rm inv}) & = & v_{2}^{\rm Sig} \frac{\rm Sig}{\rm Sig+Bg}(M_{\rm inv}) \nonumber \\ 
& + & v_{2}^{\rm Bg}(M_{\rm inv})\frac{\rm Bg}{\rm Sig+Bg}(M_{\rm inv}) \label{finvmass_v2_fit_B} \\
v_{2}^{\rm Bg}(M_{\rm inv}) &  = & p_{0}+p_{1}M_{\rm inv}+p_{2}M_{\rm inv}^{2} \nonumber \\
& + & p_{3}M_{\rm inv}^{3}
\label{finvmass_v2_fit_C}
\end{eqnarray}
The invariant mass method was tested for various particle species and directly compared to the results from the event plane method. For particles with large signal-to-background ratios in the invariant mass distribution, for instance $\Lambda$ and $\Xi$, no systematic differences were found. The present results are generally based on the event plane method. However, both methods were evaluated only for the $\phi$ meson, which shows a significantly lower signal-to-background ratio compared to all other particles. Small differences between the two methods were taken into account in the systematic uncertainties. 

\subsection{Event plane resolution correction for 0--80\%}
\label{EP_res_corr}
The event plane resolution was calculated for nine centrality bins as shown in Fig.~\ref{fEP_res}. For the integrated 0--80\% centrality bin, a new method was used to correct the observed $v_{2}^{\rm obs}$ signals. The yields of the reconstructed particles were weighted event-by-event with the inverse event plane resolution for the corresponding centrality bin. This ensured a correction which was not biased by the bin width. The $v_{2}$ signals were normalized with the mean inverse event plane resolution for the 0--80\% centrality bin. A detailed description of the method can be found in Ref.~\cite{Masui:2012zh}.

\subsection{Systematic uncertainties}
\label{syst_errors}
The systematic uncertainties were evaluated by varying the methods and parameters used to determine the event plane angles and particle yields. For the ${\it V^{0}}$ particle analyses, {\it e.g} $\Lambda$ and $\Xi$, twenty different combinations of the topology cuts listed in Section~\ref{sec_sig_V0} were applied. Each of these topology cut combinations had a significance similar to the reference cuts which were optimized for the best significance. The same number of combinations were used for the $\phi$ meson analysis, but in this case it was the $K^{\pm}$ particle identification cuts that were varied, such as the  $n\sigma_{K}$ range. 

In addition to the variation of particle identification cuts, two methods to extract the $v_{2}$ values and two ways to determine the particle yields were used as described in Section~\ref{sec_v2_extract}.
The $\pi$ and $K$ analyses depend primarily on the initial fit parameters and fit ranges as pointed out in Section~\ref{sec_pid_pikp}.  The parameters were varied in combination with two values of the proton separation ${\it m}^{2}$ cuts, resulting in a total of eighteen different combinations. In the case of the proton analysis, three different combinations of dca and $n\sigma_{p}$ cuts were studied.

The point-by-point systematic uncertainties, which consist of the variations of the particle identifications cuts and the two methods of signal extraction, were evaluated for all combinations (40 for all ${\it V^{0}}$ particles and the $\phi$ meson, 36 for $\pi$ and $K$, and 6 for protons) by calculating the root-mean-squared value for each data point. For all energies, both flattening methods for the event plane angle (which were described in Section~\ref{sec_EP_rec}) were compared. The mean value of the point-by-point differences between the two methods was defined as the global systematic uncertainty for each particle species. The mean point-by-point systematic uncertainties varied for $p$, $\pi$ and $K$ in a range of 0.0001--0.001, and for $V^{0}$ particles and the $\phi$ meson in a range of 0.0005--0.007. The mean global systematic uncertainties for all particles were in the range of 0.0005--0.003.
In addition to these studies of the systematic uncertainties, independent analyses for most of the particle species were performed.  A cross check to the previously published 62.4 GeV data~\cite{Abelev:2010tr}, where slightly different methods were used, indicated an excellent agreement to the present results within the statistical errors. 

The data were not corrected for feed-down contributions. By varying the dca cuts for all particle species, the feed-down contributions were already partly included into the systematic uncertainties described above. Previous studies showed~\cite{Abelev:2007qg} that feed-down is only significant for pions below $p_{T}$=0.4 \GeVc. For other particles, the resulting feed-down contributions to the $v_{2}$ values are negligible.

Non-flow contributions were studied for the six beam energies by comparing different methods to extract $v_{2}$ for inclusive charged hadrons~\cite{Adamczyk:2012ku}. The four-particle cumulant $v_{2}\{4\}$ strongly suppresses non-flow contributions. It has been shown that the difference between $v_{2}(\eta$-${\rm sub})$ and $v_{2}\{4\}$ is about 10--20\% for 19.2, 27, and 39 GeV and decreases with decreasing beam energy. In the following we did not treat such non-flow contributions as systematic errors.

\section{Results}
\label{sec_results}

The $v_{2}$ results corrected for the event plane resolution in 0--80\% central Au+Au collisions are presented. All results are based on the $\eta$-sub event plane method described above. The $x$-axis values of the data points are always placed at the $p_{T}$-weighted mean values within the bin limits. The statistical errors are indicated as straight vertical lines, the point-by-point systematic uncertainties are indicated either as shaded bands or with square brackets, and the global systematic uncertainties are indicated as a horizontal shaded band on the horizontal axis. For plots with several $v_{2}$ distributions, only the statistical errors are shown. 

\subsection{Elliptic flow as a function of transverse momentum}
\label{sub_v2_pt}

\begin{figure*}[]
\centering
\resizebox{13cm}{!}{%
  \includegraphics[bb = 9.486140 2.754070 519.881961 689.874940,clip]{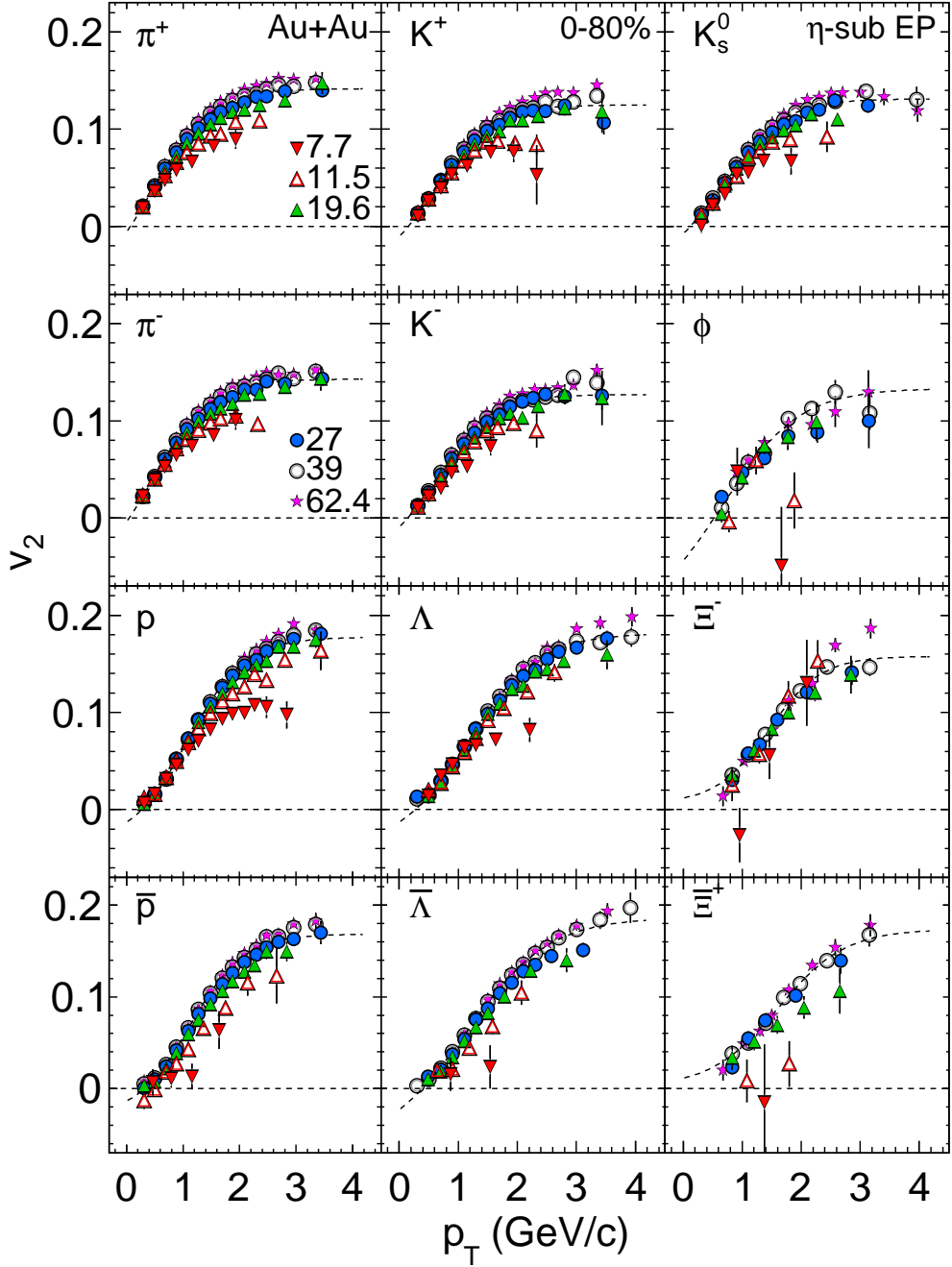}}
\caption{(Color online) The elliptic flow, $v_{2}$, as a function of the transverse momentum, $p_{T}$, from 0--80\% central Au+Au collisions for various particle species and energies. Only the statistical error bars are shown. The black dashed line is a fit to the 39 GeV data points with Eq.~(\ref{Func_v2_fit}).}
\label{fv2_pT_energy}       
\end{figure*}

\begin{figure*}[h!]
\centering
\resizebox{13cm}{!}{%
  \includegraphics[bb = 9.486140 2.754070 519.881961 689.874940,clip]{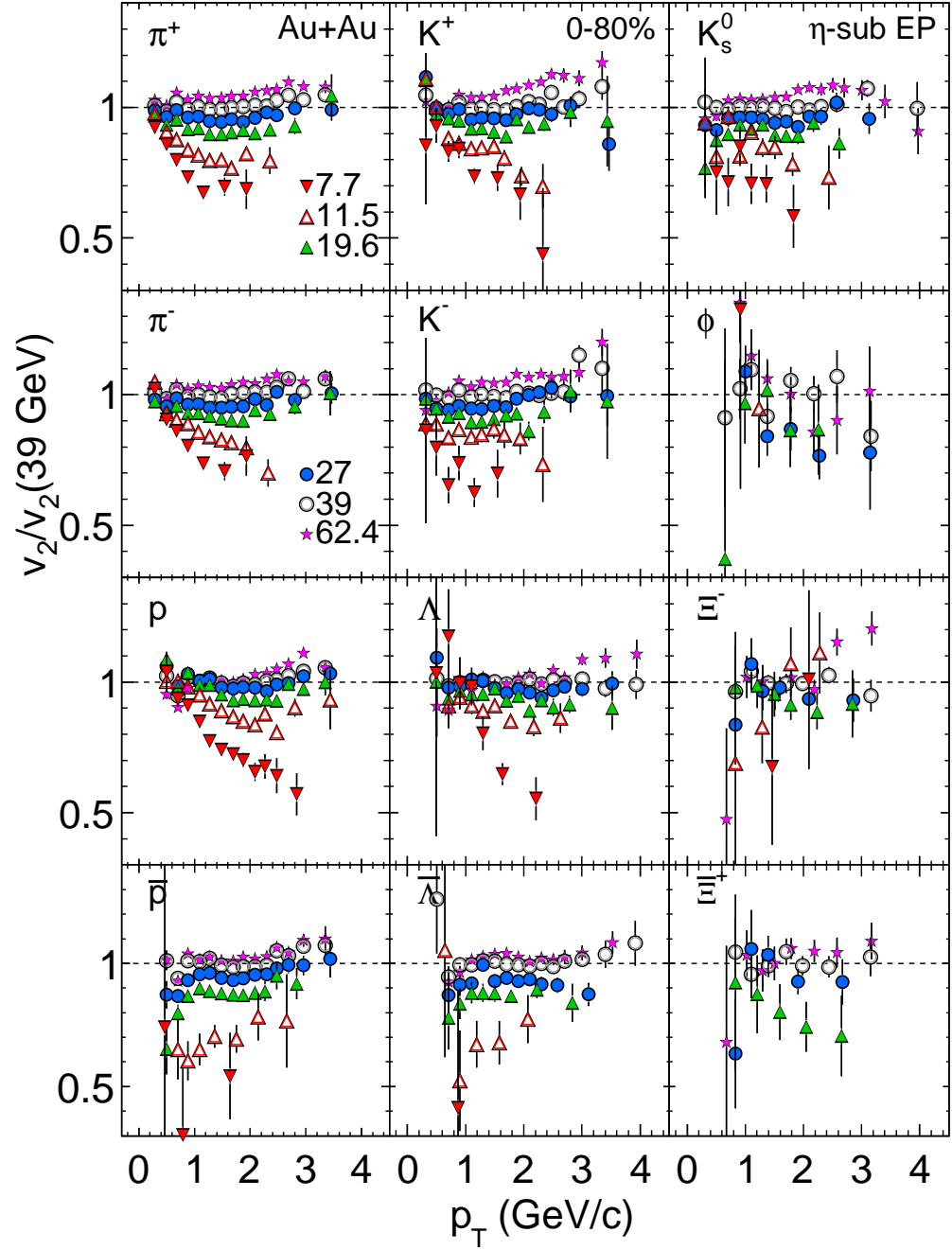}}%
\caption{(Color online) The ratio of the elliptic flow, $v_{2}(p_{T})$, relative to a fit to the 39 GeV $v_{2}(p_{T})$ data points for 0--80\% central Au+Au collisions for various particle species and energies. The error bars are statistical only. The fit and the $v_{2}(p_{T})$ data points are shown in Fig.~\ref{fv2_pT_energy}.}
\label{fv2_pT_energy_ratio}       
\end{figure*}

\begin{figure*}[]
\centering  
\resizebox{13.0cm}{!}{%
\includegraphics[bb = 13 4 505 301,clip]{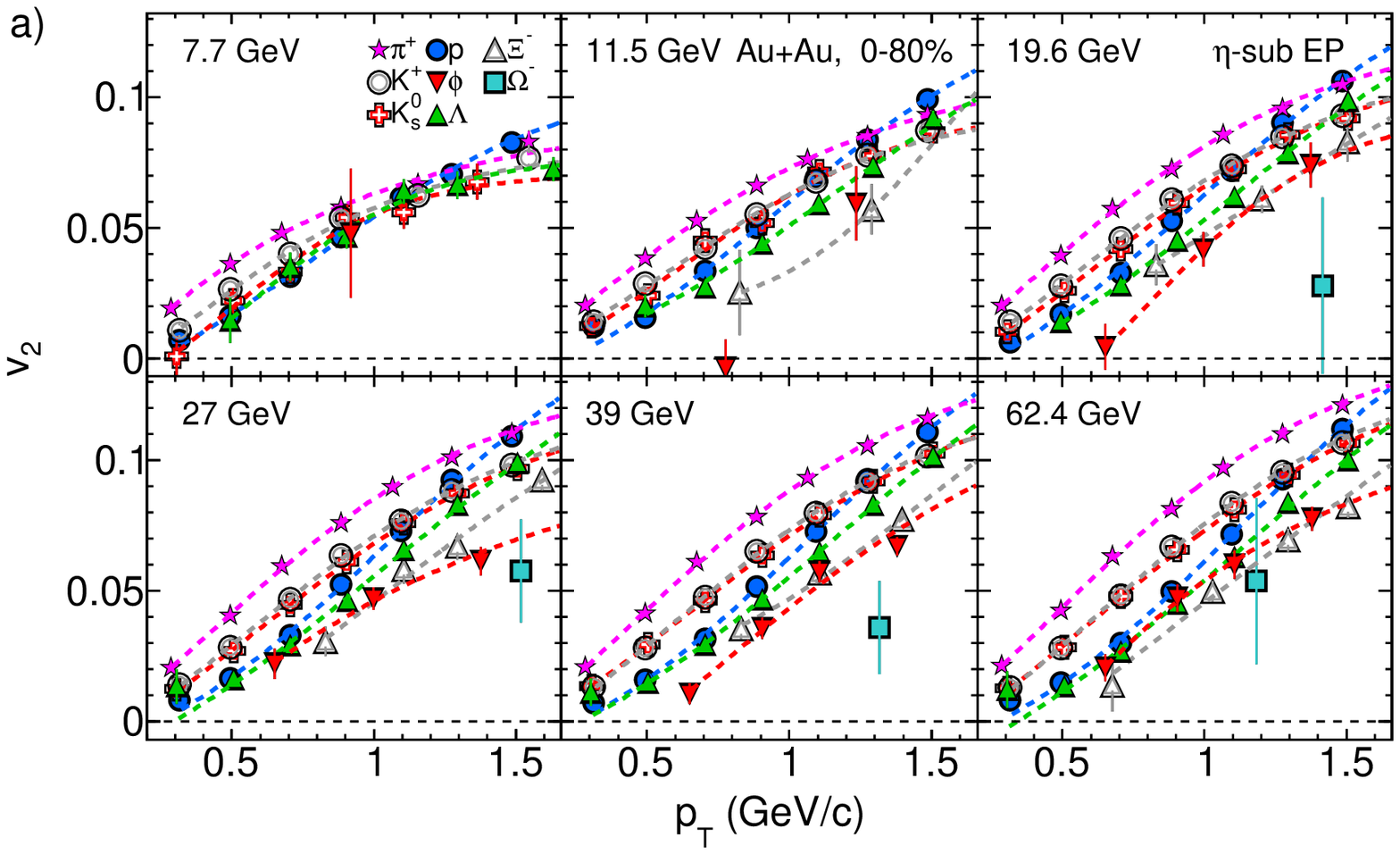}}  \\[2ex]
\resizebox{13.0cm}{!}{%
\includegraphics[bb = 13 4 505 301,clip]{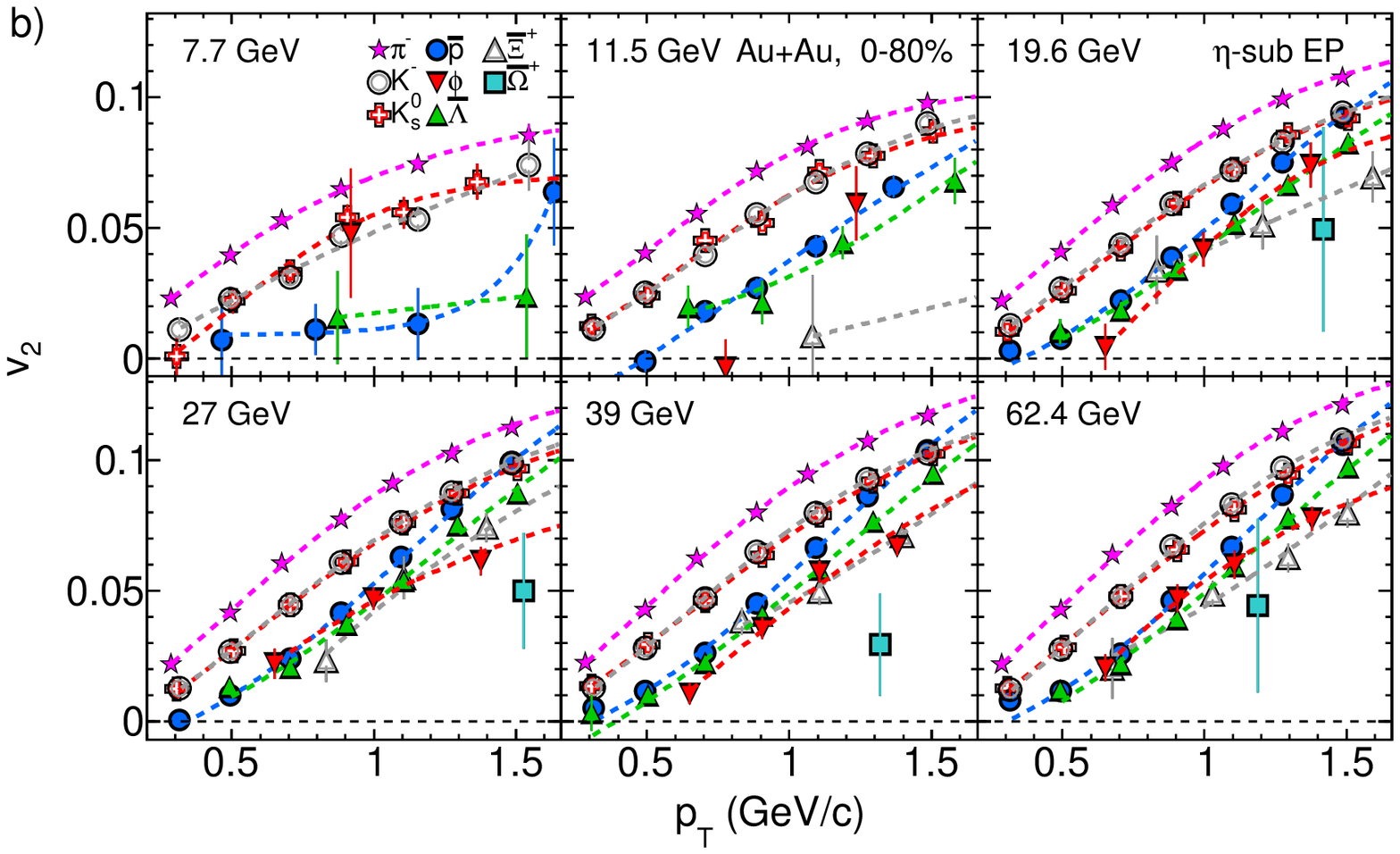}} %
 \caption{(Color online)  The elliptic flow, $v_{2}(p_{T})$, in 0--80\% central Au+Au collisions for selected particles a) and anti-particles b) (see text), plotted only for the transverse momentum range of $0.2<p_{T}<1.6$ \GeVc\ to emphasize the mass ordering at low $p_{T}$. The fit functions correspond to Eq~(\ref{Func_v2_fit}).}
 \label{fv2_vs_pT}
\end{figure*}

Figure~\ref{fv2_pT_energy} shows the energy dependence in $v_{2}(p_{T})$ for $\pi^{\pm}$, $K^{\pm}$, $p$, $\bar{p}$, $\Lambda$, $\overline{\Lambda}$, $\phi$, $K_{s}^{0}$, $\Xi^{-}$ and $\overline{\Xi}^{+}$. A similar trend of $v_{2}(p_{T})$ for all particles is observed. The $v_{2}$ increases with $p_{T}$ up to 1.5 \GeVc\ and reaches a maximum value at higher beam energies of about 0.15 for mesons and $\sim$0.2 for baryons within the measured $p_{T}$ range. The maximum values decrease with decreasing energy to about 0.07 for kaons and pions and $\sim$0.1 for protons at $\sqrt{s_{NN}}$ = 7.7 GeV. It should be noted that the $v_{2}(p_{T})$ decreases for higher $p_{T}$ values in $\sqrt{s_{NN}}$ = 200 GeV Au+Au collisions~\cite{Abelev:2008ae}. The negative anti-proton $v_{2}$ at low $p_{T}$ and at $\sqrt{s_{NN}}$ = 11.5 GeV could be due to absorption in the medium \cite{Wang:2012zzi}. A more precise picture of the energy dependence can be obtained from $v_{2}(p_{T})$ ratios. In order to define a reference, the 39 GeV data points were fitted with the following equation:
\begin{equation}
\label{Func_v2_fit}
f_{v_{2}}(n) = \frac{an}{1+e^{-(p_{T}/n-b)/c}}-dn,
\end{equation}
where {\it a}, {\it b}, {\it c} and {\it d} are fit parameters and {\it n} is the constituent-quark number of the particle~\cite{Dong:2004ve}. The corresponding ratios are shown in Fig.~\ref{fv2_pT_energy_ratio}. A non-trivial $p_{T}$ dependence is observed from the ratios of the $v_{2}(p_{T})$ values to the fits. The ratios are close to unity and nearly independent of $p_{T}$ for $\sqrt{s_{NN}}$ $>$ 19.6 GeV.  
Below 19.6 GeV, the ratios decrease (below unity) with increasing $p_{T}$ for $\pi^{\pm}$, $p$, $\Lambda$ and $K^{+}$.
At these energies the ratios are below unity, but are independent of $p_{T}$ for $\bar{p}$, $\overline{\Lambda}$, $K_{s}^{0}$ and $K^{-}$. 

As expected, the $v_{2}\{\eta \rm \text{--}sub\}$-energy ratios of charged particles~\cite{Adamczyk:2012ku} follow the same trends as 
presented here for the identified hadrons. The trends for the more abundantly produced particles presented here ($p$, $\pi$, $K$) differ from those obtained using the inclusive 
charged hadron four-particle cumulant, $v_{2}\{4\}$~\cite{Adamczyk:2012ku}.  For the lower beam energies, the values of the ratios for the inclusive charged hadron $v_{2}\{4\}$ increase with increasing $p_{T}$. As will be discussed below, the difference might be due to non-flow contributions and flow fluctuations.  For $\pi^{\pm}$, $p$, $\Lambda$ and $K^{+}$, the ratios for all energies are close to unity at low $p_{T}$ and deviate with increasing $p_{T}$, whereas the ratios for $\bar{p}$, $\overline{\Lambda}$, $K_{s}^{0}$ and $K^{-}$ seem to be independent of $p_{T}$ for all energies.

At low transverse momenta, a mass ordering was observed it Au+Au collisions at 200 GeV~\cite{Abelev:2008ae}. Lighter particles had larger $v_{2}$ values. This behavior can be qualitatively described by ideal hydrodynamics~\cite{Huovinen:2006jp}. In Fig.~\ref{fv2_vs_pT}, the $v_{2}(p_{T})$ values in the transverse momentum range of $0.2<p_{T}<1.6$ GeV/$c$ for various particle species are directly compared. For this selection of particles ($p$, $\Lambda$, $\Xi^{-}$, $\Omega^{-}$, $\pi^{+}$, $K^{+}$, $K_{s}^{0}$ and $\phi$), the mass ordering is valid for all energies. Only the $\phi$ mesons deviate from this general trend at the lower energies. Their $v_{2}(p_{T})$ values are slightly smaller compared to all of the other hadrons. Starting at 39 GeV, every $\phi$ meson $v_{2}(p_{T})$ value is smaller than the corresponding value for the heavier $\Lambda$. 

\begin{figure*}[]      
\resizebox{13cm}{!}{%
\includegraphics[bb = 7.889765 2.250000 549.131960 423.874956,clip]{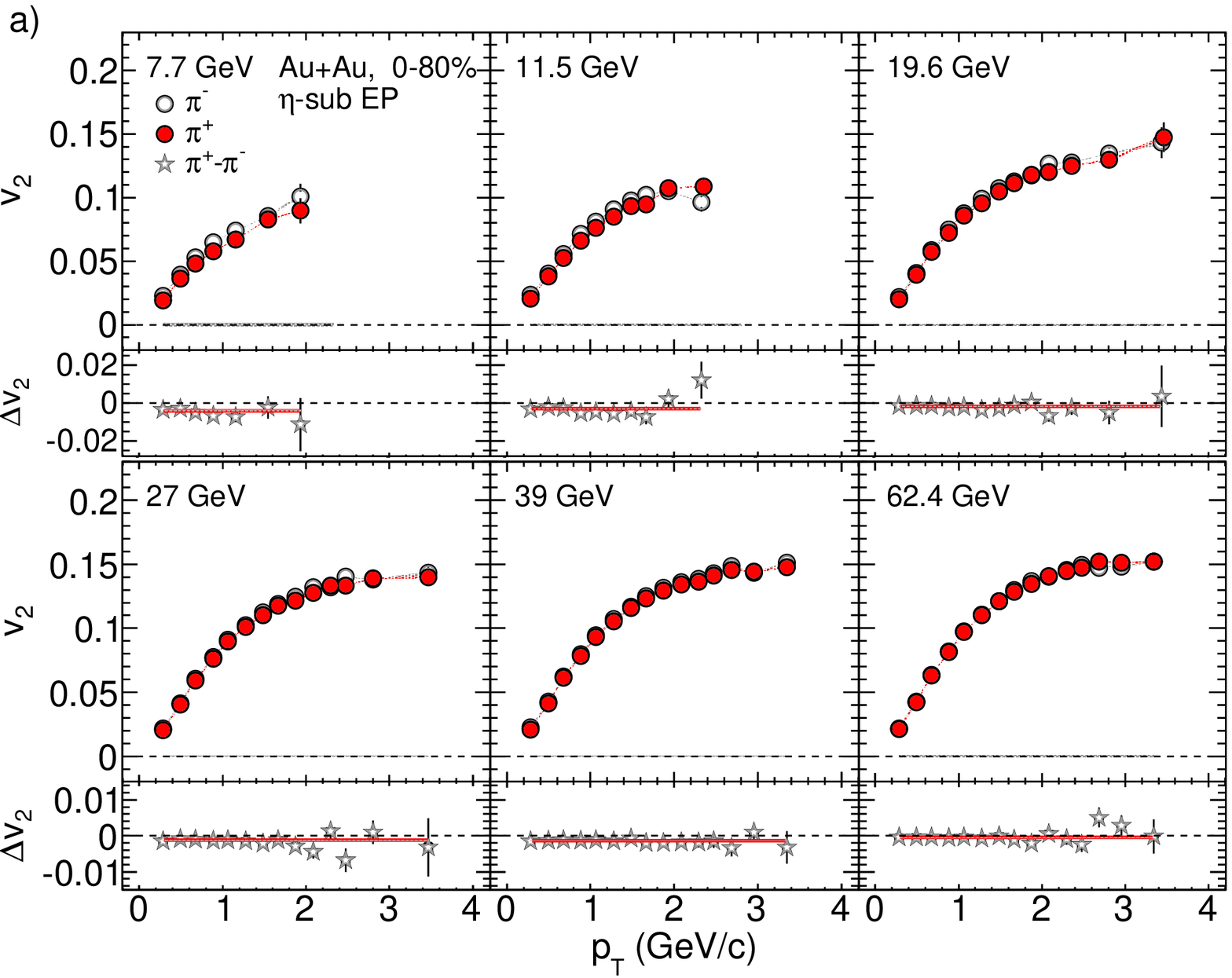}} \\[2ex]
\resizebox{13cm}{!}{%
\includegraphics[bb = 7.889765 2.250000 549.131960 423.874956,clip]{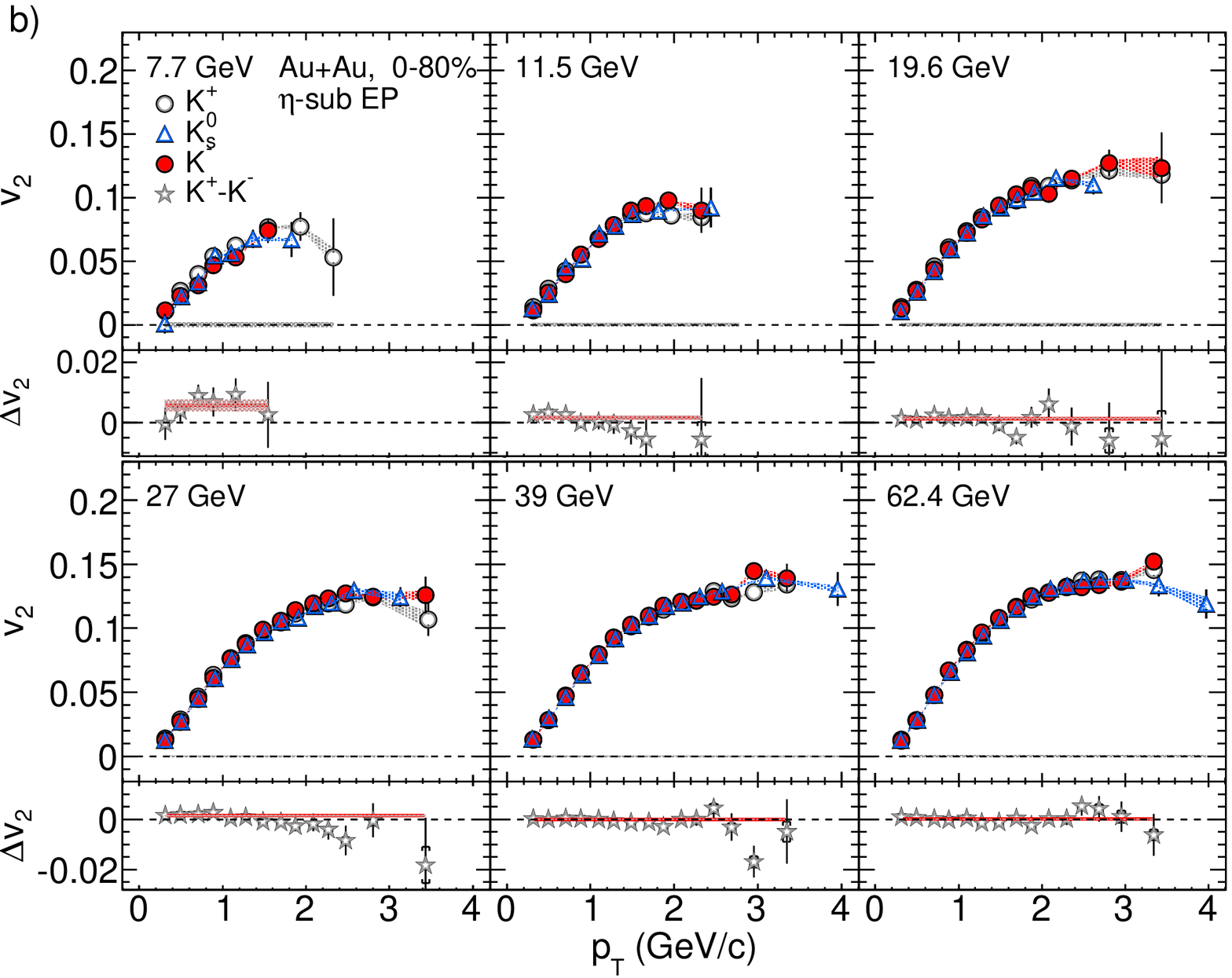}} %
 \caption{(Color online) The elliptic flow, $v_{2}$, of charged pions a) and kaons b) as a function of the transverse momentum, $p_{T}$, for 0--80\% central Au+Au collisions. The point-by-point systematic uncertainties are shown by the shaded areas, otherwise they are smaller than the symbol size. The global systematic uncertainties are shown as the shaded horizontal bars. The lower row of each panel shows the difference between a particle and corresponding anti-particle $v_{2}(p_{T})$ and a fit with a horizontal line. The red shaded area around each fit depicts the combined statistical and systematic fit errors. Different $\Delta v_{2}$ ranges were used for the upper and lower panels.}
 \label{fparticles_pion_kaon}
\end{figure*}

\begin{figure*}[]      
\resizebox{13.0cm}{!}{%
\includegraphics[bb = 7.889765 2.250000 549.131960 423.874956,clip]{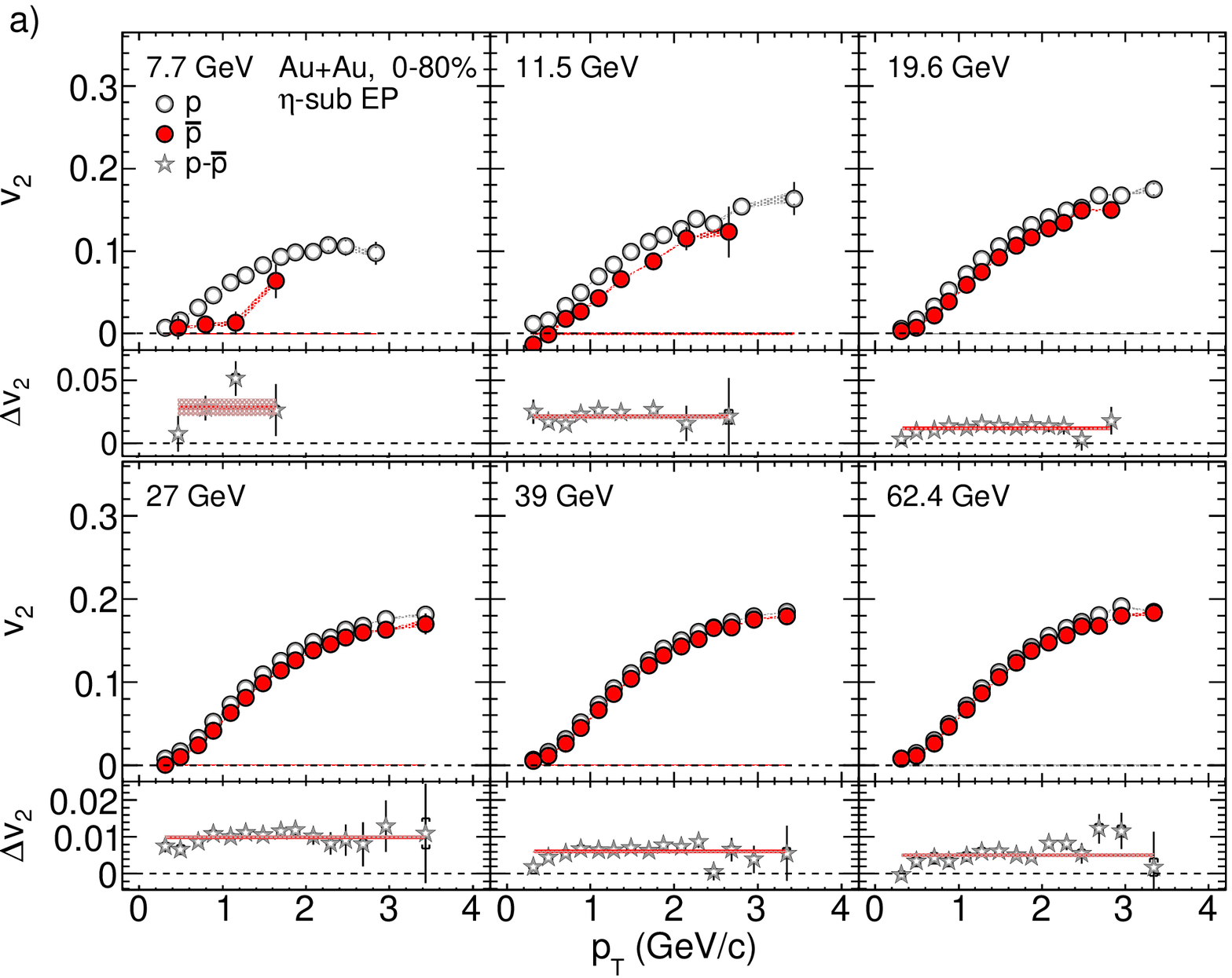}}  \\[2ex]
\resizebox{13.0cm}{!}{%
\includegraphics[bb = 7.889765 2.250000 549.131960 423.874956,clip]{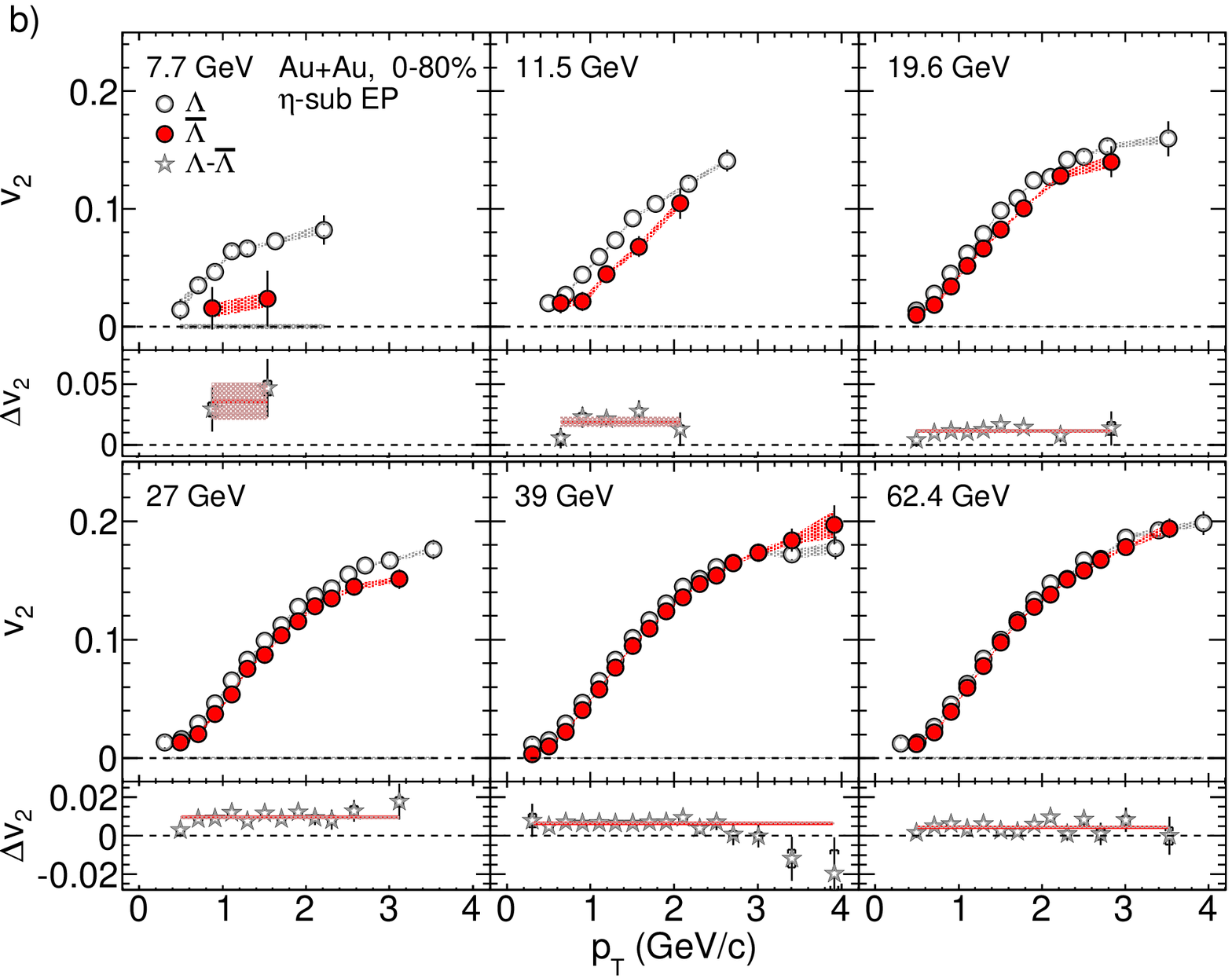}} %
 \caption{(Color online) The elliptic flow, $v_{2}$, of $p$, $\bar{p}$ a) and $\Lambda$, $\overline{\Lambda}$ b) as a function of the transverse momentum, $p_{T}$, for 0--80\% central Au+Au collisions. The point-by-point systematic uncertainties are shown by the shaded areas, otherwise they are smaller than the symbol size. The global systematic uncertainties are shown as the shaded horizontal bar. The lower row of each panel depicts the difference between a particle and corresponding anti-particle $v_{2}(p_{T})$ with a fit with a horizontal line. The red shaded area around each fit shows the combined statistical and systematic fit errors.}
 \label{fparticles_proton_lambda}
\end{figure*}

\begin{figure*}[]
\resizebox{13.0cm}{!}{%
\includegraphics[bb = 7.889765 2.250000 549.131960 423.874956,clip]{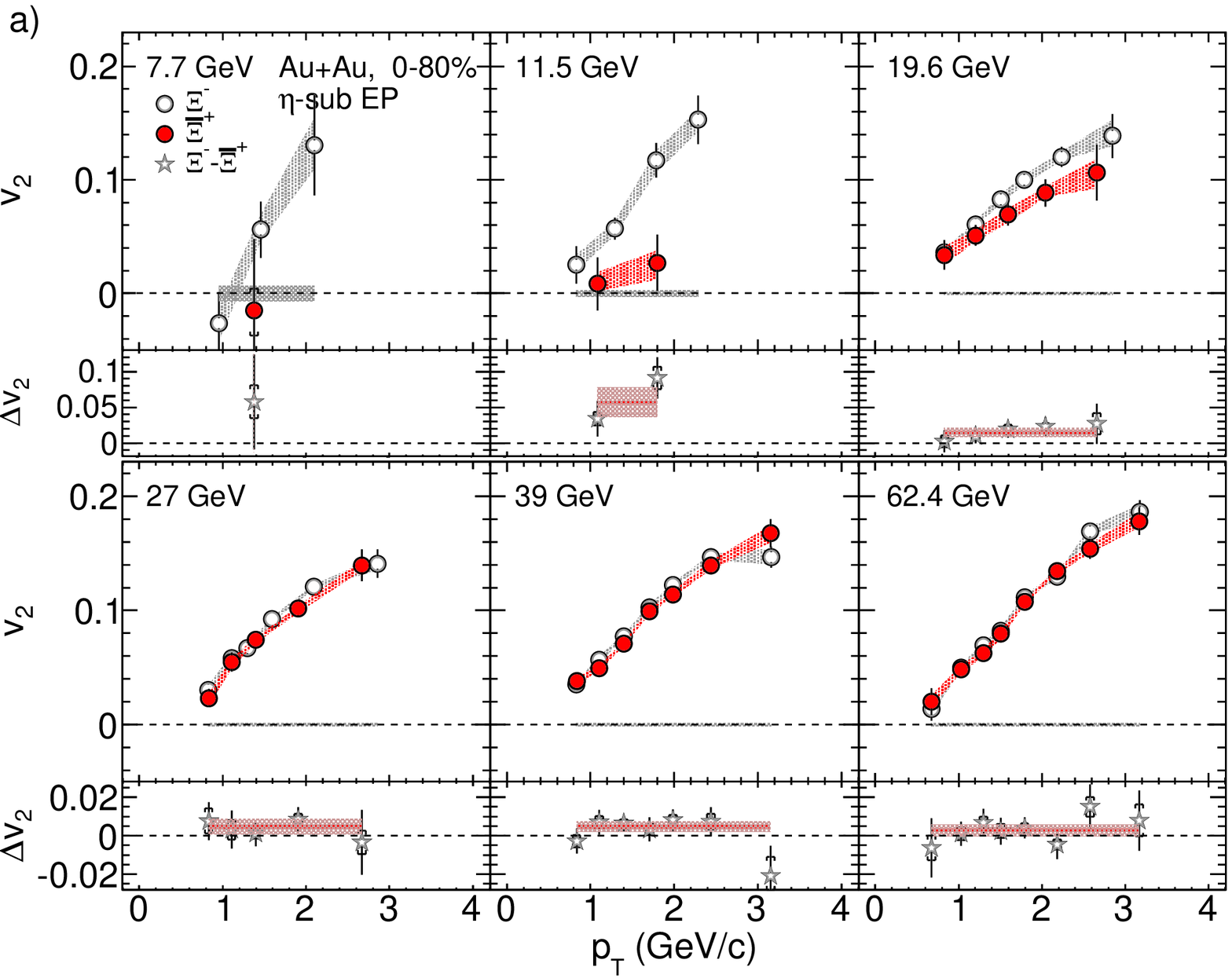}}  \\[2ex]
\resizebox{13.0cm}{!}{%
\includegraphics[bb = 7.889765 2.250000 549.131960 423.874956,clip]{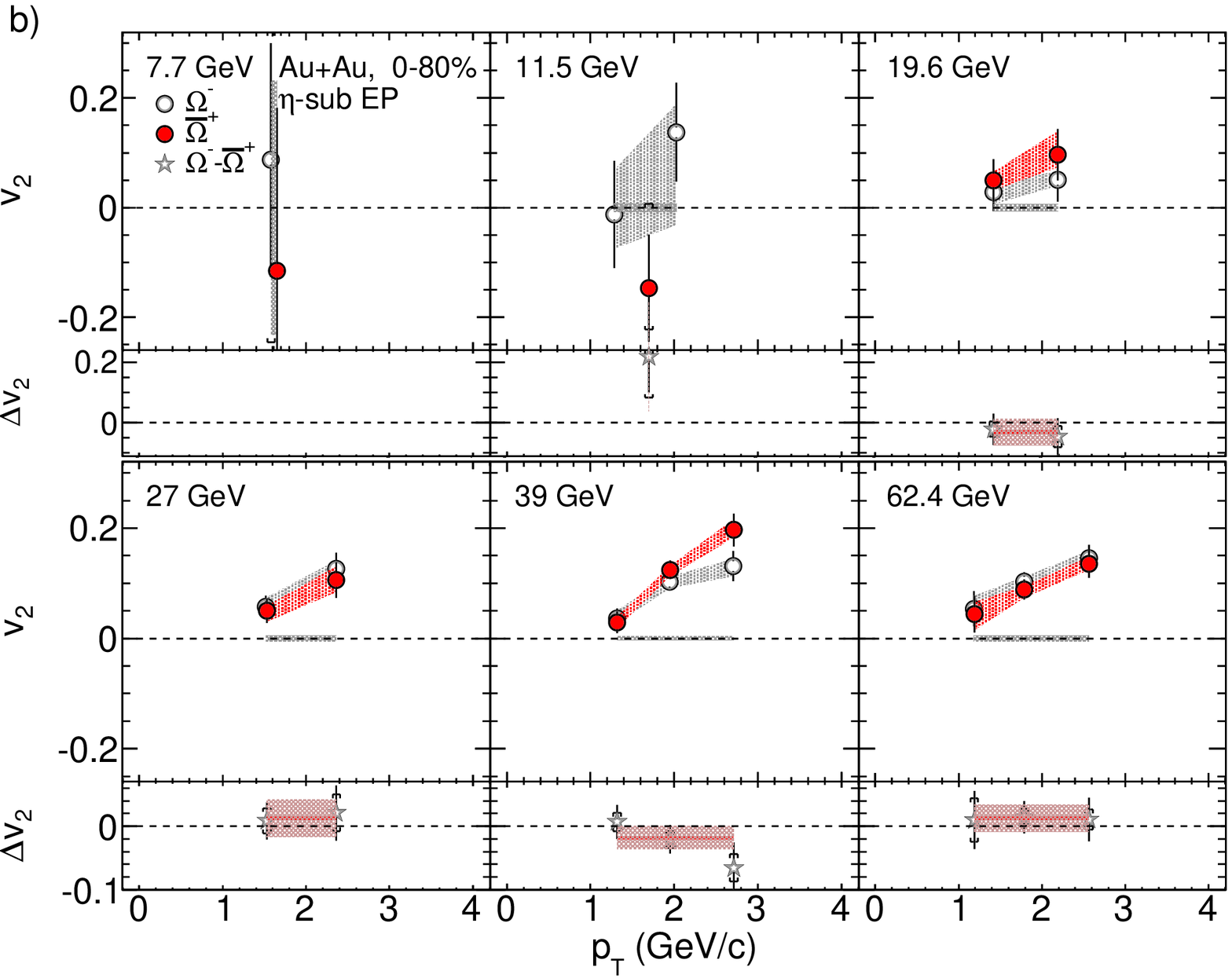}} %
 \caption{(Color online) The elliptic flow, $v_{2}$, of $\Xi^{-}$, $\overline{\Xi}^{+}$ a) and $\Omega^{-}$, $\overline{\Omega}^{+}$ b) as a function of the transverse momentum, $p_{T}$, for 0--80\% central Au+Au collisions. The point-by-point systematic uncertainties are shown by the shaded areas, while the global systematic uncertainties are shown as the shaded horizontal bar. Shown in the lower row of each panel is the difference between a particle and corresponding anti-particle $v_{2}(p_{T})$ with a fit with a horizontal line. The red shaded area around each fit shows the combined statistical and systematic fit errors.}
 \label{fparticles_Xi_Omega}
\end{figure*}

\begin{figure*}[]
\resizebox{13.0cm}{!}{%
  \includegraphics[bb = 13 4 505 301,clip]{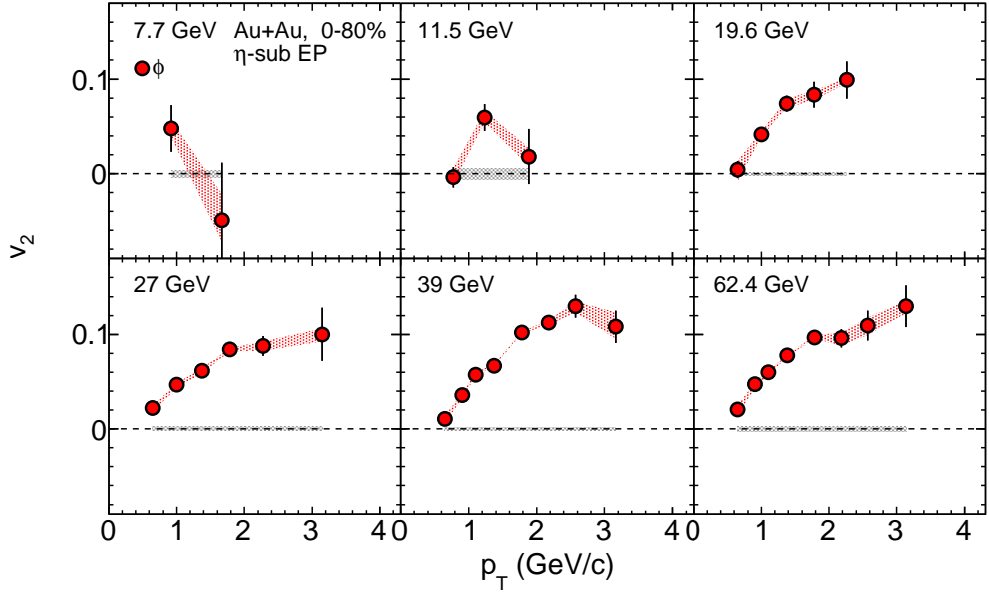}}  
\caption{(Color online) The elliptic flow, $v_{2}$, of $\phi$ mesons as a function of the transverse momentum, $p_{T}$, for 0--80\% central Au+Au collisions. The point-by-point systematic uncertainties are shown by the shaded areas, while the global systematic uncertainties are shown as the shaded horizontal bar.}
\label{fparticles_phi}       
\end{figure*}

\begin{figure*}[]
\resizebox{13.0cm}{!}{%
  \includegraphics[bb = 0.000000 1.242000 549.131960 435.672127,clip]{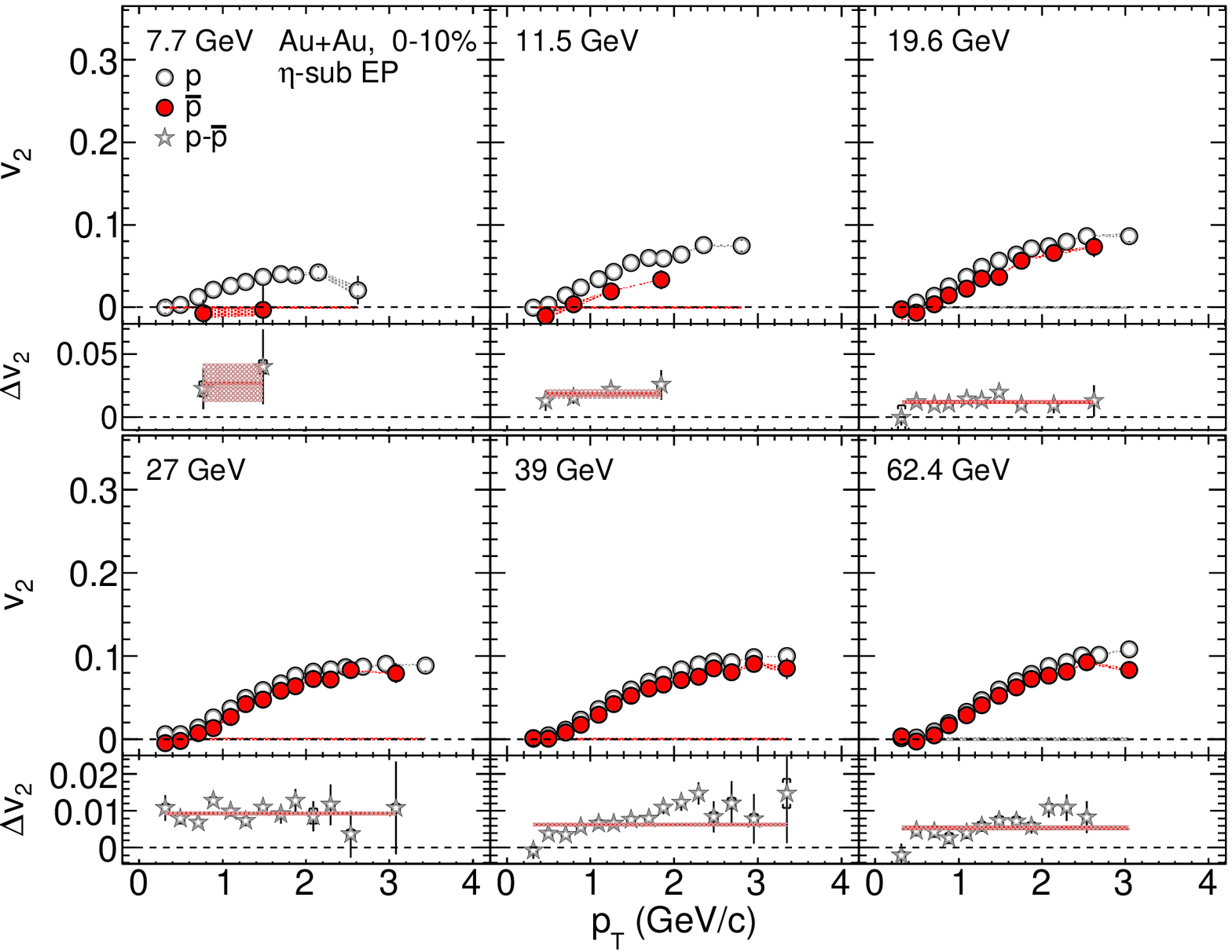}}  
\caption{(Color online) The elliptic flow, $v_{2}$, of $p$ and $\bar{p}$ as a function of the transverse momentum, $p_{T}$, for 0--10\% central Au+Au collisions. The point-by-point systematic uncertainties are shown by the shaded areas, while the global systematic uncertainties are shown as the shaded horizontal bar. Shown in the lower row of each panel is the difference between a particle and the corresponding anti-particle $v_{2}(p_{T})$ which are fit with a horizontal line. The red shaded area around each fit shows the combined statistical and systematic fit error.}
\label{fparticles_proton_cent1}       
\end{figure*}

\begin{figure*}[]
\resizebox{13.0cm}{!}{%
  \includegraphics[bb = 0.000000 1.242000 549.131960 435.672127,clip]{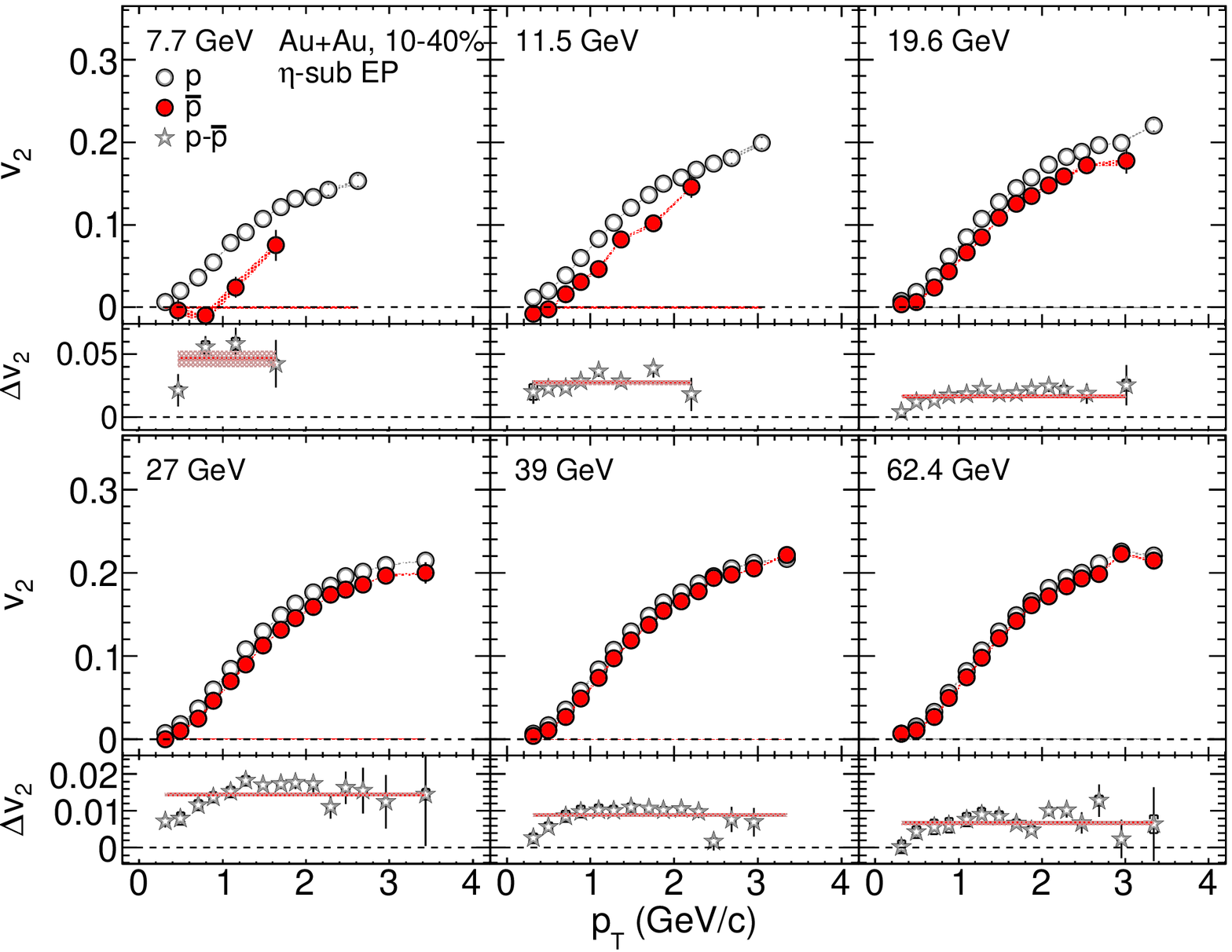}}  
\caption{(Color online) The elliptic flow, $v_{2}$, of $p$ and $\bar{p}$ as a function of transverse momentum, $p_{T}$, for 10--40\% central Au+Au collisions. The point-by-point systematic uncertainties are shown by the shaded areas, while the global systematic uncertainties are shown as the shaded horizontal bar. Shown in the lower row of each panel is the difference between a particle and the corresponding anti-particle $v_{2}(p_{T})$ which are fit with a horizontal line. The red shaded area around each fit shows the combined statistical and systematic fit error.}
\label{fparticles_proton_cent2}       
\end{figure*}

\begin{figure*}[]
\resizebox{13.0cm}{!}{%
  \includegraphics[bb = 0.000000 1.242000 549.131960 435.672127,clip]{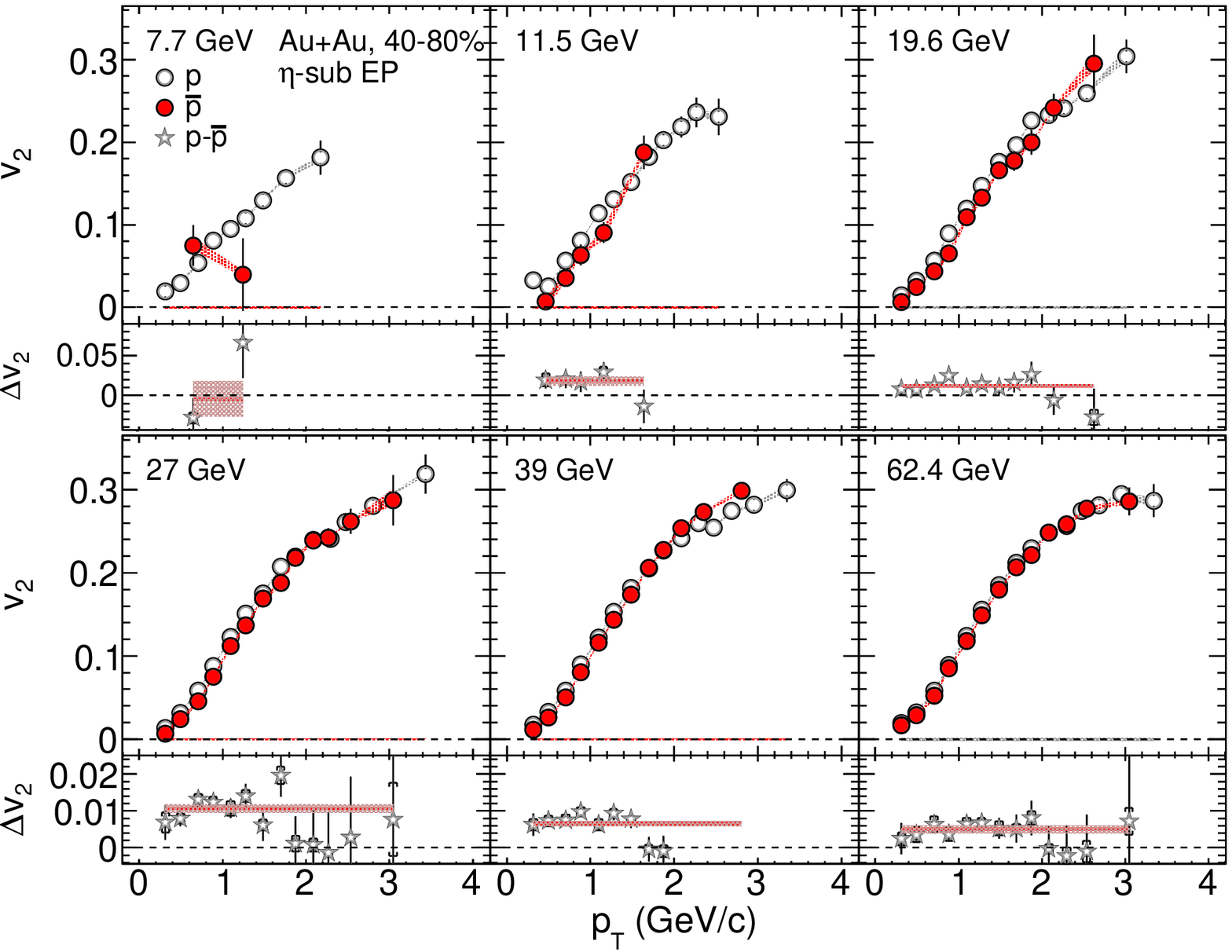}}  
\caption{(Color online) Elliptic flow, $v_{2}$, of $p$ and $\bar{p}$ as a function of transverse momentum $p_{T}$ for 40--80\% centrality Au+Au collisions. The point-by-point systematic uncertainties are shown by the shaded areas, while the global systematic uncertainties are shown as the shaded horizontal bar. Shown in the lower row of each panel is the difference between a particle and the corresponding anti-particle $v_{2}(p_{T})$ which are fit with a horizontal line. The red shaded area around each fit shows the combined statistical and systematic fit error.}
\label{fparticles_proton_cent3}       
\end{figure*}

The lower the energy, the smaller is the difference between the various particles in $v_{2}(p_{T})$ at $p_{T} < 1.5$ \GeVc. 
This could be related to a reduction of radial flow as the beam energy decreases. However, no narrowing of the spread 
of $v_{2}(p_{T})$ with beam energy is observed for the anti-particles.  At lower beam energies, the $v_{2}(p_{T})$ values
for $\bar{p}$ and $\overline{\Lambda}$ were significantly smaller than the values for their partner particles. The possible physics 
implications due to the differences in particle and anti-particle $v_{2}(p_{T})$ will be discussed in more detail in the next sections.

In Fig.~\ref{fparticles_pion_kaon}, each particle $v_{2}(p_{T})$ is directly compared, if possible, to that for its anti-particle. For the mesons the anti-particle convention from~\cite{Beringer:1900zz} is used. The point-by-point systematic uncertainties are displayed as the shaded bands which connect the data points. The global systematic uncertainties are shown as the error bands along the horizontal axis. Shown are the $v_{2}(p_{T})$ for $\pi^{+}(u\bar{d})$, $\pi^{-}(\bar{u}d)$, and $K^{+}(u\bar{s})$, $K_{s}^{0}((d\bar{s}-\bar{s}d)/\sqrt{2})$, $K^{-}(\bar{u}s)$.  At the higher energies of 27, 39 and 62.4 GeV, the charged pion $\pi^{+}$ and $\pi^{-}$ $v_{2}(p_{T})$ values show almost identical shapes and amplitudes, as expected from particles with the same mass and number of quarks. At lower energies, an increasing difference between $v_{2}(\pi^{+})$ and $v_{2}(\pi^{-})$ is observed, where $v_{2}(\pi^{-})$ is larger than $v_{2}(\pi^{+})$ for all $p_{T}$ values. In the lower rows of each panel in Fig.~\ref{fparticles_pion_kaon}, the difference in $v_{2}(p_{T})$ between particles and anti-particles is shown. The red line shows a horizontal line fit to the $\Delta v_{2}$ which will be used below (Section~\ref{sub_sec_energy_dep}) to study the energy dependence of the difference. The fit range was varied to estimate the systematic uncertainty for the fit and to test the assumption of a constant difference as a function of $p_{T}$. 

The fact that $v_{2}(\pi^{-})$ is larger than $v_{2}(\pi^{+})$ could be due to the Coulomb repulsion of $\pi^{+}$ by 
the mid-rapidity protons or to the chiral magnetic effect in finite baryon density matter
produced in the collisions~\cite{Burnier:2011bf}.
The charged kaons show an opposite trend compared to the charged pions. The $v_{2}(p_{T})$ values of $K^{+}$ are larger compared to 
$K^{-}$. The size of the difference in $v_{2}$ and the energy dependence is comparable to that of the pions. The neutral $K_{s}^{0}$ approximately follow the trends of the $v_{2}(p_{T})$ values of the $K^{-}$. 

In contrast to the charged pions and kaons, a significant difference in the $v_{2}(p_{T})$ values between $p(uud)$ and $\bar{p}(\bar{u}\bar{u}\bar{d})$ was already observed at 62.4 GeV, as shown in Fig.~\ref{fparticles_proton_lambda} a). The difference in $v_{2}$ is nearly constant as a function of $p_{T}$ and, as for the pions and kaons, the difference  increases with decreasing energy. Compared to the kaons and pions, the relative difference is at least a factor of three larger. The plots in Fig.~\ref{fparticles_proton_lambda} b) show the corresponding $v_{2}(p_{T})$ for $\Lambda(uds)$ and $\overline{\Lambda}(\bar{u}\bar{d}\bar{s})$.  The shapes and magnitudes of $v_{2}(p_{T})$ for all energies are almost identical between $p$ and $\Lambda$ and the same between $\bar{p}$ and $\overline{\Lambda}$. Hence, the difference in $v_{2}(p_{T})$ between the (anti)$\Lambda$ particles and the (anti)protons is observed. It appears that the exchange of a u-quarks with an s-quark has no influence on the difference in $v_{2}(p_{T})$. 

Figure~\ref{fparticles_Xi_Omega} a) shows the $v_{2}(p_{T})$ of $\Xi^{-}(dss)$ and $\overline{\Xi}^{+}(\bar{d}\bar{s}\bar{s})$ and b) shows the $v_{2}(p_{T})$ of $\Omega^{-}(sss)$ and $\overline{\Omega}^{+}(\bar{s}\bar{s}\bar{s})$. Within the statistical and systematic uncertainties, $\Xi^{-}$ and $\overline{\Xi}^{+}$ are indistinguishable in $v_{2}(p_{T})$ at 62.4 GeV. At 39 and 27 GeV, only a slightly larger $v_{2}(p_{T})$ of $\Xi^{-}$ with respect to $\overline{\Xi}^{+}$ is observed, whereas at 19.6 and 11.5 GeV the difference is significant and comparable to that of the protons and $\Lambda$. Due to the larger error bars, no significant effect is observed for the $\Omega^{-}$ and $\overline{\Omega}^{+}$ at any energy. 

As mentioned above, the $\phi(s\bar{s})$ meson $v_{2}(p_{T})$ is of particular interest. The hadronic cross section of $\phi$ mesons is much smaller compared to that of other hadrons~\cite{Shor:1984ui,Sibirtsev:2006yk,vanHecke:1998yu,Cheng:2003as}. This would result in a smaller $v_{2}(p_{T})$ for a fireball evolution in the hadron gas phase. The results are shown in Fig.~\ref{fparticles_phi}. At 19.6 to 62.4 GeV, the typical $v_{2}(p_{T})$ shape is seen, whereas at 7.7 and 11.5 GeV, the $v_{2}$ values at the highest measured $p_{T}$ bins are close to zero. Also, there is a significant decrease in the energy dependence of $v_{2}(p_{T})$ ({\it cf.} Fig.~\ref{fv2_pT_energy}) at transverse momenta of about 1.5--2 \GeVc. 

In~\cite{Adamova:2012md} $v_{2}(p_{T})$ studies at $\sqrt{s_{NN}}$ = 17.3 GeV for pions, kaons, and strange particles are presented for mid-central Pb+Au collisions. Due to the different centrality selection, a direct comparison was not performed.

\subsubsection{Centrality dependence of proton and anti-proton $v_{2}$}
The elliptic flow shows a strong centrality dependence which is driven by the changing initial spatial eccentricity. The present results are an average over a wide (0--80\%) centrality range. Even if the $v_{2}(p_{T})$ values for protons and anti-protons would be identical for all collision centralities, one would observe a difference in the $v_{2}$ values if the centrality dependency of the production rates would be very different. To study this possibility, Fig.~\ref{fparticles_proton_cent1}, Fig.~\ref{fparticles_proton_cent2}, and Fig.~\ref{fparticles_proton_cent3} show the proton and anti-proton $v_{2}(p_{T})$ values for the centrality ranges of 0--10\%, 10--40\%, and 40--80\% for the six beam energies. In all three of these narrower centrality ranges, a significant difference between the $p$ and $\bar{p}$ $v_{2}(p_{T})$ values is seen. For the most peripheral centrality bin (40--80\%), the elliptic flow is the largest, but the absolute difference $\Delta v_{2}(p_{T})$ is smaller compared to the mid-central bin (10--40\%) and is comparable to the most central bin (0--10\%). It is concluded that $\Delta v_{2}(p_{T})$ shows a clear centrality dependence for protons and anti-protons, and that the difference in $v_{2}(p_{T})$ remains when restricted to narrower centrality ranges.

\subsection{Elliptic flow as a function of transverse mass}
\begin{figure*}[htc!]
\resizebox{13.0cm}{!}{%
\includegraphics[bb = 13 4 505 301,clip]{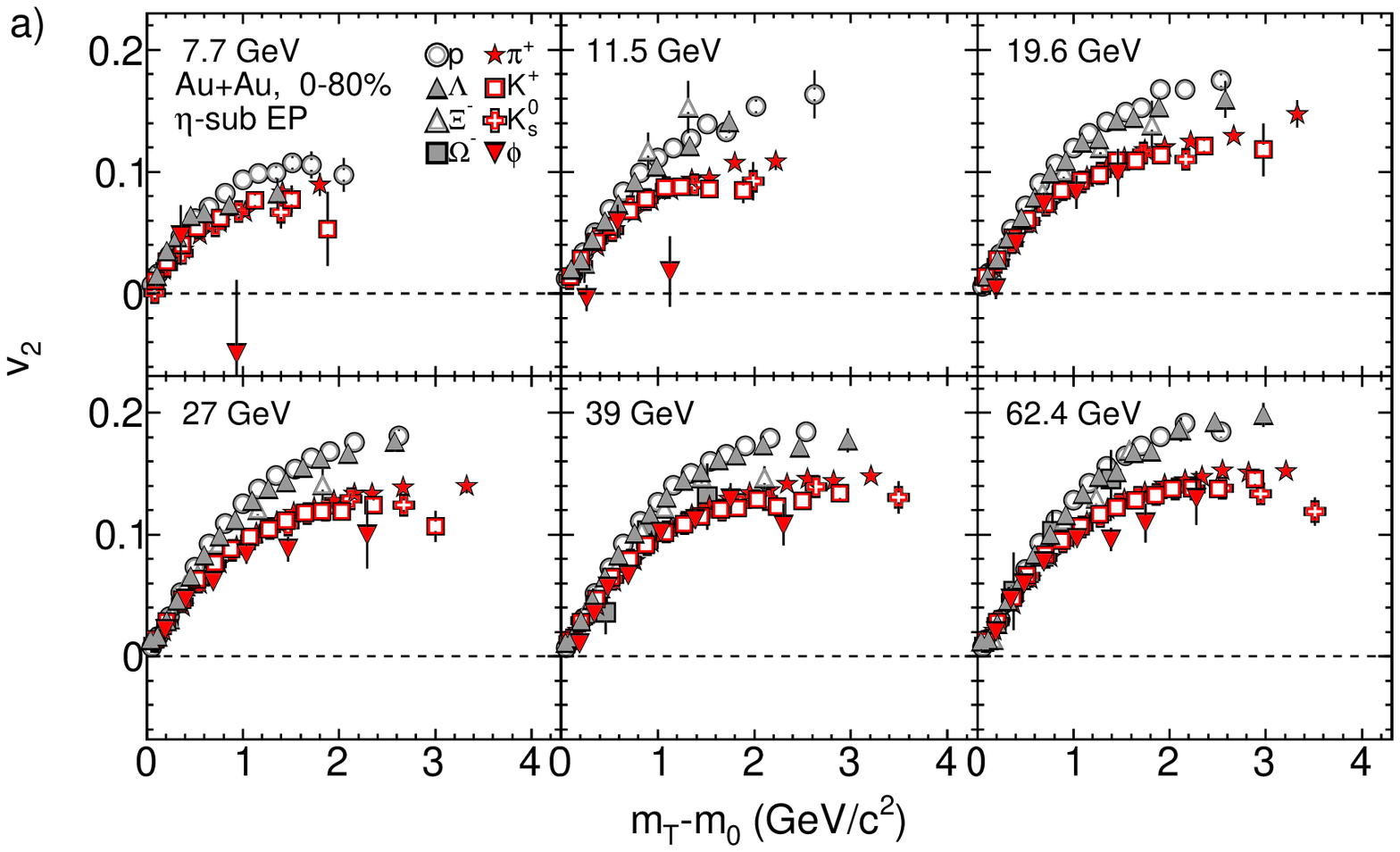}} \\[2ex]
\resizebox{13.0cm}{!}{%
\includegraphics[bb = 13 4 505 301,clip]{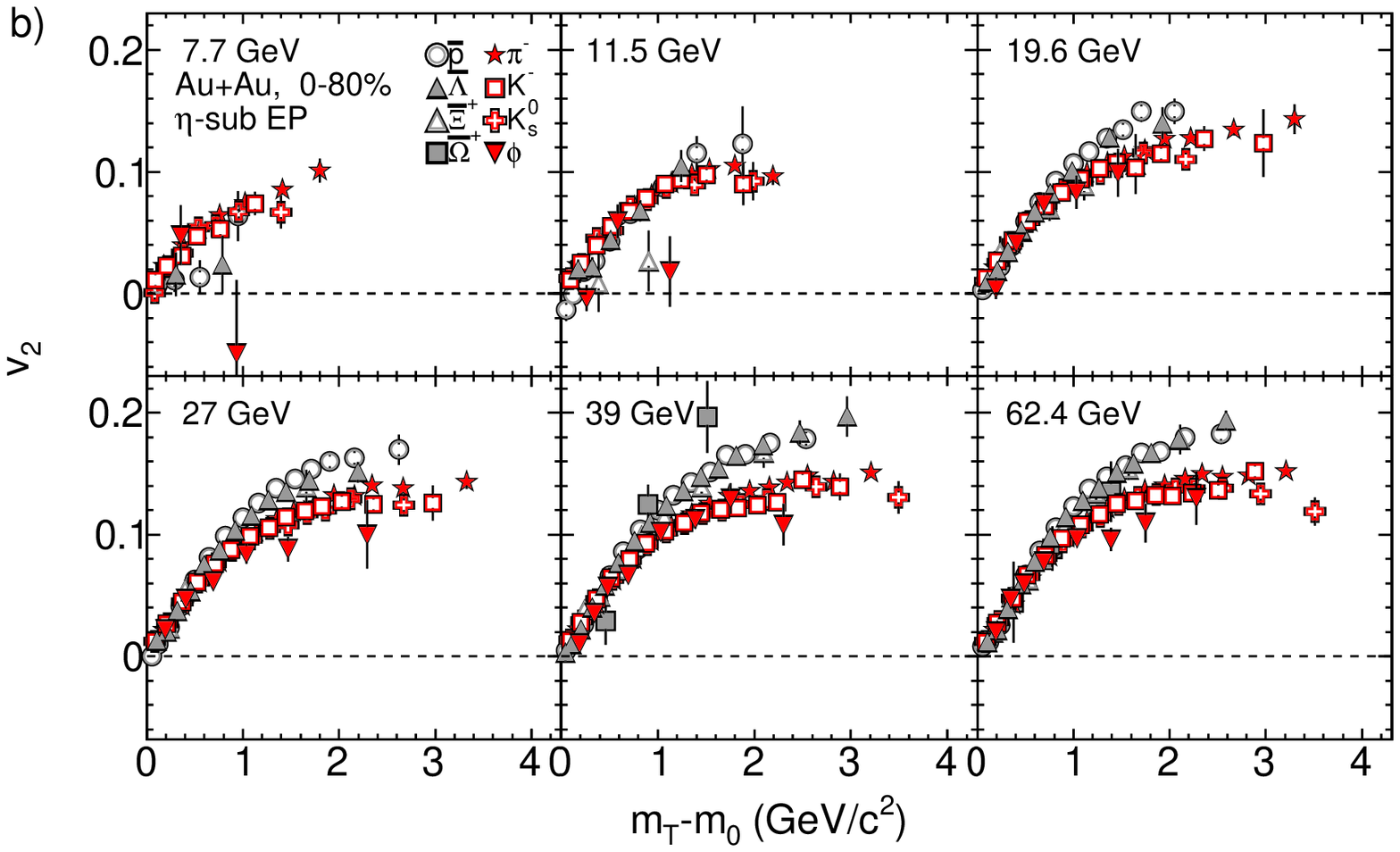}} %
 \caption{(Color online) The elliptic flow, $v_{2}$, of 0--80\% central Au+Au collisions as a function of the reduced transverse mass, $m_{T}-m_{0}$, for selected particles a) and anti-particles b). A significant splitting between the baryons (grey) and mesons (red) is observed at the higher energies. The splitting becomes smaller at 7.7 GeV. At lower energies, the baryons and mesons are consistent with each other within the measured $p_{T}$ range for the particles shown in b).}
 \label{f_v2_mt}
\end{figure*}

The $v_{2}$ values as a function of the reduced transverse mass, $m_{T}-m_{0}$, shows a clear splitting between baryons and mesons for larger $m_{T}-m_{0}$ values at $\sqrt{s_{NN}}=$200 GeV~\cite{Huovinen:2001cy}. The particle mass, charge, and strangeness content are not the driving factors. Only the number of constituent quarks separates the results into the two branches. This observation is an indication that the results are sensitive to the particle internal degrees of freedom, {\it i.e.} the quarks in the QGP phase of the collision. After hadronization, the flow of the quarks is carried by the measured particles. In a coalescence picture, this will result in the $v_{2}$ values of the baryons being a factor of 1.5 larger than the $v_{2}$ values of the mesons~\cite{Huovinen:2001cy}. Figure~\ref{f_v2_mt} shows the $v_{2}(m_{T}-m_{0})$ values for all six BES energies and the same selection of particles a) and corresponding anti-particles b) as presented above. The baryons and mesons are clearly separated in Fig.~\ref{f_v2_mt} a) above $(m_{T}-m_{0}) > 1\  \rm{GeV}/c^{2}$. The separation at 7.7 GeV between protons and $\pi^{+}$, $K^{+}$ is significantly smaller than that at all of the other energies. The $\Lambda$ hyperons follow the meson branch at 7.7 GeV. 

The anti-particles at 39 and 62.4 GeV show a similar behavior as the particles, and at all lower energies the meson and baryon branches approach each other. At 11.5 GeV, a difference between the baryons and mesons is no longer observed, and at 7.7 GeV the anti-proton and $\overline{\Lambda}$ $v_{2}(m_{T}-m_{0})$ are below the meson branch in the measured $m_{T}-m_{0}$ range.
The trend observed is a decrease in the baryon-meson splitting in $v_{2}(m_{T}-m_{0})$ for ($m_{T}-m_{0}$) $>$ 1 GeV/$c^2$
as the energy is lowered, both for the particle and anti-particle groups. 

\subsection{Number-of-constituent quark scaling of $v_{2}$}
The splitting in $v_{2}(m_{T}-m_{0})$ between the mesons and baryons at transverse mass values above $1\ \rm{GeV}/c^{2}$ shown in Fig.~\ref{f_v2_mt} implies a dependence of the $v_{2}$ values on the number of constituent quarks, $n_{q}$. The NCQ scaling was originally predicted for $v_{2}(p_{T})$ at intermediate transverse momenta~\cite{Voloshin:2002wa}. A scaling of $p_{T}$ and $v_{2}$ with $1/n_{q}$ was suggested. Indeed, the scaled $v_{2}$ values for all particles at 200 GeV Au+Au collisions collapse to a common single trend at intermediate $p_{T}$ values~\cite{Abelev:2007rw,Abelev:2010tr,Adams:2005zg,Abelev:2007qg,Adare:2012vq}. This is interpreted as a possible signature for partonic degrees of freedom (quarks and gluons) in the initial stage of the system, where most of the elliptic flow develops. This scaling should vanish in a hadron gas system at lower energies. Thus, the break down of NCQ scaling would be a necessary signature for a QCD phase transition from partonic to hadronic matter. 

\begin{figure*}[htc!]
\resizebox{13.0cm}{!}{%
\includegraphics[bb = 13 4 505 301,clip]{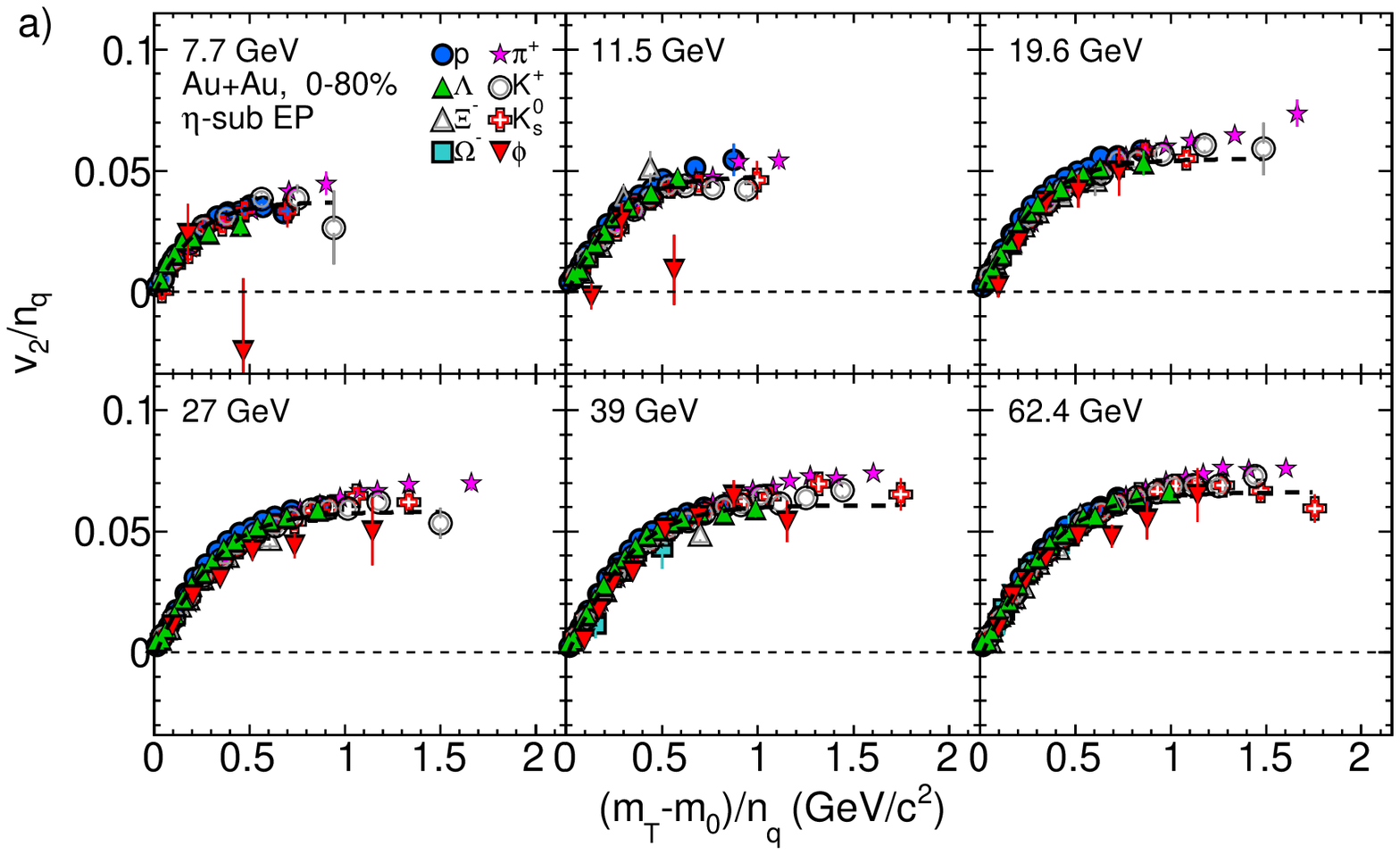}} \\[2ex]
\resizebox{13.0cm}{!}{%
\includegraphics[bb = 13 4 505 301,clip]{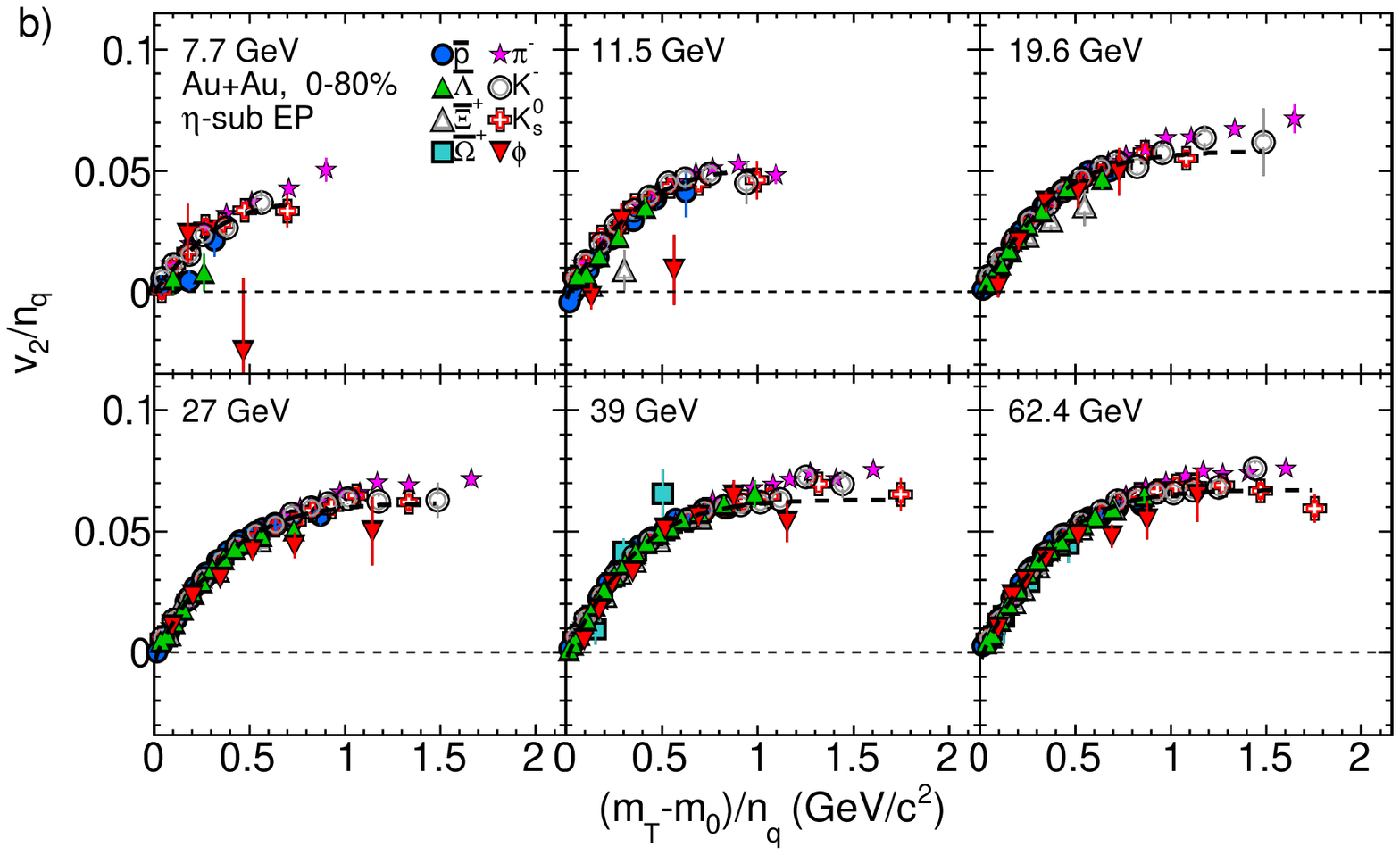}} %
 \caption{(Color online) The Number-of-Constituent Quark (NCQ) scaled elliptic flow, $v_{2}/n_{q}$ versus $(m_{T}-m_{0})/n_{q}$, for 0--80\% central Au+Au collisions for selected particles a) and corresponding anti-particles b). The dashed lines show the results of simultaneous fits with Eq.~(\ref{Func_v2_fit}) to all particles except the pions.}
 \label{fNCQ_mT}
\end{figure*}

Since particles and anti-particles have the same number of quarks, the NCQ scaling transformation of $v_{2}$ does not change their relative separation. This means that the difference in $v_{2}(p_{T})$ for particles and corresponding anti-particles observed in Section~\ref{sub_v2_pt} constitutes a violation of this NCQ scaling. Possible physics causes for this difference will be discussed below.
In the following, NCQ scaling will be shown separately for a selection of particles and anti-particles. Since 
a better agreement between the different particles (even at low $(m_{T}-m_{0})/n_{q}$ values) is achieved
with	 the $(v_{2}/n_{q})((m_{T}-m_{0})/n_{q})$ scaling compared to the $(v_{2}/n_{q})(p_{T}/n_{q})$ scaling,
Fig.~\ref{fNCQ_mT} presents the scaled distributions versus $(m_{T}-m_{0})/n_{q}$. The corresponding scaled plots for $v_{2}(p_{T})$ are shown in Fig.~\ref{fNCQ_pT} of the Appendix.

\begin{figure*}[htc!]
\resizebox{13.0cm}{!}{%
\includegraphics[bb = 13 4 505 301,clip]{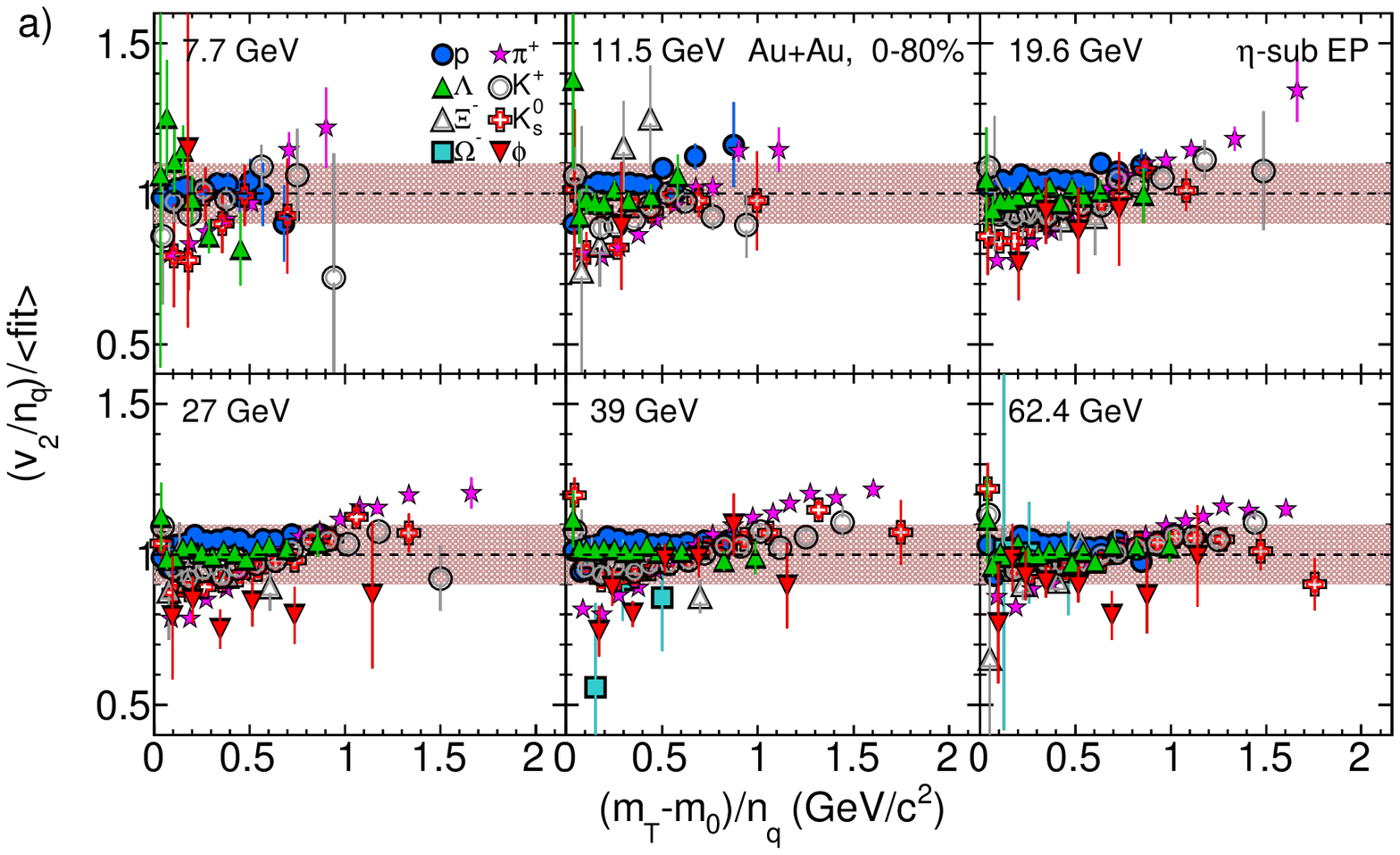}} \\[2ex]
\resizebox{13.0cm}{!}{%
\includegraphics[bb = 13 4 505 301,clip]{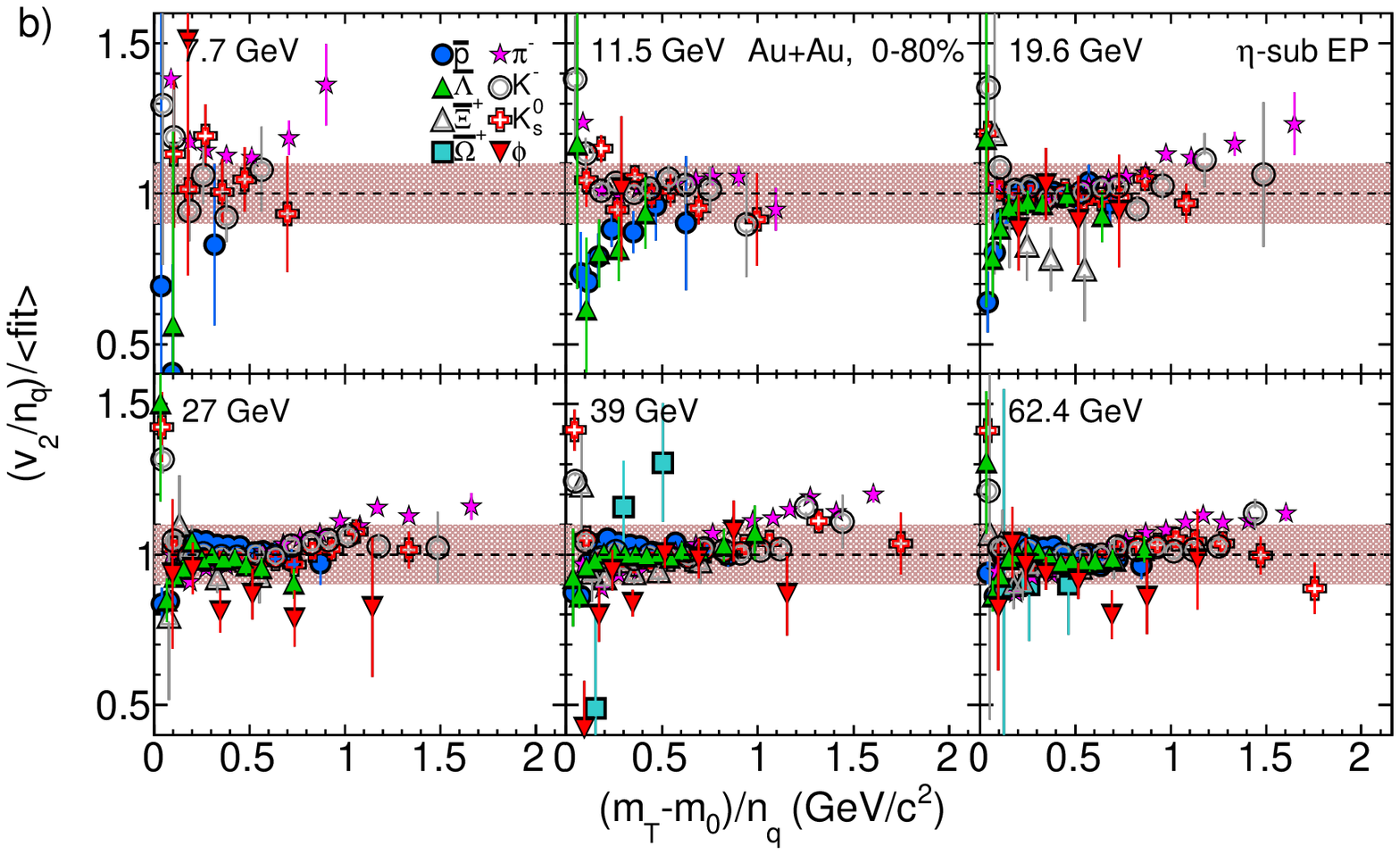}} %
 \caption{(Color online) The Number-of-Constituent Quark (NCQ) scaled elliptic flow, $v_{2}/n_{q}$ versus $(m_{T}-m_{0})/n_{q}$, ratio to the fit function (see text) for 0--80\% central Au+Au collisions for selected particles a) and corresponding anti-particles b). Most of the data points at the larger $(m_{T}-m_{0})/n_{q}$ values are within a $\pm$10\% interval around unity, which is shown as the shaded region to guide the eye. Some data points for $\phi$ and $\overline{\Xi}^{+}$ are outside of the plot axis range.}
 \label{fNCQ_mT_ratio}
\end{figure*}

The NCQ scaling should only hold in the transverse momentum range of $1.5<p_{T}<5$ \GeVc\ \cite{Huovinen:2001cy,Fries:2003kq}. For the corresponding scaled transverse mass and transverse momentum range, a fair agreement for most of the particles and energies is observed. Only the $\phi$ mesons deviate from the trend at 7.7 and 11.5 GeV, with the maximum measured $p_{T}/n_{q}$ value just reaching the lower edge of the expected NCQ scaling range. The values deviate from those for the other particles and anti-particles at the highest ($m_{T}-m_{0}$) values at $\sqrt{s_{NN}}$ = 7.7 and 11.5 GeV by 1.8$\sigma$ and 2.3$\sigma$, respectively. The values for $\overline{\Xi}^{+}$ at 11.5 GeV were similar and show a relatively small $v_{2}$ compared to the other hadrons. This could be related to the lower hadronic cross sections of particles containing multiple strange quarks. These observations may indicate that hadronic interactions become more important than partonic effects for the systems formed at collision energies $\lesssim$ 11.5 GeV~\cite{Mohanty:2009tz,Nasim:2013fb}. It is noted that recent results of elliptic flow measurements of $J/\Psi$ mesons at 
$\sqrt{s_{NN}}=$ 200 GeV also show smaller $v_{2}(p_{T})$ values compared to those for other hadrons~\cite{Adamczyk:2012pw}. However, in
this representation of NCQ scaling, both particles and anti-particles appear to follow the scaling. In the previous sub-section, 
an absence of the baryon-meson splitting of $v_{2}(m_{T}-m_{0})$ for ($m_{T}-m_{0}$) $>$ 1 GeV/$c^2$ is observed.

For each energy, simultaneous fits with Eq.~(\ref{Func_v2_fit}) were applied to all particles except the pions, which are biased by resonance decays~\cite{Dong:2004ve}. The ratios of the data to the fits are shown in Fig.~\ref{fNCQ_mT_ratio} and for the transverse momentum in Fig.~\ref{fNCQ_pT_ratio} of the Appendix. Most of the data points in the high transverse momentum range, agree within the uncertainties within a $\pm$10\% interval around unity.  At lower values of $m_{T}-m_{0}$, larger
deviations from unity are observed.

\FloatBarrier

\subsection{Energy dependence of the particle and anti-particle $v_{2}$ difference}
\label{sub_sec_energy_dep}

\begin{figure}[htc!]
\includegraphics[width=0.48\textwidth]{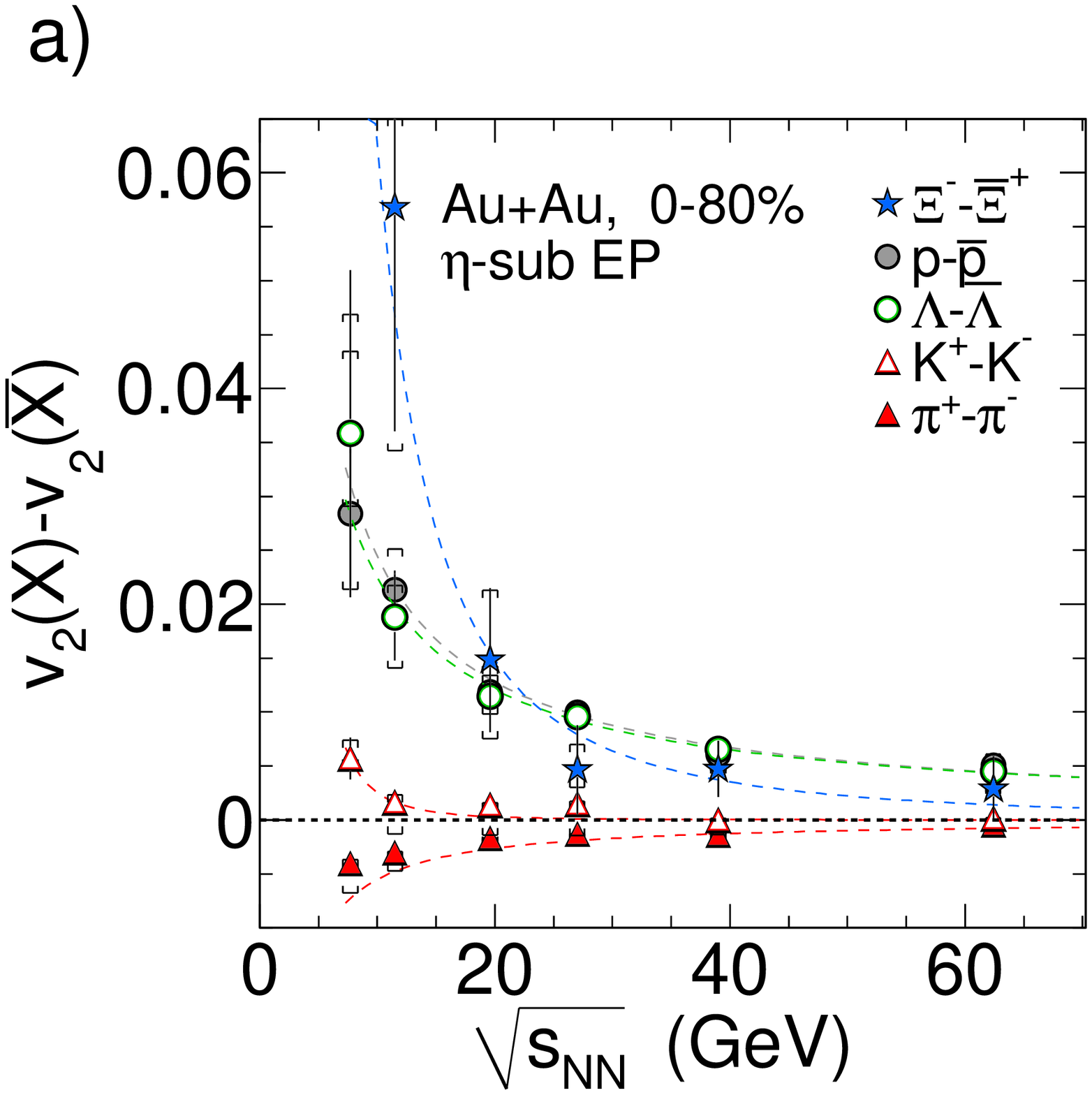} %
\includegraphics[width=0.48\textwidth]{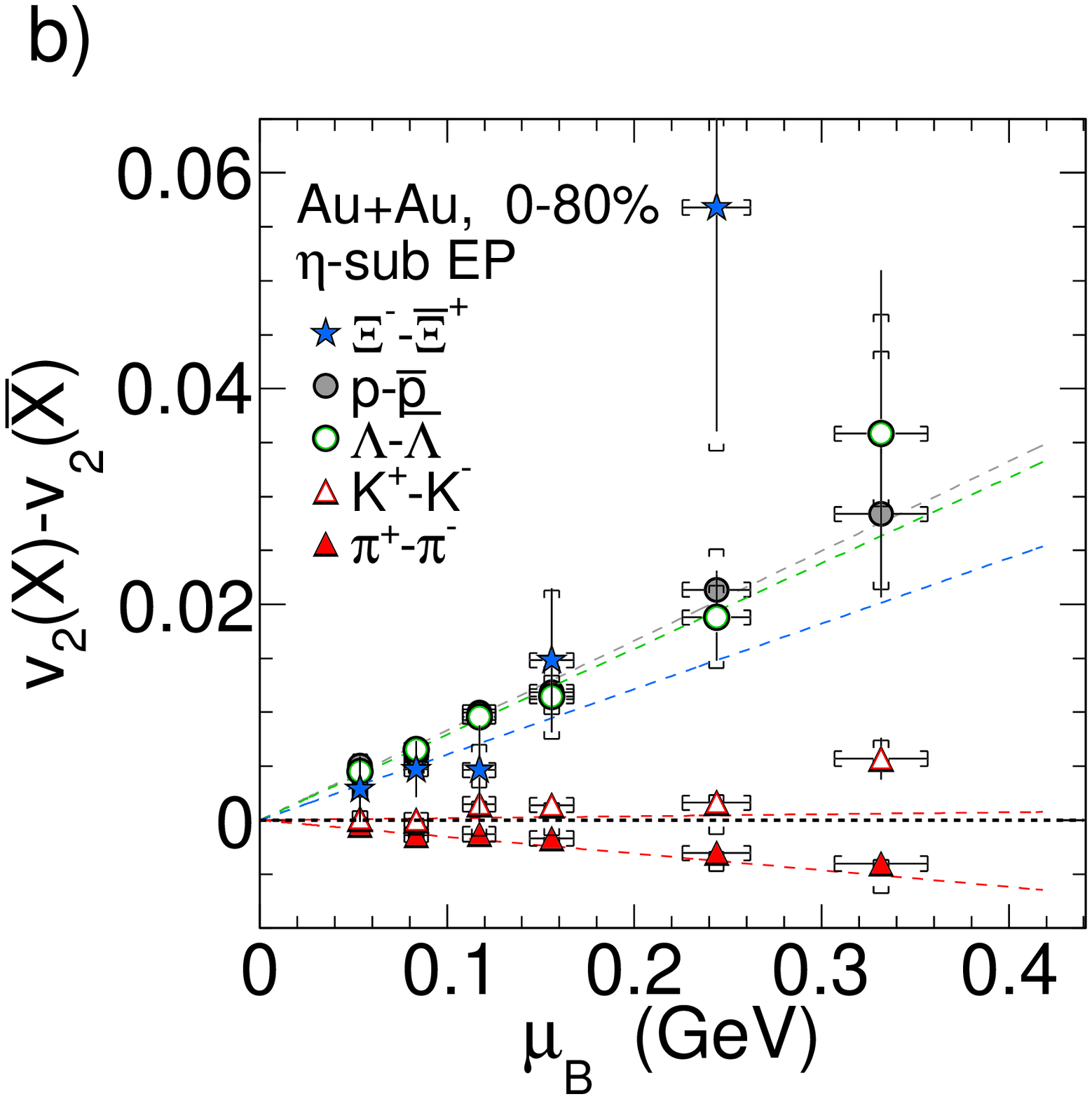} %
 \caption{(Color online) The difference in the $v_{2}$ values between a particle $X$ and its corresponding anti-particle (${\rm \overline{X}}$) (see legend) as a function of $\sqrt{s_{NN}}$ a) and $\mu_{B}$ b) for 0--80\% central Au+Au collisions. The dashed lines in plot a) are fits using Eq.~(\ref{fDelta_v2}), and lines through the origin are shown for plot b). The values of $\mu_{B}$ are from the parametrization of Ref.~\cite{Tiwari:2011km} (see text for details).}
 \label{fDiff_v2_sNN_muB}
\end{figure}

\begin{table*}[htc]  
\caption{\label{table_2}Fit parameters $a$ and $b$ of Eq.~(\ref{fDelta_v2}) and the slopes $m$ of the straight line fits shown in Fig.~\ref{fDiff_v2_sNN_muB}  b) for the different particle species. The first $\chi^{2}$ per Number-of-Degrees of Freedom (NDF) value corresponds to the fit with Eq.~(\ref{fDelta_v2}), the second to the straight line fits. Both include statistical and systematic uncertainties.}  
\footnotesize\rm  
\begin{tabular}{ >{\centering\arraybackslash}m{1in} >{\centering\arraybackslash}m{1in}  >{\centering\arraybackslash}m{1in} >{\centering\arraybackslash}m{1in} >{\centering\arraybackslash}m{1in} >{\centering\arraybackslash}m{1in}}  
\hline  
Particle & $a$ & $b$ & $\chi^{2}/NDF$ & $m$ & $\chi^{2}/NDF$ \\  
\hline \hline  
$\pi^{+}-\pi^{-}$ &$ -0.064 \pm 0.025 $&$ 1.068 \pm 0.106 $& 22.5 / 4 &$ -0.0155 \pm 0.0005 $& 22.8 / 5 \\  
$K^{+}-K^{-}$&$ 3.219 \pm 10.068 $&$ 3.104 \pm 1.4440 $& 0.4 / 4 &$ 0.0018 \pm 0.0017 $& 7.3 / 5 \\  
$p-\bar{p}$ &$ 0.209 \pm 0.099 $&$ 0.9329 \pm 0.143 $& 1.8 / 4 &$ 0.0831 \pm 0.0039 $& 2.0 / 5 \\  
$\Lambda-\overline{\Lambda}$ &$ 0.177 \pm 0.086 $&$ 0.896 \pm 0.139 $& 0.6 / 4 &$ 0.0794 \pm 0.0040 $& 0.7 / 5 \\  
$\Xi^{-}-\overline{\Xi}^{+}$ &$ 7.363 \pm 18.997 $&$ 2.072 \pm 0.825 $& 1.0 / 3 &$ 0.0607 \pm 0.0210 $& 2.5 / 4 \\  
\hline  
\end{tabular}  
\end{table*}  

In this sub-section, the energy dependence of the $v_{2}$ difference between particles $X$ ($p$, $\Lambda$, $\Xi^{-}$, $\pi^{+}$, $K^{+}$) and anti-particles $\overline{X}$ ($\bar{p}$, $\overline{\Lambda}$, $\overline{\Xi}^{+}$, $\pi^{-}$, $K^{-}$) is studied. Figure~\ref{fDiff_v2_sNN_muB} a) shows a fit to the $\Delta v_{2}(p_{T})$ values from Figs.~\ref{fparticles_pion_kaon}, \ref{fparticles_proton_lambda}, and \ref{fparticles_Xi_Omega}. This difference is denoted in the following as $v_{2}(X)-v_{2}(\overline{X})$ and is shown as a function of the beam energy $\sqrt{s_{NN}}$. At 62.4 GeV, the $v_{2}$ difference for mesons is close to zero, whereas the baryons show a difference of 0.003 to 0.005. The difference increases for all particle species as the energy decreases. It reaches values of about 0.03 for $\Lambda$ and protons and 0.004-0.005 for kaons and pions at 7.7 GeV. The baryons show a steeper rise compared to the mesons. The pions and kaons show a similar trend, but opposite with respect to their charge. Also, the protons and $\Lambda$ are very similar at all energies. Compared to the protons and $\Lambda$, the $\Xi$ show a slightly smaller difference at higher energies, but a larger difference at lower energies. One should note that the $\Xi$ result at 11.5 GeV covers a much smaller $p_{T}$ range compared to all of the other data points. This could cause additional systematic effects which are not included in the error bars. The difference in $v_{2}(\sqrt{s_{NN}})$ shown in Fig.~\ref{fDiff_v2_sNN_muB} a) was parametrized with:
\begin{equation}
f_{\Delta v_{2}}(\sqrt{s_{NN}}) = a \cdot s_{NN}^{-b/2}.
\label{fDelta_v2}
\end{equation}
The fit results of the parameters {\it a} and {\it b} are listed in Table~\ref{table_2}.  

In Fig.~\ref{fDiff_v2_sNN_muB} b), the $v_{2}$ difference is shown as a function of the baryonic chemical potential, $\mu_{B}$. A parametrization from~\cite{Tiwari:2011km} was used to determine the $\mu_{B}$ values for each beam energy. Since this parametrization was done for the most central collisions, a correction has to be applied to take into account the difference to the minimum bias collisions. To do this, the measured centrality dependence of $\mu_{B}$ for Au+Au collisions at $\sqrt{s_{NN}}$ = 62.4 and 200 GeV from~\cite{Abelev:2008ab} was used. The ratio between the mean $\mu_{B}$ over all centralities, which is an approximation for the 0--80\% central $\mu_{B}$ values, and the most central values is $0.83\pm0.06$ for 62.4 GeV and $0.84\pm0.14$ for 200 GeV. The minimum bias $\mu_{B}$ values were calculated for all energies by multiplying this factor with the obtained values from the parametrization under the assumption that these ratios do not change with energy. The resulting $\mu_{B}$ values with errors are shown in Fig.~\ref{fDiff_v2_sNN_muB} b). 

Each particle data set was fitted with a straight line that passes through the origin; the slope parameters, $m$, are listed in Table~\ref{table_2}. A linear increase of the $v_{2}$ difference with $\mu_{B}$ is observed for all particle species from 62.4 GeV down to 7.7 GeV. Only at 11.5 GeV a ~2$\sigma$ deviation for the $\Xi$ and at 7.7 GeV a deviation for the kaons was found. This linear scaling behavior suggests that the baryon chemical potential is directly connected to the difference in $v_{2}$ between particles and anti-particles.

\section{Discussion}
\label{sec_discuss}
Comparisons of the data to transport and other models are described.
\subsection{Transport model comparisons}  
\label{sec_model}
\begin{figure*}[]
\centering
\resizebox{12.5cm}{!}{%
  \includegraphics{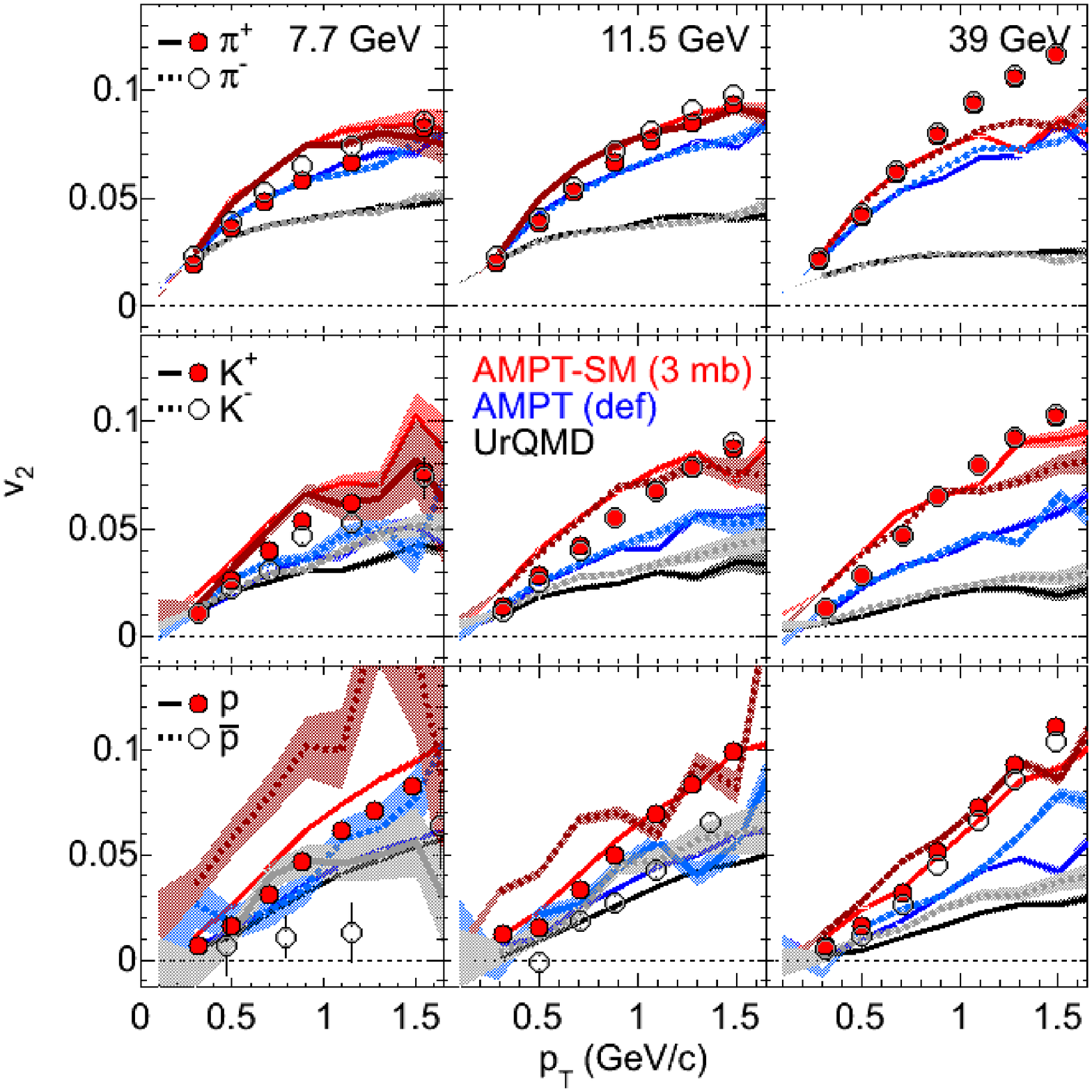}}
\caption{(Color online) The elliptic flow, $v_{2}$, of $\pi^{\pm}$, $K^{\pm}$, $p$ and $\bar{p}$ as a function of the transverse momentum, $p_{T}$, for 0--80\% central Au+Au collisions at $\sqrt{s_{NN}}=$ 7.7, 11.5 and 39 GeV. The symbols depict the data while the lines show the model results from AMPT with default settings (blue), AMPT with the string melting (SM) option and a hadronic cross section of 3 mb (red), and from UrQMD (black). The solid and dashed lines represent positively and negatively charged particles, respectively.}
\label{fModel_v2}       
\end{figure*}

In Fig.~\ref{fModel_v2}, the measured elliptic flow of $\pi^{\pm}$, $K^{\pm}$, $p$ and $\bar{p}$ for 0--80\% central Au+Au collisions at $\sqrt{s_{NN}}$ = 7.7, 11.5 and 39 GeV is compared with model calculations. The models used were UrQMD (Ultra-relativistic Quantum Molecular Dynamics), version 2.3~\cite{urqmd}, and AMPT (A Multi-Phase Transport), version 1.11~\cite{ampt}.  In order to be consistent with the analysis of the data, the number of charged particles within the pseudorapidity range $|\eta| <0.5 $ was used for the centrality definition in the model calculations. The $v_{2}$ values of the (anti-)particles were calculated relative to the true event plane. The difference between the true event (or participant) and the reaction plane (as inferred in the data) could bias the results as discussed in Ref.~\cite{Steinheimer:2012bn}. In total, about one million events were generated from each model at each energy.

The UrQMD model~\cite{urqmd} is based on a microscopic transport theory where the phase-space description of the collisions plays the central role. It allows for the covariant propagation of all hadrons on classical trajectories including stochastic binary scattering, color string formation, and resonance decay. This model includes more than fifty baryon and forty-five meson species and incorporates baryon-baryon, meson-baryon and meson-meson interactions. A comparison of the data with the UrQMD model can provide information about the contributions to the elliptic flow from the hadronic phase.
 
The AMPT model~\cite{ampt} has Glauber-based initial conditions which are the same as those used in the HIJING (Heavy Ion Jet Interaction Generator)~\cite{hijing} model. In this model, mini-jet partons are created and scatter before they fragment into hadrons. The String Melting version (AMPT-SM) of the AMPT model is based on the idea that, for energy densities beyond a critical value of about 1 GeV/fm$^{3}$, strings and partons cannot coexist. Therefore, the strings are melted into partons by converting the hadrons into their valence quarks. The Zhang's Parton Cascade (ZPC) model~\cite{ampt} was used to describe the scattering between the quarks. Once the interactions have stopped, the partons hadronize through the mechanism of parton coalescence. In the default AMPT model, partons are recombined with their parent string when they stop interacting, and the resulting strings are converted to hadrons using the Lund string fragmentation model. The interactions between the mini-jet partons in the default AMPT model and those between partons in the AMPT-SM model could give rise to substantial elliptic flow. The AMPT-SM calculations would thus indicate the contributions to the measured $v_{2}$ from the partonic interactions. The parton-parton interaction cross section in the string-melting version of the AMPT model was taken to be 3 mb.
  
The $v_{2}(p_{T})$ values obtained from all of these models were nearly identical for $\pi^{+}$ and $\pi^{-}$, and $K^{+}$ and $K^{-}$, respectively. Only the anti-protons, compared to the protons, showed a significantly larger $v_{2}(p_{T})$ in the UrQMD and AMPT-SM models. This is in clear contradiction to the observations from the data described here. The UrQMD model generally under-predicts the $v_{2}(p_{T})$ values. Only at 7.7 and 11.5 GeV are the anti-proton $v_{2}(p_{T})$ values close to or below the UrQMD values. As was pointed out above, a purely hadronic system (as described by the UrQMD model) does not appear to explain the relatively large flow of the particles at these energies. As seen in Fig.~\ref{fModel_v2}, the AMPT-SM model provides the best description of the data, except for $p$ and $\bar{p}$ at 7.7 GeV. In all other cases, the AMPT default calculations, and more so the UrQMD calculations, under-predict the $v_{2}(p_{T})$ values. 

\subsection{Interpretations from models} 
\label{sec_int_from_others}

Several interpretations have been suggested for the possible physical causes for the difference in the $v_{2}$ values for particles and their corresponding anti-particles based on  preliminary results. The process involved was to create or modify a model to qualitatively describe the difference in the $v_{2}$ values between particles and corresponding anti-particles that is shown in Fig.~\ref{fDiff_v2_sNN_muB}. In Ref.~\cite{Dunlop:2011cf}, it was argued that the effect results from quark transport from the projectile nucleons to mid-rapidity. The authors assumed that the elliptic flow of transported quarks is larger than that from produced quarks. Thus, the asymmetry of quarks and anti-quarks in the particles and corresponding anti-particles leads to a systematically larger flow of the particles compared to the anti-particles. The energy dependence was explained by the increase of nuclear stopping in heavy ion collisions with decreasing energy. The resulting patterns for $\pi$, $K$, $p$ and $\Lambda$ is qualitatively in agreement with the data. However, a similar difference in $v_{2}$ for mesons ($\pi^{\pm}$, $K^{\pm}$) and for baryons (($p$,$\bar{p}$), ($\Lambda,\overline{\Lambda}$), and ($\Xi^{-}$, $\overline{\Xi}^{+}$)) is observed which is not expected in this picture.

In Ref.~\cite{Xu:2012gf}, an AMPT model calculation for $\sqrt{s_{NN}}=$ 7.7, 11.5 and 39 GeV was presented. The authors included mean-field potentials in the hadronic stage of that model. As a consequence of these potentials, particles like $K^{-}$ and $\bar{p}$ are attracted by the hadronic matter and are trapped in the system whereas $K^{+}$ and protons feel a repulsive force and have the tendency to leave the system along the participant plane. The observed pattern shown in Fig.~\ref{fDiff_v2_sNN_muB} cannot be explained by a default AMPT calculation without hadronic potentials, as discussed in Section~\ref{sec_model}. With the potentials included, a fair qualitative agreement was achieved. However, the difference in $v_{2}$ between $K^{+}$ and $K^{-}$ in the calculation is close to the difference for $p$ and $\bar{p}$, in clear contradiction to the present experimental results. The authors noted that further investigations are important to understand these effects in more detail.

Similar studies were performed for the data collected by the KaoS collaboration at SIS (Schwerionensynchroton at GSI) at energies of 1--2 AGeV for $K^{\pm}$ mesons~\cite{Uhlig:2004ue}. In this case, the IQMD (Isospin Quantum Molecular Dynamics) transport model was used for the comparison to the KaoS results. The trends for the data and model calculations observed at those (very low) energies are opposite to those reported in this paper.

In Ref.~\cite{Steinheimer:2012bn}, a hybrid (hydrodynamical plus UrQMD) calculation was performed. Qualitatively, the trend for $\Delta v_{2}$ can be described for protons, $\Lambda$, and $\pi$, whereas the trend for kaons is opposite to the present observations. The effect for the protons primarily resulted from the treatment of a non-zero net baryon number density and chemical
potential. The results are slightly changed by using the UrQMD afterburner which describes the final stage interactions. Another effect discussed in this paper~\cite{Steinheimer:2012bn} is related to the event plane calculation. It was claimed that fluctuations in this calculation can bias the event plane to be rotated towards the most abundantly produced particles. This would, for example, increase the $v_{2}$ values for protons and reduce them for anti-protons.

\begin{figure}[]
\resizebox{8cm}{!}{%
  \includegraphics[bb = 12.385054 3.492070 487.877962 458.879963,clip]{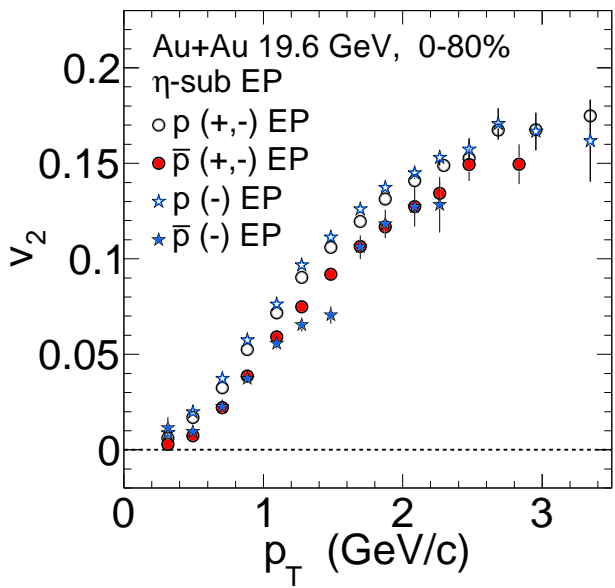}}
\caption{(Color online) The proton and anti-proton elliptic flow for 0--80\% central Au+Au collisions at $\sqrt{s_{NN}}$ = 19.6 GeV, where ``(+,-) EP" refers to the event plane reconstructed using all of the charged particles, and ``(-) EP" refers to the event plane reconstructed using only the negatively charged particles. The error bars are statistical only.}
\label{fProton_v2_negEP}       
\end{figure}

In Fig.~\ref{fProton_v2_negEP}, a study to explore this possibility is presented.  The elliptic flow for protons and anti-protons as a function of $p_{T}$, for 0--80\% central Au+Au collisions at $\sqrt{s_{NN}}$ = 19.6 GeV is shown for two different kinds of reconstructed event planes. The event plane reconstructed using all of the charged particles is denoted by
(+,-) EP, while the event plane reconstructed using only the negatively-charged particles is denoted by (-) EP. 
The $v_{2}(p_{T})$ values for protons using the (-) EP method are slightly, but systematically larger than those from 
the standard (+,-) EP method. The anti-proton $v_{2}(p_{T})$ values are essentially unchanged. A reduced $v_{2}(p_{T})$ value would be expected for protons if such baryon number fluctuations caused such a bias. The increased $v_{2}(p_{T})$ for protons may be due to non-flow. For example, resonance decays could cause a larger change in non-flow contributions to proton $v_{2}$ than to anti-proton $v_{2}$ between the two different event planes. More detailed studies from theory and experiment are needed to investigate the event-by-event baryon fluctuations and their possible effects on the event plane reconstruction.

A recent calculation based on the Nambu-Jona-Lasinio (NJL) model can also qualitatively explain the differences between $p$-$\bar{p}$, $\Lambda$-$\overline{\Lambda}$, and $K^{+}$-$K^{-}$ using the vector mean field potential~\cite{Song:2012cd}. The vector potential is repulsive for quarks and attractive for anti-quarks, which results in different flow patterns. To calculate the flow for the hadrons, a coalescence model was used.

A different approach was followed in Ref.~\cite{Greco:2012hh} by assuming simplified rapidity distributions for u- and d-quarks that are different from those for $s$, $\overline{\rm u}$,  $\overline{\rm d}$, and $\overline{\rm s}$ quarks.
It is claimed that under these initial conditions a breakdown of the $v_{2}$ NCQ scaling would not necessarily be a consequence of a phase transition, but rather the result of the different rapidity distributions of the valence and produced quarks. Ref.~\cite{Greco:2012hh} also notes that the model results in a difference between particles and anti-particles that is opposite to that presented here.

\subsection{Conclusions}
\label{sec_conclude}

The strong energy dependence of the difference in $v_{2}(p_{T})$ between particles and their corresponding anti-particles 
is a new observation in the field of relativistic heavy ion collisions. It cannot be explained in a purely hydrodynamic approach 
since particles and anti-particles have the same mass. It is also incompatible with a scenario where the flow is only produced 
in a thermalized and equilibrated QGP without any additional quark potentials. Other effects, such as hadronic interactions, or the scenarios involving transported quarks that were discussed in Section~\ref{sec_int_from_others}, could be responsible for the present observations.
However, the agreement of the data with the transport based models is at present only qualitative. 
The energy dependence of $v_{2}(X)-v_{2}(\overline{X})$ suggests a strong dependence on the values of the baryon chemical potential $\mu_{B}$.

The NCQ scaling was observed at $\sqrt{s_{NN}}=$ 200 GeV for all particles and anti-particles as they have the same number 
of quarks. The observed breakdown of such a scaling with decreasing energy could be interpreted as the emerging dominance of hadronic interactions over partonic interactions in the systems formed in the collisions.  The observed difference in the $v_2$ values demonstrates that the particles and anti-particles are no longer consistent with a single NCQ scaling law. The additional splitting between the particles and corresponding anti-particles at the lower beam energies breaks NCQ scaling. Even amongst the particles and 
anti-particles separately, an absence of the baryon-meson splitting is observed at  $\sqrt{s_{NN}}=$ 7.7  and 11.5 GeV in the representation $v_{2}(m_{T}-m_{0})$ 
for ($m_{T}-m_{0}$) $>$ 1 GeV/$c^2$. However, the corresponding NCQ scaling shows no significant deviation from the scaling in the appropriate intermediate $p_{T}$ range.

It is observed that $\phi$ mesons at $\sqrt{s_{NN}}=$ 7.7 and 11.5 GeV indicate a different trend at the highest $p_{T}$ values. This would be in agreement with the picture that the $\phi$ mesons have a lower $v_{2}(p_{T})$ in a hadronic environment compared to other hadrons due to their lower hadronic cross section~\cite{Shor:1984ui,Sibirtsev:2006yk,vanHecke:1998yu,Cheng:2003as}. 
Larger event samples are needed at these energies in order to make more quantitative conclusions.  The corresponding anti-particles show a similar NCQ scaling trend as the particles at energies larger than $\sqrt{s_{NN}}=$ 11.5 GeV. At $\sqrt{s_{NN}}=$ 7.7 and 11.5 GeV, the event sample sizes for most of the anti-baryons need to be increased in order to make quantitative statements on the validity of NCQ scaling.

At energies larger than $\sqrt{s_{NN}}=$ 11.5 GeV, NCQ scaling holds independently for particles and anti-particles, while at lower energies significant differences appear. The strong increase of the difference in $v_{2}$ between the particles and corresponding anti-particles with decreasing energy warrants further experimental and theoretical investigation.

\section{Summary}
\label{sec_summary}
Results on the mid-rapidity elliptic flow $v_{2}(p_{T})$ for $\pi^{\pm}$, $K^{\pm}$, $K_{s}^{0}$, $p$, $\bar{p}$, $\phi$, $\Lambda$, $\overline{\Lambda}$, $\Xi^{-}$, $\overline{\Xi}^{+}$, $\Omega^{-}$ and $\overline{\Omega}^{+}$ from Au+Au collisions at $\sqrt{s_{NN}}=$ 7.7, 11.5, 19.6, 27, 39 and 62.4 GeV were presented. For all of the particle species, $v_{2}$ increases with increasing energy at high transverse momenta, whereas $v_{2}$ at low $p_{T}$ values depend on the particle species. A significant difference in  $v_{2}(p_{T})$  between the particles and corresponding anti-particles was observed. At energies above 39 GeV, the difference was approximately constant with energy, while the difference increased as the energy decreased. Hence, a significant dependence of $v_{2}(X)-v_{2}(\overline{X})$ on the baryon chemical potential, $\mu_{B}$, is indicated. The difference $v_{2}(X)-v_{2}(\overline{X})$ was larger for baryons than for mesons. This difference cannot be reproduced by transport models in their standard configuration. Only the AMPT model with an included hadronic potential shows a similar pattern. Other models show the same qualitative trends. 

The NCQ scaling that was observed for all particles and anti-particles at $\sqrt{s_{NN}}=$ 200 GeV, no longer holds at the lower beam
energies of $\sqrt{s_{NN}}=$ 11.5 and 7.7 GeV. This is seen as an increase of $v_{2}(X)-v_{2}(\overline{X})$ with decreasing beam energy.
The baryon-meson splitting of $v_{2}(m_{T}-m_{0})$ for ($m_{T}-m_{0}$) $>$ 1 GeV/$c^2$, which formed the basis of NCQ scaling observation at 200 GeV, was not observed for  anti-particles at the lower energies. In the representation
of $v_{2}(m_{T}-m_{0})$/$n_{q}$ {\it vs.} ($m_{T}-m_{0}$)/$n_{q}$, no significant deviations from NCQ scaling were observed for particles and anti-particles separately at energies above $\sqrt{s_{NN}}=$ 11.5 GeV.  At $\sqrt{s_{NN}}=$ 7.7 and 11.5 GeV, the anti-baryons and the $\phi$ meson indicate a different trend. At the highest ($m_{T}-m_{0}$) data points at $\sqrt{s_{NN}}$ = 7.7 and 11.5 GeV, these particles deviate from the other hadrons by 1.8$\sigma$ and 2.3$\sigma$, respectively.

\section{Acknowledgements}
We thank the RHIC Operations Group and RCF at BNL, the NERSC Center at LBNL and the Open Science Grid consortium for providing resources and support. This work was supported in part by the Offices of NP and HEP within the U.S. DOE Office of Science, the U.S. NSF, the Sloan Foundation, the DFG cluster of excellence ‘Origin and Structure of the Universe’ of Germany, CNRS/IN2P3, FAPESP CNPq of Brazil, Ministry of Ed. and Sci. of the Russian Federation, NNSFC, CAS, MoST, and MoE of China, GA and MSMT of the Czech Republic, FOM and NWO of the Netherlands, DAE, DST, and CSIR of India, Polish Ministry of Sci. and Higher Ed., Korea Research Foundation, Ministry of Sci., Ed. and Sports of the Rep. of Croatia, and RosAtom of Russia, and VEGA of Slovakia.
\clearpage

\onecolumngrid
\section*{Appendix: NCQ scaling of $v_{2}(p_{T})$}

\begin{figure*}[h!]
\resizebox{13.0cm}{!}{%
\includegraphics[bb = 13 4 505 301,clip]{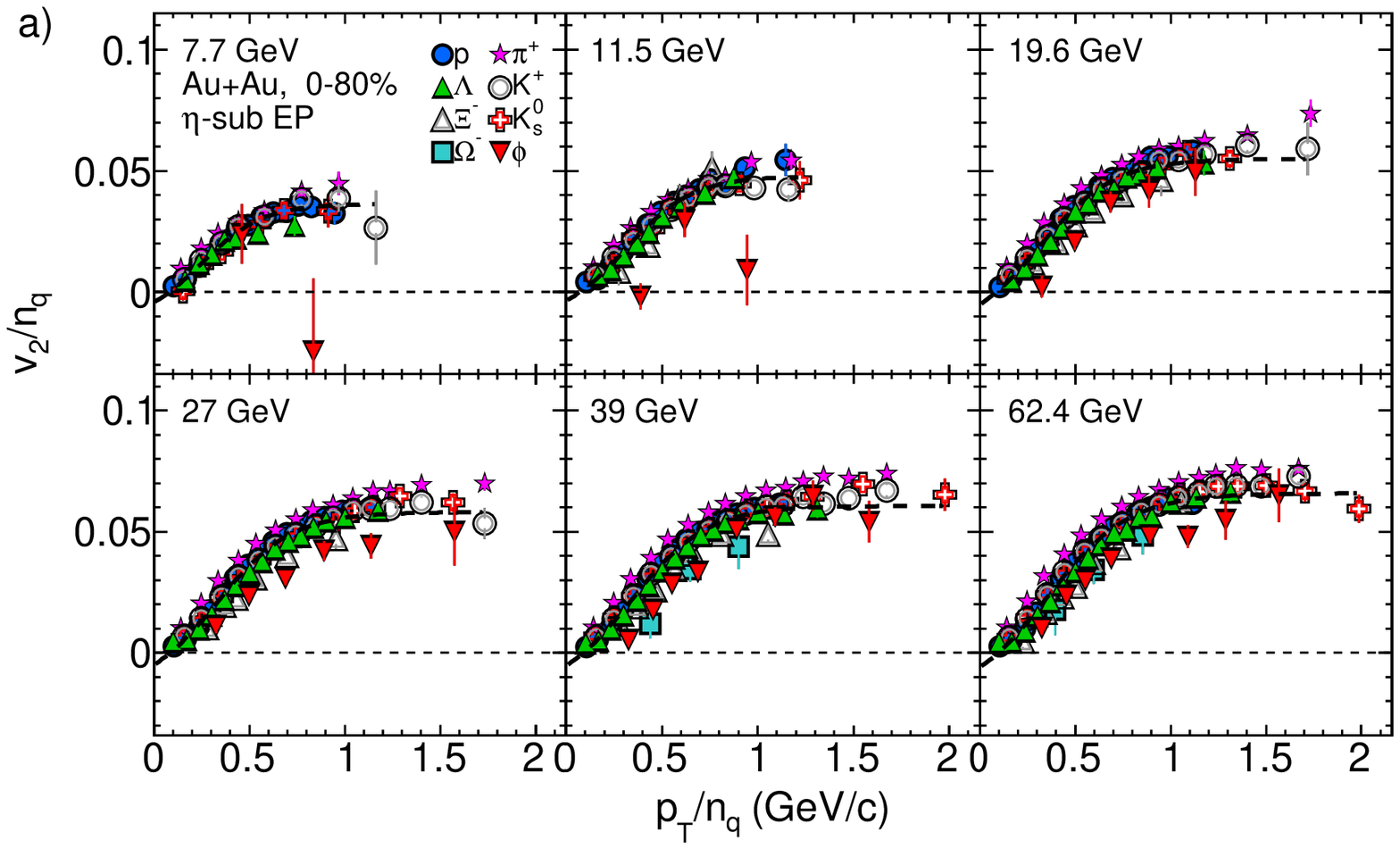}} \\[2ex]
\resizebox{13.0cm}{!}{%
\includegraphics[bb = 13 4 505 301,clip]{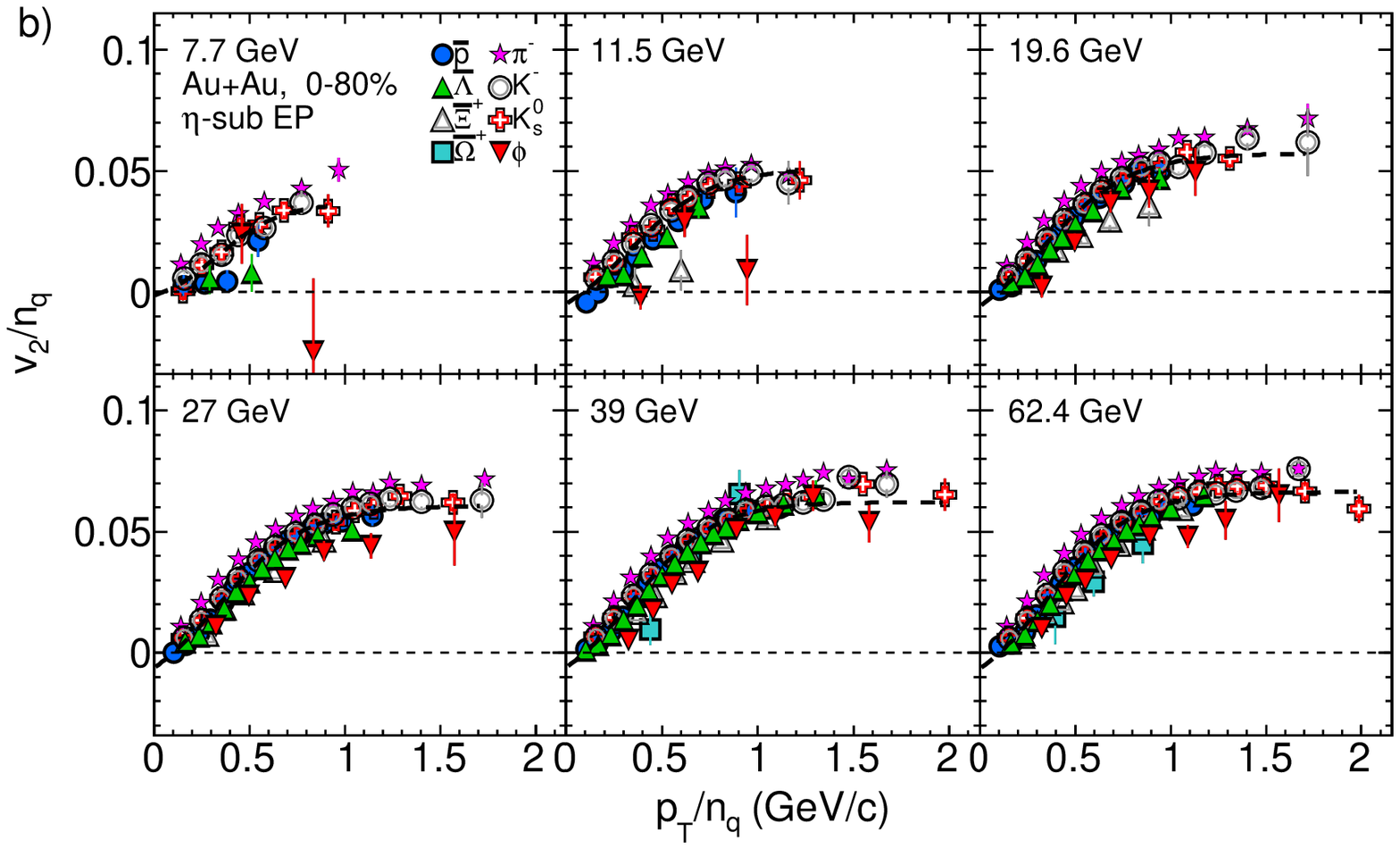}} %
 \caption{(Color online) The number-of-Constituent Quark (NCQ) scaled elliptic flow, $v_{2}/n_{q}$ versus $p_{T}/n_{q}$, for 0--80\% central Au+Au collisions for selected particles a) and corresponding anti-particles b). The dashed lines show the results of simultaneous fits with Eq.~(\ref{Func_v2_fit}) to all particles except the pions.}
 \label{fNCQ_pT}
\end{figure*}

The NCQ scaling of $v_{2}$ is shown in Fig.~\ref{fNCQ_pT}. Plotted there is $v_{2}/n_{q}$ versus
the scaled transverse momentum, $p_{T}/n_{q}$, where $n_q$ is the number of quarks in the particle. 
The same data is shown versus $(m_{T}-m_{0})/n_{q}$ in Fig.~\ref{fNCQ_mT}.
There is a wider variation of the scaled $v_{2}$ when plotted versus $p_{T}/n_{q}$ as compared
to $(m_{T}-m_{0})/n_{q}$.

\begin{figure*}[]
\resizebox{13.0cm}{!}{%
\includegraphics[bb = 13 4 505 301,clip]{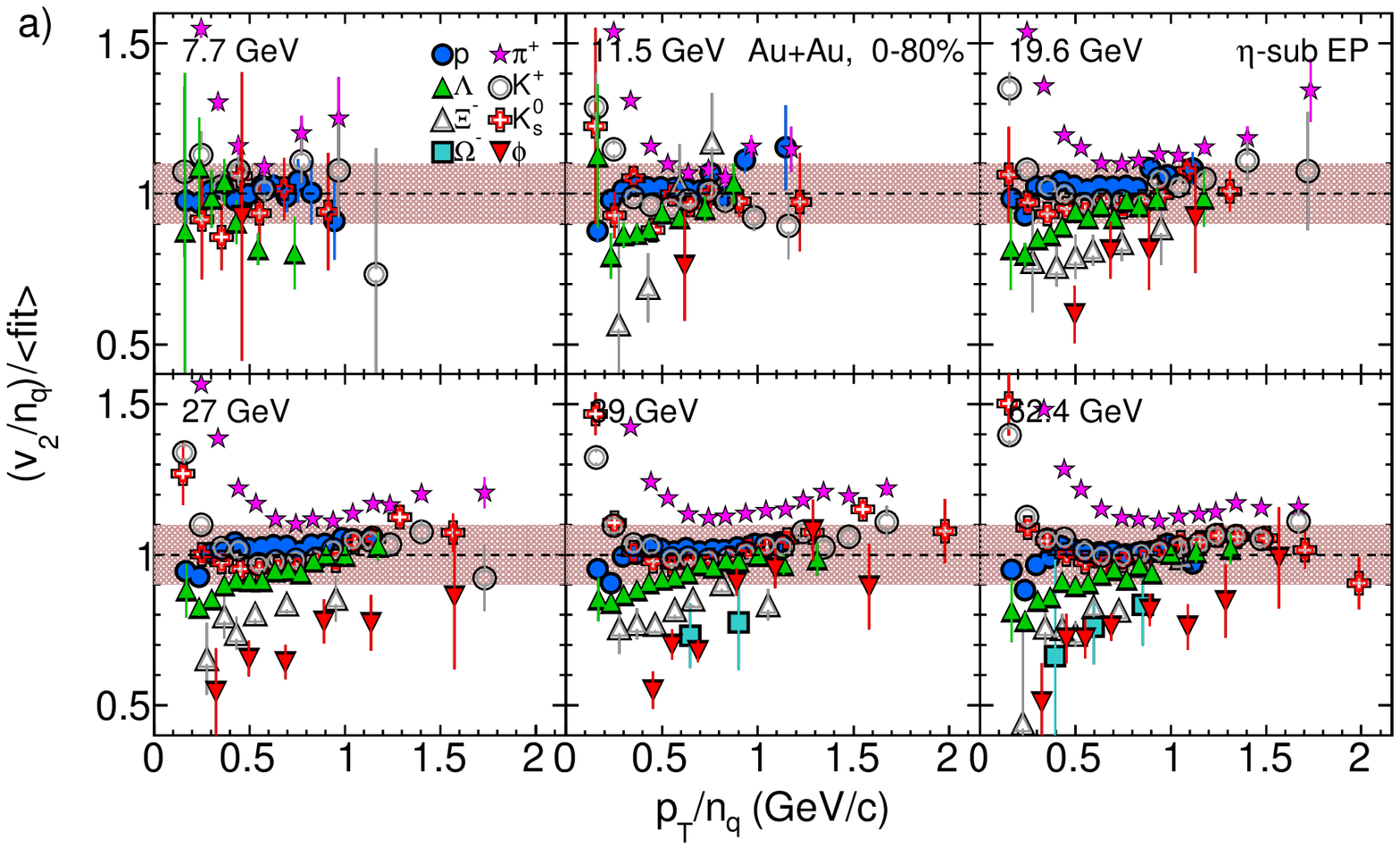}} \\[2ex]
\resizebox{13.0cm}{!}{%
\includegraphics[bb = 13 4 505 301,clip]{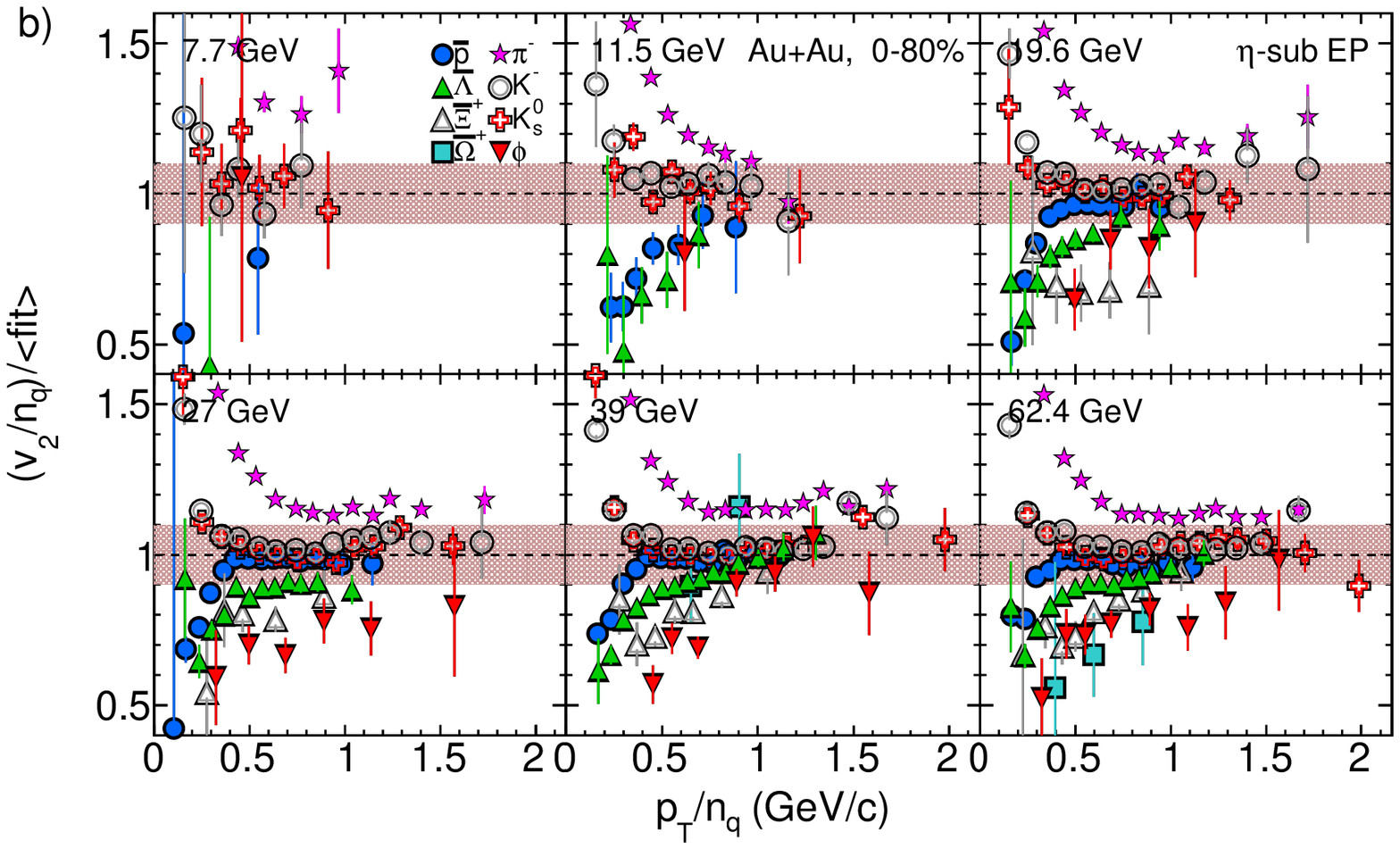}} %
 \caption{(Color online)  The Number-of-Constituent Quark (NCQ) scaled elliptic flow, $v_{2}(p_{T})/n_{q}$, ratio to a fit function (see text) for 0--80\% central Au+Au collisions for selected particles a) and corresponding anti-particles b). Most of the data points at the larger $p_{T}/n_{q}$ values are consistent with unity to $\pm$10\%, which is shown as the shaded areas to guide the eye. Some of the data points for $\phi$ and $\overline{\Xi}^{+}$ are outside of the plot range.}
 \label{fNCQ_pT_ratio}
\end{figure*}

Simultaneous fits to all of the $v_{2}/n_{q}$ versus $p_{T}/n_{q}$ values, except for those 
for the pions, were performed. In similarity to Fig.~\ref{fNCQ_mT_ratio}, shown in 
Fig.~\ref{fNCQ_pT_ratio} is the ratio of the data points in Fig.~\ref{fNCQ_pT} to 
the simultaneous fits as a function of $p_{T}/n_{q}$. Most of the data points are 
within 10\% of the fit function at $p_{T}/n_{q}$ values larger than 1 GeV/$c$.  
At lower momenta, the $v_{2}/n_{q}$ versus $p_{T}/n_{q}$ values diverge due to the mass splitting
that was shown in Fig.~\ref{fv2_vs_pT}.

\clearpage

\twocolumngrid


\end{document}